\documentclass[]{aa}
\usepackage[nolist]{acronym}
\usepackage[mathlines]{lineno}

\usepackage{graphicx}
\usepackage{txfonts}
\usepackage{natbib,twoopt}
\bibpunct{(}{)}{;}{a}{}{,} 
\usepackage{hyperref}
\usepackage{nameref}
\usepackage{cleveref}
\Crefname{figure}{Fig.}{Figs.}
\Crefname{figure}{Fig.}{Figs.}
\usepackage{nccmath}
\usepackage{subcaption}
\usepackage{siunitx}
\usepackage{booktabs} 
\usepackage{multirow} 
\usepackage{threeparttable} 
\usepackage{tabularx} 
\usepackage{array} 
\usepackage{bm} 
\usepackage{float}
\usepackage[hyphenbreaks]{breakurl}
\usepackage{scalerel}
\usepackage{tikz}
\usetikzlibrary{svg.path}

\definecolor{orcidlogocol}{HTML}{A6CE39}
\tikzset{
  orcidlogo/.pic={
    \fill[orcidlogocol] svg{M256,128c0,70.7-57.3,128-128,128C57.3,256,0,198.7,0,128C0,57.3,57.3,0,128,0C198.7,0,256,57.3,256,128z};
    \fill[white] svg{M86.3,186.2H70.9V79.1h15.4v48.4V186.2z}
                 svg{M108.9,79.1h41.6c39.6,0,57,28.3,57,53.6c0,27.5-21.5,53.6-56.8,53.6h-41.8V79.1z M124.3,172.4h24.5c34.9,0,42.9-26.5,42.9-39.7c0-21.5-13.7-39.7-43.7-39.7h-23.7V172.4z}
                 svg{M88.7,56.8c0,5.5-4.5,10.1-10.1,10.1c-5.6,0-10.1-4.6-10.1-10.1c0-5.6,4.5-10.1,10.1-10.1C84.2,46.7,88.7,51.3,88.7,56.8z};
  }
}

\newcommand\orcidicon[1]{\href{https://orcid.org/#1}{\mbox{\scalerel*{
\begin{tikzpicture}[yscale=-1,transform shape]
\pic{orcidlogo};
\end{tikzpicture}
}{|}}}}

\newcommand{\orcididDario}{0000-0001-9282-9462}
\newcommand{\orcididIgnas}{0000-0003-1624-3667}
\newcommand{\orcididSam}{0000-0003-4760-6168}
\newcommand{\orcidNatalie}{0009-0009-6634-1741}
\newcommand{\orcididJB}{0000-0003-2233-4821}

\definecolor{cobalt}{rgb}{0.06, 0.2, 0.65}
\hypersetup{
  colorlinks,
  citecolor=cobalt,
  linkcolor=[rgb]{0.8, 0.2, 1.0},
  urlcolor=cobalt,
}

\definecolor{lightgray}{gray}{0.95} 

\makeatletter
\renewcommand*\aa@pageof{, page \thepage{} of \pageref*{LastPage}}

\newcommand{\teff}{$T_{\rm eff}$}
\newcommand{\logg}{\ensuremath{\log g}}

\def\kms{\ifmmode{\rm km\th s^{-1}}\else km\th s$^{-1}$\fi}
\def\th{\thinspace}
\newcommand{\Mjup}{M$_{\text{Jup}}$}
\newcommand{\Rjup}{R$_{\text{Jup}}$}

\newcommand{\thirteenCO}{\textsuperscript{13}CO}
\newcommand{\thirteenC}{\textsuperscript{13}C}

\newcommand{\CeighteenO}{C\textsuperscript{18}O}
\newcommand{\CseventeenO}{C\textsuperscript{17}O}

\newcommand{\eighteenO}{\textsuperscript{18}O}

\newcommand{\pRT}{\texttt{petitRADTRANS}}
\newcommand{\ultranest}{\texttt{ultranest}}
\newcommand{\micron}{$\mu$m}

\newcommand{\Teff}{$T_{\text{eff}}$}

\newcommand{\Cratio}{\textsuperscript{12}C/\textsuperscript{13}C}
\newcommand{\Oratio}{\textsuperscript{16}O/\textsuperscript{18}O}

\newcommand{\water}{H$_2$O}
\newcommand{\methane}{CH$_4$}
\newcommand{\COtwo}{CO$_2$}

\newcommand{\eighteenOwater}{H\textsubscript{2}\textsuperscript{18}O}

\newcommand{\isomethane}{\textsuperscript{13}CH\textsubscript{4}}
\newcommand{\isoCOtwo}{\textsuperscript{13}CO\textsubscript{2}}
\newcommand{\thirteenmethane}{\textsuperscript{13}CH\textsubscript{4}}
\newcommand{\thirteenCOtwo}{\textsuperscript{13}CO\textsubscript{2}}
\newcommand{\dmethane}{CH\textsubscript{3}D}

\newcommand{\twelveCOtwo}{\textsuperscript{12}CO\textsubscript{2}}

\newcommand{\pyrox}{\texttt{pyROX}}

\newcommand{\leiden}{Leiden Observatory, Leiden University, P.O. Box 9513, 2300 RA, Leiden, The Netherlands}

\newcommand{\ucsd}{Department of Astronomy \& Astrophysics, University of California, San Diego, La Jolla, CA 92093, USA}
\newcommandtwoopt{\citeads}[3][][]{\href{http://adsabs.harvard.edu/abs/#3}%
{\def\hyper@linkstart##1##2{}%
    \let\hyper@linkend\@empty\citealp[#1][#2]{#3}}}
\newcommandtwoopt{\citepads}[3][][]{\href{http://adsabs.harvard.edu/abs/#3}%
{\def\hyper@linkstart##1##2{}%
    \let\hyper@linkend\@empty\citep[#1][#2]{#3}}}

\makeatother

\begin{document} 

\title{JWST high-contrast spectroscopy with speckle modelling: Atmospheric retrievals of the T dwarf companion HD 19467 B}

\titlerunning{JWST high-contrast spectroscopy with speckle modelling}
\author{D. Gonz\'{a}lez Picos       \inst{\ref{leiden}\orcidicon{\orcididDario}}
           \and T. van der Post \inst{\ref{leiden}}
           \and S. de Regt          \inst{\ref{leiden}\orcidicon{\orcididSam}}
           \and J.-B. Ruffio        \inst{\ref{ucsd}\orcidicon{\orcididJB}}
           \and N. Grasser          \inst{\ref{leiden}\orcidicon{\orcidNatalie}}
           \and I.A.G. Snellen      \inst{\ref{leiden}\orcidicon{\orcididIgnas}}
           }

\institute{
    \leiden\\ \email{picos@strw.leidenuniv.nl}\label{leiden}
    \and
    \ucsd\label{ucsd}
}

   \date{Received YYYY-MM-DD; accepted YYYY-MM-DD}

\abstract
{High-contrast, medium-resolution spectroscopy with JWST can resolve molecular and isotopic features in cool substellar atmospheres, but for close-in companions the extracted spectra can be biased by wavelength-dependent residual stellar contamination.}
{We assess the impact of residual speckles on atmospheric inference for the T-dwarf companion HD~19467~B and compare the results to the field T dwarf 2MASS~J0415$-$0935.}
{We analyse JWST/NIRSpec G395H spectra (2.87--5.2~\micron; \(R\sim2700\)) and perform Bayesian atmospheric retrievals with \pRT{} coupled to nested sampling using \texttt{ultranest}. We use a flexible, parameterised pressure--temperature profile with free, constant-with-altitude molecular abundances. For HD~19467~B we fit the PSF-subtracted spectrum with a linear model that includes the atmospheric model and a set of speckle spectra from the integral field unit.}
{We detect \water, \methane, CO, \COtwo, and NH$_3$ in both atmospheres and measure carbon isotopic ratios from CO isotopologues, finding \Cratio$=154^{+19}_{-17}$ for HD~19467~B and \Cratio$=85\pm5$ for 2MASS~J0415$-$0935. Speckle contamination primarily affects the low-frequency spectral shape at 3.0--3.7~\micron\ and can affect retrieved abundances if not accounted for. We obtain seemingly constrained posteriors for some additional species (e.g. SiO and H$_2$S) in some cases, but treat these as tentative because cross-correlation does not yield significant detections; PH$_3$ is not detected in either target.}
{Joint fitting of the atmospheric spectrum and the speckle contamination enables native-resolution retrievals of the high-contrast companion HD 19467 B with JWST/NIRSpec without continuum subtraction. Over 2.87--5.2~\micron, medium-resolution spectroscopy constrains elemental and isotopic composition; both objects exhibit near-solar metallicity and subsolar C/O ratios.}
  
\maketitle
%
\section{Introduction}\label{sec:introduction}

Brown dwarfs, with masses below the hydrogen-burning limit, bridge the gap between giant planets and low-mass stars. As brown dwarfs evolve and cool, their atmospheres become rich in molecules and condensates, which shape their spectra across the M, L, T, and Y classes \citep{kirkpatrickDwarfsCoolerDefinition1999,burrowsTheoryBrownDwarfs2001,cushingInfraredSpectroscopicSequence2005}. The overlap in effective temperatures between brown dwarfs and directly imaged exoplanets makes them a valuable laboratory for the physics and chemistry that govern planetary atmospheres \citep{fahertyPOPULATIONPROPERTIESBROWN2016}.

During the brown dwarf cooling sequence, the L/T transition marks a major change in atmospheric appearance, which is commonly linked to cloud structure and dynamics \citep{burgasserEvidenceCloudDisruption2002,gaoSedimentationEfficiencyCondensation2018}. Older, higher-gravity objects often exhibit largely cloud-free photospheres, with clouds thought to sink to deeper atmospheric layers \citep{allardLimitingEffectsDust2001,saumonEvolutionDwarfsColorMagnitude2008}. Conversely, younger, lower-gravity objects often exhibit cloudier atmospheres, with condensates present at or above the photosphere \citep{ackermanPrecipitatingCondensationClouds2001,burrowsDwarfModelsTransition2006}. A further hallmark is the increase in CH$_4$ at the expense of CO as temperatures fall below \(\sim 1500\)~K (e.g. \citealt{zahnleMethaneCarbonMonoxide2014}). Disequilibrium chemistry can inhibit CO$\rightarrow$CH$_4$ conversion, leaving detectable CO in cool T dwarfs and providing a diagnostic of atmospheric mixing time-scales (e.g. \citealt{fegleyAtmosphericChemistryBrown1996,mosesDisequilibriumCarbonOxygen2011}).

T dwarfs show prominent H$_2$O, CH$_4$, and NH$_3$ absorption in the infrared, with frequent evidence for CO (e.g. \cite{burgasserSpectraDwarfsNearInfrared2002,burrowsDwarfsTheoreticalSpectra2003,calamariAtmosphericRetrievalBrown2022}). The presence of CO in the spectra of T dwarfs indicates disequilibrium chemistry, as was first identified in Jupiter \citep{prinnCarbonMonoxideJupiter1977,bezardCarbonMonoxideJupiter2002}. Historically, most spectroscopy of brown dwarfs was obtained from the ground between 1 and 2.5~$\mu$m. The 3--5~$\mu$m window, which carries critical information on CH$_4$, H$_2$O, NH$_3$, CO, and CO$_2$, is challenging from the ground due to high thermal background, but was accessed by space-based facilities such as Spitzer \citep{cushingInfraredSpectroscopicSequence2005} and AKARI \citep{sorahanaRadiiBrownDwarfs2013}. The advent of JWST has transformed the field of brown dwarf atmospheres. Medium-resolution spectroscopy beyond 3~$\mu$m at unprecedented sensitivity and stability \citep{rigbySciencePerformanceJWST2023} is now possible for objects that are too faint for ground-based facilities. Recent observations of brown dwarfs with NIRSpec and MIRI demonstrate the remarkable data quality achievable at these wavelengths \citep{milesJWSTEarlyreleaseScience2023,hoodHighPrecisionAtmosphericConstraints2024,hochJWSTTSTHighContrast2024}.

Despite JWST’s stability and sensitivity, characterising close-in companions remains challenging due to high contrast and structured point-spread functions. Robust modelling and subtraction of the host-star PSF are essential to unlock medium-resolution spectroscopy with the NIRSpec IFU at small angular separations. Building on moderate-resolution high-contrast methods developed from the ground, \citealt{ruffioJWSTTSTHighContrast2023} demonstrated advanced PSF modelling and companion extraction that improve sensitivity at small to moderate separations (i.e. 300--2000~mas). These strategies are directly relevant to systems such as HD~19467, where high-contrast techniques are essential to obtain spectroscopy of companions at small angular separations \citep{maireOrbitalSpectralCharacterization2020,ruffioJWSTTSTHighContrast2023}. The imperfect subtraction of diffracted starlight can leave a significant imprint in direct-imaging datasets, often referred to as speckles \citep{currieDirectImagingSpectroscopy2023}. A central goal of current high-contrast instrumentation and post-processing is therefore to mitigate and quantify speckle residuals to enable reliable characterisation of faint companions at progressively smaller separations \citep{ruffioDataPostprocessingGain2026}.

The wealth of spectral features in brown dwarf atmospheres enables detailed characterisation of atmospheric composition, structure, and dynamics. The analysis of chemical abundances is key to understanding the formation pathways of brown dwarfs and planets. Elemental ratios and bulk metallicity encode the relative contributions of gas and solids accreted during formation and may be linked to the birth location \citep{obergEFFECTSSNOWLINESPLANETARY2011,madhusudhanExoplanetaryAtmospheresKey2019,molliereInterpretingAtmosphericComposition2022,zhangAtmosphericRegimesTrends2020,fortneyCARBONTOOXYGENRATIOMEASUREMENT2012}. Recent work on carbon, oxygen and sulfur ratios in a multiplanetary system revealed super-stellar ratios akin to Jupiter, suggesting that the enrichment pathways of Solar System gas giants might be applicable to super-Jupiters like the HR 8799 planets \citep{ruffioJupiterlikeUniformMetal2026}.
Isotope ratios have been proposed as complementary tracers of formation pathways \citep{molliereDetectingIsotopologuesExoplanet2019,zhang13COrichAtmosphereYoung2021}. Isotopologues of carbon monoxide such as $^{13}$CO and C$^{18}$O have been measured for a number of brown dwarfs and directly imaged planets \citep{zhang12CO132021,gandhiJWSTMeasurements132023,xuanAreThesePlanets2024,grasserESOSupJupSurvey2025}. These measurements provide valuable insights into birth environment and potential accretion of solids into the atmosphere, and may be used to disentangle formation pathways of planets and brown dwarfs \citep{zhang13COrichAtmosphereYoung2021}. The distinct carbon isotope ratios in the HR 8799 planets \citep{ruffioJupiterlikeUniformMetal2026} suggest that isotopic abundances may vary as a function of birth location within the same system.

The large number of molecular features in the spectra of substellar objects requires detailed modelling of the temperature and chemistry of the atmosphere. Traditionally, grids of forward models incorporating radiative-convective thermochemical equilibrium, condensation, and non-equilibrium chemistry are used to predict spectra for a small set of fundamental parameters \citep{allardModelsVerylowmassStars2012,marleySonoraBrownDwarf2021}. However, these models are limited by the number of free parameters and the assumptions made about the chemistry and physics of the atmosphere. A complementary approach is to use data-driven retrievals to infer molecular abundances and bulk properties directly from the observations, as originally developed for planetary sciences \citep{rodgersInverseMethodsAtmospheric2000,irwinNEMESISPlanetaryAtmosphere2008a}. 

The forward models used in atmospheric retrievals typically incorporate minimal physical constraints, and are able to adjust the temperature structure and composition of the atmosphere to fit the observations (e.g. \cite{lineUniformAtmosphericRetrieval2015,burninghamRetrievalAtmosphericProperties2017,molliereRetrievingScatteringClouds2020a}). Retrievals are typically performed using a Bayesian framework, and the resulting posterior distributions of the free parameters are used to interpret the properties of the atmosphere. The number of free parameters in the retrieval is typically much larger than the number of free parameters in the forward model (e.g. \citealt{regtESOSupJupSurvey2024}). Despite this, atmospheric retrievals have been very successful at fitting the near-infrared spectra of brown dwarfs to high precision, even in situations where additional physical processes were needed to explain the observations (e.g. thermal inversions; \citealt{fahertyMethaneEmissionCool2024}).

\section{Data}
We analyse spectra of two well-studied T dwarfs: the high-contrast companion HD~19467~B and the nearest known isolated T dwarf 2MASS~J0415$-$0935. The comparative analysis of the two objects probes the impact of residual speckles in the 2.9--5.3~\micron{} spectrum of HD~19467~B and highlights modelling limitations for T-dwarf atmospheres.

\subsection{HD 19467 B}

HD 19467 B is a brown dwarf companion to the G3V star HD 19467, discovered through the TRENDS high-contrast imaging survey \citep{creppTRENDSHIGHCONTRASTIMAGING2014}. This system has been extensively characterised through multiple observational campaigns, providing one of the most comprehensive datasets for a substellar companion.

\subsubsection{System properties and previous studies}

The companion was first detected through Keck/NIRC2 coronagraphic imaging combined with long-baseline radial velocity measurements, establishing a firm dynamical lower limit on the companion mass \citep{creppTRENDSHIGHCONTRASTIMAGING2014}. Low-resolution ground-based spectroscopy confirmed the T-dwarf nature with a spectral classification of T5.5 $\pm$ 1 \citep{creppDIRECTSPECTRUMBENCHMARK2015}.

Subsequent high-contrast imaging with VLT/SPHERE and NaCo, combined with archival data and joint orbital fitting, refined the dynamical mass estimates \citep{maireOrbitalSpectralCharacterization2020}. The most recent analysis incorporating JWST astrometry and orbital constraints yields an updated dynamical mass of 71.6$^{+5.3}_{-4.6}$ M$_\mathrm{J}$ \citep{hochJWSTTSTHighContrast2024}. The orbital characterisation by \citet{maireOrbitalSpectralCharacterization2020} showed that HD 19467 B is in a long-period, high-eccentricity orbit.

The host star age is old (8--10~Gyr; \citealt{maireOrbitalSpectralCharacterization2020,greenbaumFirstObservationsBrown2023}), making HD~19467~B a rare target for high-contrast spectroscopy of an old, cool brown dwarf.

\subsubsection{JWST observations and data reduction}

\begin{figure*}[htbp]
    \centering
    \includegraphics[width=\textwidth]{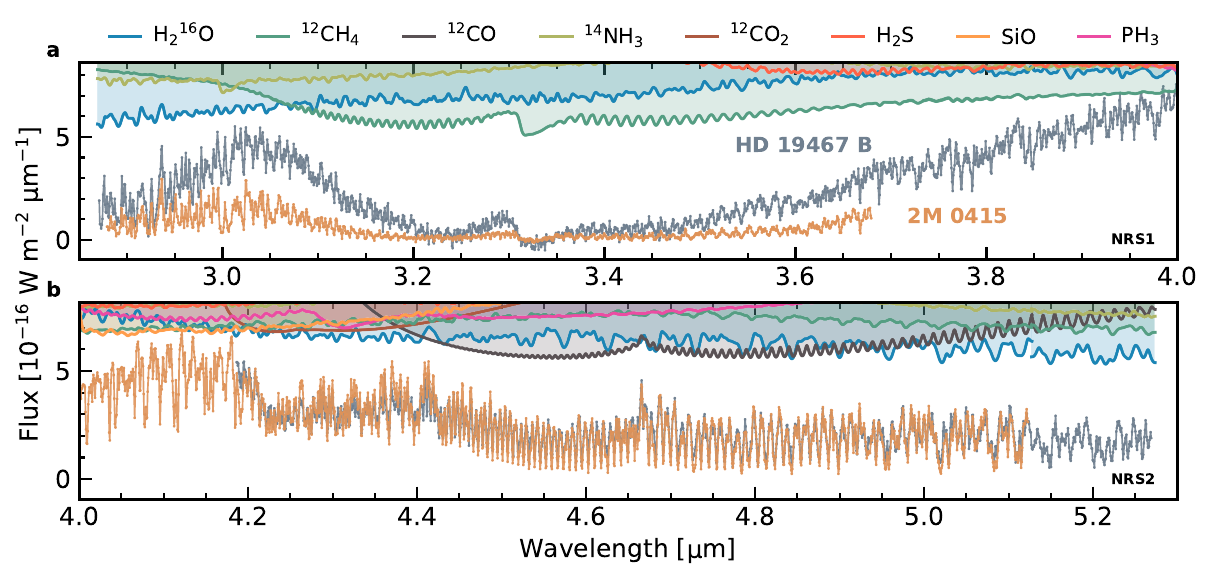}
    \caption{JWST NIRSpec spectrum of HD 19467 B and 2MASS J0415$-$0935. The flux has been scaled for plotting purposes to a reference spectrum of a 1 R$_\mathrm{Jup}$ object at 10 pc. Data for detectors NRS1 (panel a) and NRS2 (panel b) is shown. The different wavelength coverage is due to the different observing setups for the two objects (i.e. IFU and fixed slit). Molecular opacity is overlaid for a reference temperature and pressure of 900 K and 2 bar. The opacity is calculated using the retrieved volume mixing ratios for the detected species and for non-detections we use reference values from the models (i.e. PH3).}
    \label{fig:hd19467b_spectrum}
\end{figure*}

JWST/NIRSpec IFU observations of HD~19467~B were obtained as part of Cycle~1 GTO programme 1414 (PI: Marshall Perrin) targeting high-contrast companions. In this work, we use the companion spectrum as extracted and processed by \citet{ruffioJWSTTSTHighContrast2023}. Using a reference-star PSF subtraction (reference differential imaging; RDI), they extracted the first moderate-resolution spectrum of HD~19467~B between 2.9--5.3~\micron\ at a signal-to-noise ratio (S/N) of \(\sim 10\) in the continuum. The reduction used the JWST calibration pipeline (v1.10.2) to generate flux-calibrated detector images, with stellar-PSF subtraction and flux extraction performed directly on the detector images rather than reconstructed spectral cubes \citep{ruffioJWSTTSTHighContrast2023}; the same extracted spectrum was analysed with atmospheric grids by \citet{hochJWSTTSTHighContrast2024}.
The extracted spectra of both targets are shown in \Cref{fig:hd19467b_spectrum}.

\subsection{2MASS J0415-0935}

2MASS J04151954$-$0935066 (2M~0415 hereafter) is a T8 dwarf discovered by \citealt{burgasserSpectraDwarfsNearInfrared2002} that has become a standard for spectral classification of late-type, cool T dwarfs. This object has been extensively studied across multiple wavelength regimes, making it a useful reference object for atmospheric modelling (e.g. \citealt{leggettPhysicalSpectralCharacteristics2007,yamamuraAKARIOBSERVATIONSBROWN2010,hoodHighPrecisionAtmosphericConstraints2024,merchanDiversityColdWorlds2025}).

\subsubsection{System properties and previous studies}

Discovered in the 2MASS survey, 2M~0415 is a nearby ($5.71 \pm 0.06$ pc; \citealt{dupuyHAWAIIINFRAREDPARALLAX2012}) late T dwarf (T8). Its near-infrared magnitudes are \(J=15.34\) and \(H=15.67\), and its near-infrared spectrum shows the strong \methane\ and \water\ absorption characteristic of late-T dwarfs \citep{burgasserSpectraDwarfsNearInfrared2002}.

Mid-infrared spectroscopy with Spitzer/IRS (5--20~\micron) and Subaru/IRCS (2.9--4.1~\micron) provided key constraints on the spectral energy distribution \citep{cushingInfraredSpectroscopicSequence2005,saumonEvolutionDwarfsColor2008}. Model fits to these data yielded the first comprehensive estimates of the fundamental parameters, later broadly confirmed by subsequent studies \citep{saumonEvolutionDwarfsColor2008,filippazzoFUNDAMENTALPARAMETERSSPECTRAL2015}. More recently, near-complete SED analyses of 2M~0415 have delivered some of the most precise fundamental-parameter constraints for any T dwarf by combining the same G395H dataset analysed here with additional data spanning \(\sim\)1--20~\micron\ \citep{hoodHighPrecisionAtmosphericConstraints2024,merchanDiversityColdWorlds2025}.

\subsubsection{JWST observations and previous analyses}

JWST/NIRSpec observations of 2MASS~J0415$-$0935 were obtained in Cycle~1 GO programme 2124 (PI J. Faherty) using the G395H grating and F290LP filter, covering 2.87--5.14~\micron\ at \(R\approx2700\). The data and a detailed analysis were presented by \citet{hoodHighPrecisionAtmosphericConstraints2024}, who performed grid-based fits and combined the NIRSpec spectrum with lower-resolution spectra (i.e. IRTF/SpeX and Spitzer/IRS) in a full-SED retrieval. The resulting parameters agree well with earlier determinations and have substantially reduced uncertainties (see \Cref{sec:discussion}).

The NIRSpec spectrum shows prominent bands of \water, \methane, \COtwo, and CO. In particular, strong CO absorption at 4.6--5.0~\micron\ indicates disequilibrium chemistry, consistent with earlier lower-resolution evidence from AKARI \citep{yamamuraAKARIOBSERVATIONSBROWN2010}. The spectrum also covers the \(\nu_1\) band of NH$_3$ near 3~\micron, previously identified in JWST low-resolution data of the Y0 dwarf WISE~J035934.06$-$540154.6 \citep{beilerFirstJWSTSpectral2023}.

\begin{figure}[htbp]
    \centering
    \includegraphics[width=\columnwidth]{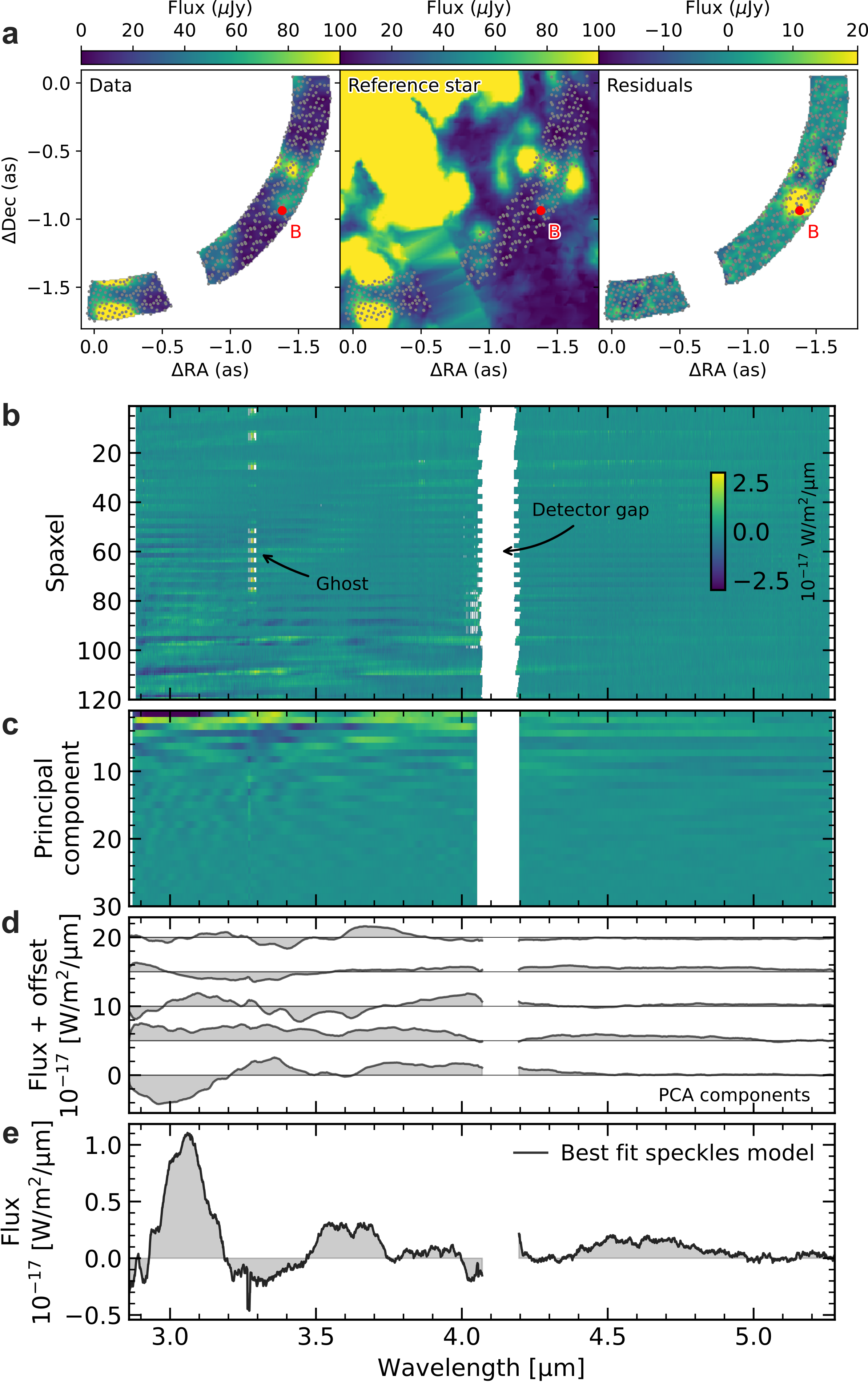}
    \caption{Speckle spectral field for HD~19467~B. a. Annular slice of the integral field unit used to extract the speckle model spectra, adopted from \citealt{ruffioJWSTTSTHighContrast2023}. b. Spectra from individual pixels in the annular slice of the IFU ($n=120$ pixels). c. The PCA reconstruction of the speckle spectra with 30 components, from top to bottom the components with the largest fitted amplitudes. d. The amplitude of the five speckle components with the largest amplitude from the joint atmospheric and speckle fit. e. Combined speckle model from the best-fit retrieval.}
    \label{fig:speckle_field}
\end{figure}

\section{Methods}

\subsection{Atmospheric modelling}\label{sec:atmospheric_modelling}

We compute high-resolution emission spectra with \pRT{} v3.1 \citep{mollierePetitRADTRANSPythonRadiative2019,blainSpectralModelHighresolutionFramework2024}. Line opacities are taken from \pRT{} where available, or generated with \pyrox\footnote{\url{https://py-rox.readthedocs.io/}} \citep{regtPyROXRapidOpacity2025} using the latest line lists \citep{gordonHITRAN2020MolecularSpectroscopic2022,tennyson2024ReleaseExoMol2024}. In this work we use the following line lists for \textsuperscript{12}\methane\ \citep{yurchenkoExoMolLineLists2024}, \textsuperscript{13}\methane\ \citep{gordonHITRAN2020MolecularSpectroscopic2022}, CH$_3$D \citep{rothmanHITRAN2012MolecularSpectroscopic2013}, H$_2^{16}$O \citep{polyanskyExoMolMolecularLine2018}, H$_2^{18}$O \citep{polyanskyExoMolMolecularLine2017}, CO isotopologues \citep{rothmanHITEMPHightemperatureMolecular2010,liROVIBRATIONALLINELISTS2015}, NH$_3$ \citep{colesExoMolMolecularLine2019}, $^{12}$CO$_2$ and $^{13}$CO$_2$ \citep{hargreavesUpdatingCarbonDioxide2025}, SiO \citep{yurchenkoExoMolLineLists2022}, H$_2$S \citep{azzamExoMolMolecularLine2016,chubbMarvelAnalysisMeasured2018}, and PH$_3$ \citep{sousa-silvaExoMolLineLists2015}.

Model spectra are generated at a resolving power of R=100,000 and subsequently convolved to the instrumental resolution of NIRSpec using a wavelength-dependent Gaussian kernel whose full-width at half-maximum (FWHM) is set by the NIRSpec calibration files (see also \citealt{gonzalezpicosDisentanglingDiscAtmospheric2025}\footnote{\url{https://github.com/DGonzalezPicos/broadpy}}). Recent in-flight measurements suggest the actual resolving power may be 1--24\% higher (\citealt{shajibAccurateMeasurementSpectral2025}). The resolving power could alternatively be parameterised and fitted (\citealt{gandhiJWSTMeasurements132023,ruffioJupiterlikeUniformMetal2026}); we nevertheless fix it to the calibration values for simplicity. This choice reproduces the observed line shapes well, as evidenced by the quality of the fit and the shape of the cross-correlation functions (see \Cref{fig:hd19467b_best_fit_spec,fig:cross_correlation_molecules}).

We adopt a flexible atmospheric retrieval framework in which we retrieve a parameterised pressure--temperature profile together with constant-with-altitude molecular abundances. Model spectra are generated with \pRT{} using the same line lists and resolving power throughout.

The free parameters are radius, mass or \logg, radial velocity (one per detector; Appendix~\ref{sec:radial_velocity}), an error-inflation parameter, molecular abundances, and isotopologue ratios. For HD~19467~B we use the dynamical mass $71.6^{+5.3}_{-4.6}$~\Mjup\ \citep{hochJWSTTSTHighContrast2024} as a Gaussian prior (truncated at 2-$\sigma$ with \texttt{scipy.stats.truncnorm}\footnote{\url{https://docs.scipy.org/doc/scipy/reference/generated/scipy.stats.truncnorm.html}}); this is our default configuration throughout the paper. Surface gravity is then derived from the retrieved radius and mass. The dynamical constraint breaks the well-known degeneracy between \logg{} and metallicity in medium-resolution retrievals of cool T dwarfs (see \Cref{sec:fundamental_parameters} for a discussion of the unconstrained case). For 2M~0415, which lacks a dynamical mass, we retrieve \logg{} directly as a free parameter. The complete list of retrieved parameters and priors is given in Table~\ref{tab:HD19467B_J0415_parameters}.

The pressure--temperature profile is parameterised with free temperature gradients and pressure levels, following \citet{gonzalezpicosDisentanglingDiscAtmospheric2025} based on \citet{zhangELementalAbundancesPlanets2023a}. The temperature at each atmospheric layer $T_j$ is calculated from the temperature gradients, which are linearly interpolated from the gradients at the reference points $\nabla_i$ and the pressure levels $P_i$ with $i=0,1,\ldots,6$:
\begin{equation}
T_j = T_{j-1} \cdot \left(\frac{P_j}{P_{j-1}}\right)^{\nabla_j},
\end{equation}
where $T_0$ is the surface temperature at 100 bar. The pressure levels are fixed at the bottom ($P_0 = 100$ bar) and top ($P_6 = 10^{-4}$ bar) of the atmosphere; the intermediate levels are set by the free parameters $\log P_{\rm RCE}$ and $\Delta \log P$ via $\log P_i = \log P_{\rm RCE} + x(i) \cdot \Delta \log P$ with $x(i) = -2,-1,0,1,2$ for $i=1,\ldots,5$. The composition is modelled with constant-with-altitude volume mixing ratios, with a free parameter for each molecule (except minor isotopologues that are fitted via ratios with the main isotopologues).

From the individual abundances, we calculate the gas-phase C/O ratio as:
\begin{align}
(\mathrm{C/O})_{\mathrm{gas}} &= \frac{\mathrm{CH_4} + \mathrm{CO} + \mathrm{CO_2}}{\mathrm{H_2O} + \mathrm{CO} + 2\,\mathrm{CO_2}}.
\end{align}
Gas-phase abundances neglect any elemental reservoirs in condensates (e.g. oxygen sequestration into silicates; \citealt{burrowsChemicalEquilibriumAbundances1999,lineUniformAtmosphericRetrieval2017}). As a result, \((\mathrm{C/O})_{\mathrm{gas}}\) can overestimate the bulk C/O ratio when condensation is important. To facilitate comparisons to bulk compositions, we estimate bulk C/O and O/H using empirical corrections \citep{calamariPredictingCloudConditions2024,kothariComprehensiveAtmosphericRetrieval2026}. Specifically, we adopt an oxygen condensation fraction \(f_{\mathrm{O}}=0.371\) \citep{calamariPredictingCloudConditions2024}, such that
\begin{align}
    (\mathrm{C/O})_{\mathrm{bulk}} &= \frac{(\mathrm{C/O})_{\mathrm{gas}}}{1 + f_{\mathrm{O}}\,(\mathrm{C/O})_{\mathrm{gas}}},\\
    (\mathrm{O/H})_{\mathrm{bulk}} &= (\mathrm{O/H})_{\mathrm{gas}} \left[1 - f_{\mathrm{O}}\,(\mathrm{C/O})_{\mathrm{bulk}}\right].
\end{align}
We note that a typo in Equation~11 of \citet{kothariComprehensiveAtmosphericRetrieval2026} is corrected here. When not explicitly stated, abundances should represent the bulk composition throughout the present work. We report elemental ratios with respect to the solar values as
\begin{align}
\left[\mathrm{X/H}\right] &= \log_{10}(\mathrm{X/H}) - \log_{10}(\mathrm{X/H})_{\odot},
\end{align}
with abundances from \citet{asplundChemicalMakeupSun2021}.
To derive effective temperatures from the posteriors, we compute a low-resolution spectrum ($\lambda/\Delta\lambda =100$) for each posterior sample over 0.3--28~\micron\ and integrate it to obtain the bolometric flux,
\begin{equation}
F_{\rm bol} = \int_{\lambda_1}^{\lambda_2} F_{\lambda}\,\mathrm{d}\lambda,
\end{equation}
with \(\lambda_1 = 0.3~\mu\mathrm{m}\) and \(\lambda_2 = 28~\mu\mathrm{m}\). Outside the G395H window we include additional opacity sources relevant at shorter wavelengths (TiO, VO, FeH, HCN, Na, K), adopting chemical-equilibrium abundances at the corresponding bulk C/O and metallicity (using [C/H] as a proxy). We then compute \(T_{\rm eff} = \left(F_{\rm bol}/\sigma_{\rm SB}\right)^{1/4}\). Additional opacity sources not included here---in particular oxides and hydrides at bluer optical and near-infrared wavelengths (e.g. MgO, CrH)---could bias \(F_{\rm bol}\) if they are significant absorbers and thus shift \Teff\ to lower values. We therefore report \Teff\ primarily as a consistency check but note that the most robust constraints on fundamental parameters come from full-SED observations and analyses.

\subsection{Speckle modelling for HD 19467 B}\label{sec:speckle_modelling}
Due to imperfect PSF subtraction, residual starlight (also referred to as speckles) is present in the companion spectrum. This residual signal exhibits a small non-zero mean, which is subtracted from the companion spectrum as described in \citealt{ruffioJWSTTSTHighContrast2023}. Residual contamination from the host star is therefore not fully removed at the companion position and may bias the retrieval of atmospheric parameters if it is not accounted for. In some cases, stray light leaks into the instrument and produces bright stripes across the detector, also referred to as ghosts (see panel b of \Cref{fig:speckle_field} and \citet{ruffioJWSTTSTHighContrast2023} for more details).

For HD 19467 B, we implement a speckle-modelling approach using a linear model framework that includes the atmospheric signal and a set of speckle model spectra. The speckle spectra used here are provided by \citealt{ruffioJWSTTSTHighContrast2023}\footnote{\url{https://zenodo.org/records/11391741}} and are extracted from an annular region (0.3--0.4 arcseconds) around the companion. This is an alternative to using a covariance matrix to model the speckle structure \citep{grecoMEASUREMENTTREATMENTIMPACT2016}, which is implemented in \citealt{ruffioJWSTTSTHighContrast2023}.

We calculate the dominant modes of the speckle spectra using principal component analysis (PCA) decomposition (see \Cref{fig:speckle_field}, also \citealt{hoeijmakersMediumresolutionIntegralfieldSpectroscopy2018}). The speckle model is defined by smoothing the PCA components with a Gaussian filter (51 pixels) and retaining \(N=30\) components. We tested a range of filter widths and component numbers by iteratively performing linear fits on the best-fit model from an initial retrieval. We found that the selected number of components and filter width captured 90\% of the variance in the speckle spectra and that the results were not very sensitive to further optimisation.
We use the same number of components for all detectors but note that the linear fit is performed on each detector separately. We refrain from further optimisation of the speckle model (i.e. filter width and number of components for each detector) to avoid overfitting and preserve the simplicity of the model. The speckle field used to construct the speckle basis is shown in \Cref{fig:speckle_field}.

\subsection{Bayesian retrieval framework}\label{sec:bayesian_retrieval}

We employ a Bayesian retrieval framework to determine atmospheric parameters from the observed spectra. For 2M~0415, the residuals between the atmospheric model and the data are computed directly and the likelihood is evaluated with \Cref{eq:log_l}. For HD~19467~B, a few additional steps are required to account for residual speckle contamination. We use a linear model that includes the planetary signal and a set of speckle model spectra, following the approach of \citet{hochModerateresolutionKbandSpectroscopy2020,landmanPictorisEyesUpgraded2024}.

The model is expressed as:
\begin{equation}
\mathbf{d} = \mathbf{M}\boldsymbol{\phi} + \mathbf{n} = \phi_0\mathbf{M}_0 + \sum_{i=1}^{N}\phi_i\mathbf{M}_i + \mathbf{n},
\end{equation}
where $\mathbf{d}$ is the extracted spectrum at the companion position, $\mathbf{M}$ is the model matrix, $\boldsymbol{\phi}$ is the vector of linear coefficients, and $\mathbf{n}$ is Gaussian-distributed noise with zero mean and covariance matrix $\boldsymbol{\Sigma}$. The coefficient $\phi_0$ scales the companion model, while $\phi_{1\ldots N}$ scale the speckle components. The column $\mathbf{M}_0$ is the atmospheric model of the companion generated with \pRT{}, and $\mathbf{M}_{1\ldots N}$ are the speckle components derived from PCA decomposition of the annular region spectra.

The covariance incorporates an error-inflation parameter $b$ that accounts for systematic uncertainties beyond the formal errors. The effective variance for each data point is:
\begin{equation}
\sigma^2_{\text{eff},i} = 0.5 \cdot \sigma^2_i \cdot \left(1 + 10^{2b}\right) = s^2 \cdot \sigma^2_i,
\end{equation}
where $\sigma_i$ is the formal uncertainty for the $i$-th data point. We adopt a diagonal covariance, $\boldsymbol{\Sigma}^{-1} = \text{diag}(1/\sigma^2_{\text{eff}})$, which in preliminary tests captured the residual structure without additional correlated-noise parameters.

When speckle correction is applied, we solve for the optimal speckle coefficients $\boldsymbol{\phi}_{\text{speckles}}$ using a least-squares solver\footnote{\url{https://numpy.org/doc/2.3/reference/generated/numpy.linalg.lstsq.html}}:
\begin{equation}
\boldsymbol{\phi}_{\text{speckles}} = \text{solve}(\mathbf{M}^T\boldsymbol{\Sigma}^{-1}\mathbf{M}, \mathbf{M}^T\boldsymbol{\Sigma}^{-1}\mathbf{r}^T),
\end{equation}
where $\mathbf{r} = \mathbf{f} - \mathbf{s}$ denotes the residuals between the observed flux $\mathbf{f}$ and the model spectrum $\mathbf{s}$. The corrected residuals are then:
\begin{equation}
\mathbf{r}' = \mathbf{f} - (\mathbf{s} + \boldsymbol{\phi}_{\text{speckles}}^T\mathbf{M}).
\end{equation}
Because the amplitude of the atmospheric model is set by the radius, the speckle amplitudes are fitted after subtracting the atmospheric model.

The chi-squared statistic is computed as:
\begin{equation}
\chi^2 = \mathbf{r}^T\boldsymbol{\Sigma}^{-1}\mathbf{r},
\end{equation}
where $\mathbf{r}$ denotes either the original or speckle-corrected residuals $\mathbf{r}'$, and $N$ is the number of valid data points. The log-likelihood is:
\begin{equation}\label{eq:log_l}
\ln\mathcal{L} = -\frac{1}{2}\left(N\ln(2\pi) + \ln|\boldsymbol{\Sigma}| + \chi^2\right),
\end{equation}
where $|\boldsymbol{\Sigma}|$ is the determinant of the covariance matrix. The optimal linear parameters $\tilde{\boldsymbol{\phi}}$ and noise scaling factor $\tilde{s}$ are obtained by minimising $\chi^2$.

We use nested sampling for posterior exploration and evidence computation (\ultranest{}; \citealt{skillingNestedSampling2004,buchnerUltraNestRobustGeneral2021}). We adopt 1000 live points and a convergence criterion of $\Delta\ln Z = 0.5$, using a mixed-mode step sampler \citep{buchnerNestedSamplingMethods2023,buchnerComparisonStepSamplers2023}. For HD~19467~B we verified consistency with \texttt{PyMultiNest} in constant-efficiency mode (\Cref{fig:corner_comparison_hd19467b_pmn}).

\subsection{Detection of minor species}
We assess the detectability of minor species (e.g. SiO, \CeighteenO) by removing each species from the best-fit free model and quantifying the preference for inclusion using the Akaike Information Criterion (AIC; \citealt{akaikeNewLookStatistical1974}). We define
\(\Delta\mathrm{AIC} \equiv \mathrm{AIC}(\mathrm{no~molecule}) - \mathrm{AIC}(\mathrm{baseline})\),
where positive values favour inclusion. Detection significance is estimated using the criterion of \citep{thorngrenBayesianModelComparison2026}. While an evidence-based comparison would be more robust, full Bayesian retrievals for each species are unfeasible given the computational cost (\(\sim\)10$^4$ CPU hours per retrieval).

\begin{figure*}[htbp]
    \centering
    \includegraphics[width=\textwidth]{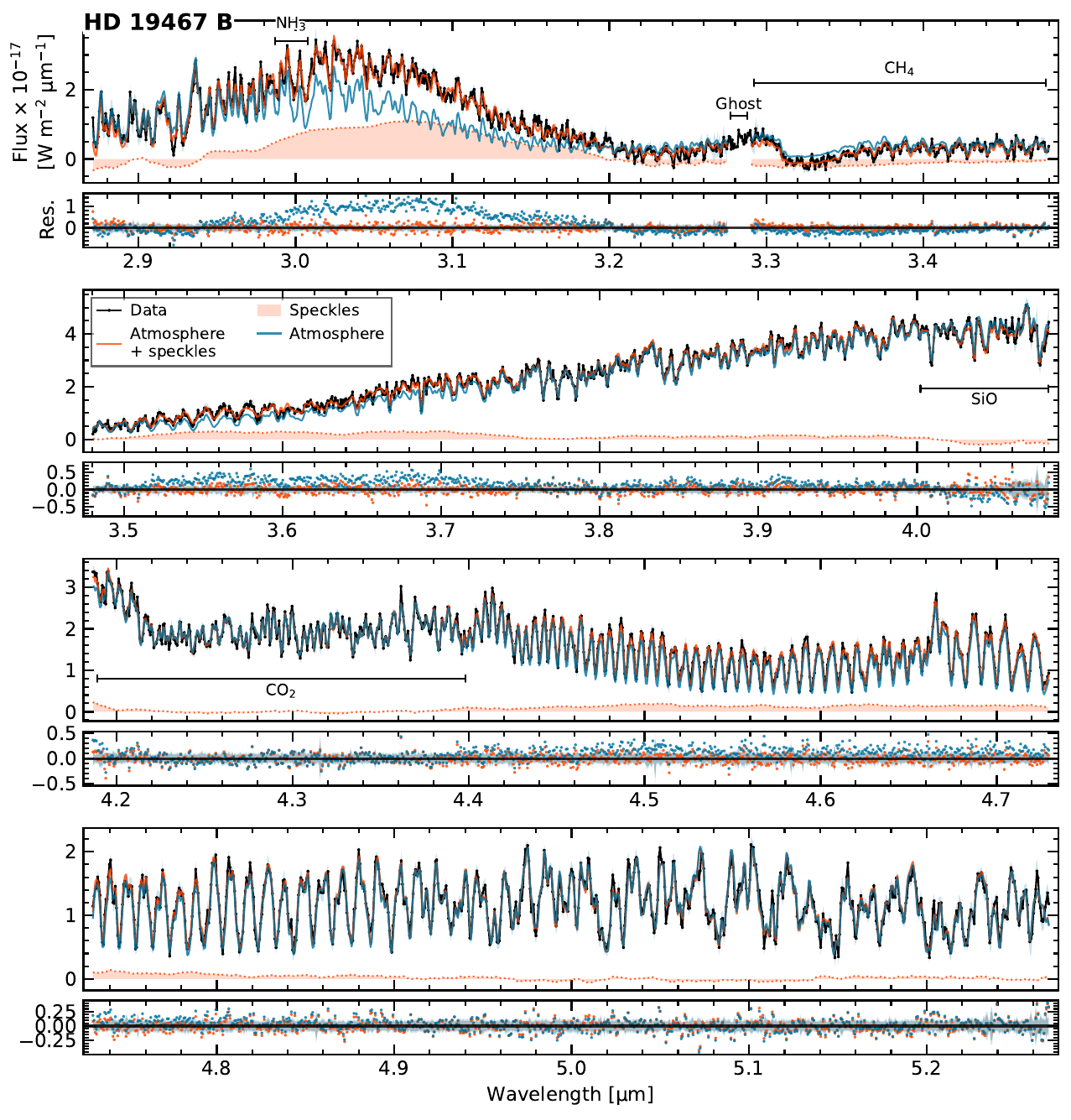}
    \caption{Best-fit model spectrum for HD~19467~B. We show the individual contributions from the atmospheric model and the speckle model. Selected opacity sources are indicated in the relevant wavelength regions. The equivalent figure for 2M~0415 is shown in \Cref{fig:best_fit_model_spectrum_2M0415}. The data products required to reproduce this figure are available on Zenodo \citep{gonzalezpicosJWSTHighcontrastSpectroscopy2026}.}
    \label{fig:hd19467b_best_fit_spec}
\end{figure*}

\begin{figure}[htbp]
    \centering
    \includegraphics[width=\columnwidth]{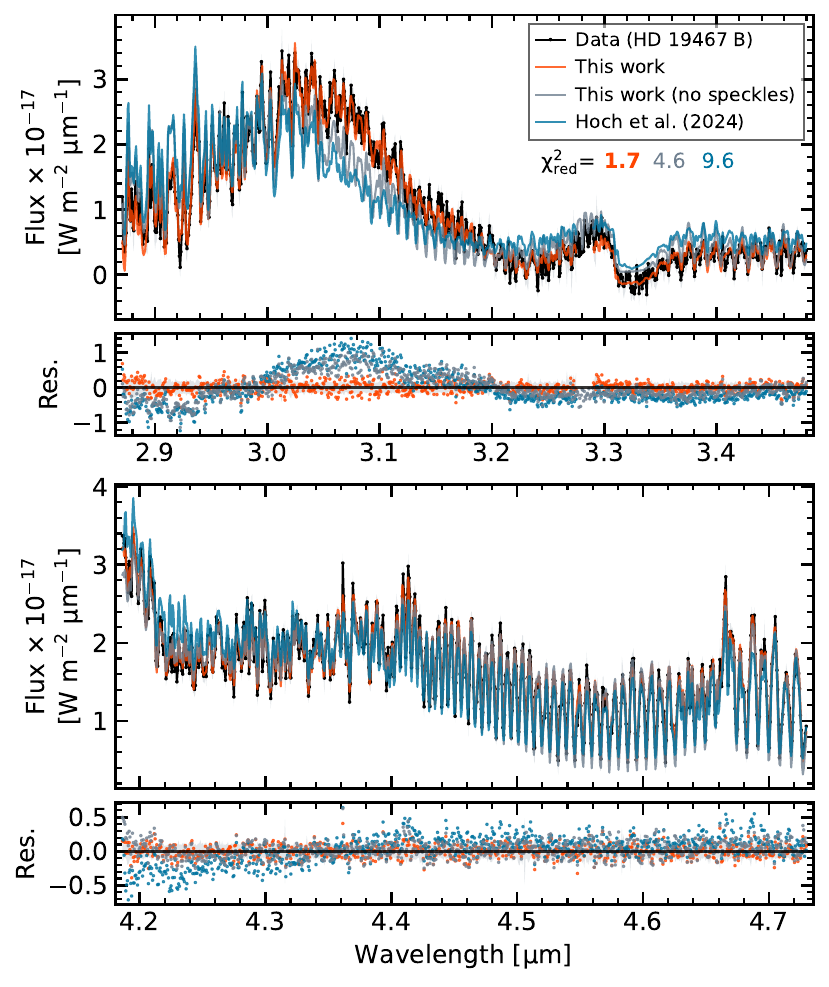}
    \caption{Comparison of our best-fit model spectrum for HD 19467 B in two selected wavelength regions with best-fit model from \citet{hochJWSTTSTHighContrast2024} using a custom \texttt{NewEra} model grid \citep{hauschildtNewEraModelGrid2025}.}
    \label{fig:hd19467b_spec_comparison_hoch24_free}
\end{figure}

\begin{figure*}[htbp]
    \centering
    \includegraphics[width=\textwidth]{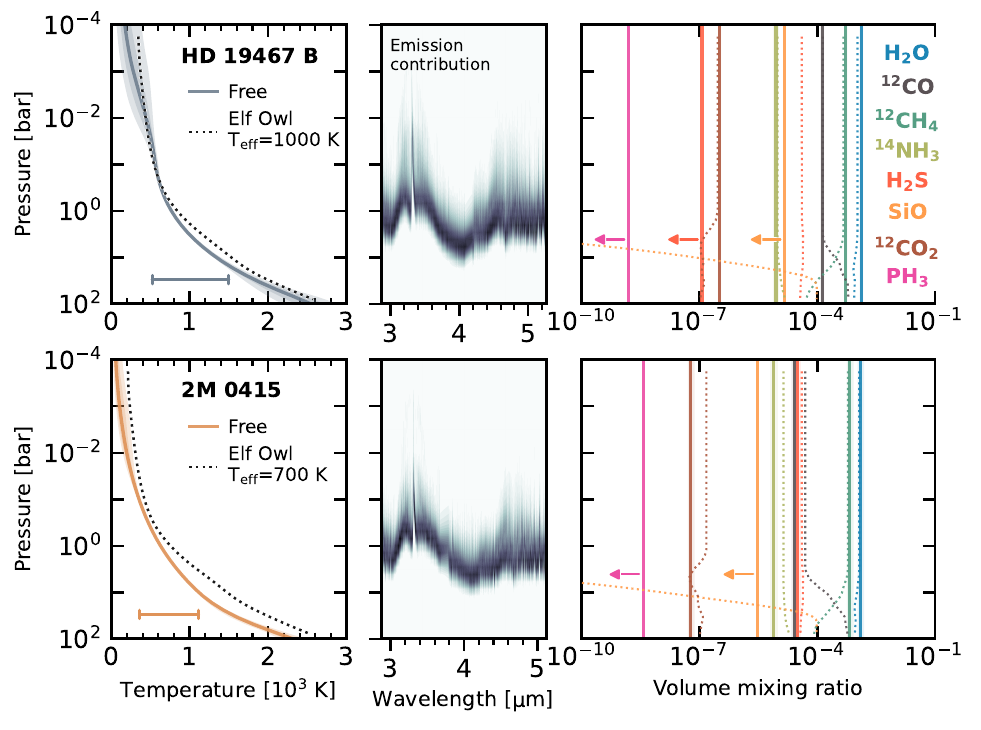}
    \caption{Atmospheric structure of HD 19467 B and 2M 0415 from the free retrievals. Left: Temperature profiles. The pressure and temperature range of the photosphere (emission contribution function \(>1\%\)) is indicated by error bars. Middle: Emission contribution function, dark regions indicate high contribution. Right: Volume mixing ratios. Arrows indicate upper limits for the abundances of non-detected species. Sonora Elf Owl models \citep{mukherjeeSonoraSubstellarAtmosphere2024,woganSonoraSubstellarAtmosphere2025} are overlaid for the temperature and volume mixing ratios assuming a common metallicity of \([\mathrm{M/H}]=0.2\), default C/O=0.458 and moderate vertical mixing, \(\log K_{zz}=3\), which yields photospheric abundances broadly consistent with the retrievals. For HD~19467~B we use \Teff{} \(=1000\)~K and \(\log g=5.5\); for 2M~0415 we use \Teff{} \(=700\)~K and \(\log g=5.0\).}
    \label{fig:pt_composition}
\end{figure*}

\begin{figure*}[htbp]
    \centering
    \includegraphics[width=\textwidth]{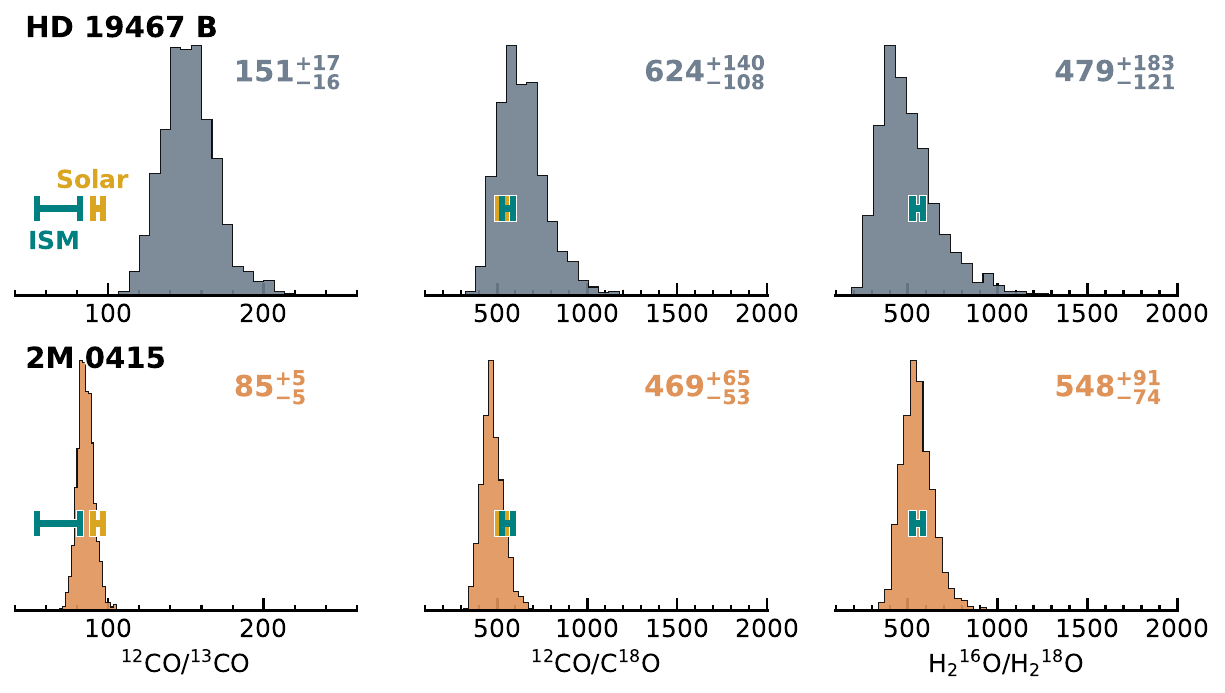}
    \caption{Posterior distributions of isotopologue ratios for HD 19467 B (top; with dynamical-mass prior) and 2M 0415 (bottom). The values of the ISM (68 $\pm$ 15; \citealt{milam1213Isotope2005}) and solar (93.5 $\pm$ 3.1; \citealt{lyonsLightCarbonIsotope2018}) are shown for reference.}
    \label{fig:isotope_ratios}
\end{figure*}

\section{Results}\label{sec:results}

We present atmospheric retrieval results for HD~19467~B and 2M~0415 using the framework described in \Cref{sec:atmospheric_modelling}. For HD~19467~B we adopt the dynamical mass of \citet{hochJWSTTSTHighContrast2024} as a Gaussian prior; this is our default configuration and the one we discuss throughout. The impact of dropping the mass prior is discussed in \Cref{sec:fundamental_parameters}. For 2M~0415 no dynamical mass is available and \logg{} is a free parameter. We focus on the spectral fits, atmospheric structure, molecular abundances, and isotopic ratios inferred from the G395H data.

\Cref{fig:hd19467b_best_fit_spec} shows the best-fit spectrum of HD~19467~B and highlights the contribution of residual speckle contamination. \Cref{fig:best_fit_model_spectrum_2M0415} shows the best-fit spectrum for 2M 0415. In \Cref{fig:hd19467b_spec_comparison_hoch24_free} we compare our best-fit model spectrum for HD~19467~B to \citet{hochJWSTTSTHighContrast2024}. \Cref{fig:pt_composition} summarises the best-fit temperature profiles, emission contribution functions, and retrieved compositions for both targets. The posterior probability distributions are shown in \Cref{fig:cornerplot_hd19467b} for HD~19467~B and in \Cref{fig:cornerplot_j0415} for 2M~0415. Isotopologue posteriors are shown in \Cref{fig:isotope_ratios}, and a full list of retrieved parameters is provided in Table~\ref{tab:HD19467B_J0415_parameters}.

\subsection{Spectral fits}\label{sec:fit_quality}

We fitted the full JWST/NIRSpec G395H spectra of both HD~19467~B and 2M~0415, achieving a residual scatter comparable to the SNR of the data with error-scaling factors of $s=1.10$ and $s=1.99$, respectively (typical uncertainty $\pm 0.02$). The joint spectrum and speckle linear model reproduces the native-resolution data for HD~19467~B without requiring continuum subtraction (see \Cref{fig:hd19467b_best_fit_spec,sec:speckles}).

\subsection{Temperature-pressure profile}\label{sec:temperature_profiles}

The temperature-pressure profiles of both objects are well constrained by the data, as shown in Figure~\ref{fig:pt_composition}.

Our pressure-range sensitivity, as represented by the emission contribution function in \Cref{fig:pt_composition}, extends from approximately $10$ to $0.01$~bar.

\subsection{Chemical composition}\label{sec:composition}

We report detections of \water, CO, \methane, NH$_3$, and \COtwo\ in the atmospheres of both HD~19467~B and 2M 0415. We find constrained posterior distributions for the abundances of all these species (see \Cref{fig:cornerplot_hd19467b,fig:cornerplot_j0415}) and significant cross-correlation peaks (SNR > 3; see \Cref{fig:cross_correlation_molecules}).

Additionally, our retrievals appear to constrain the abundance of SiO in HD~19467~B but find only weak evidence for its presence in 2M 0415. This is discussed in more detail in \Cref{sec:trace_molecules}.
In both cases, however, the cross-correlation analysis does not yield a significant detection for SiO (SNR $<3$; \Cref{fig:cross_correlation_molecules}), and we therefore treat any apparent constraints on SiO as tentative.
We also constrain the abundances of minor isotopologues including \thirteenCO\, \CeighteenO\ and tentative evidence for \eighteenOwater. We assess the potential presence of \isomethane\ and \isoCOtwo{} in \Cref{sec:isotopes}.

Figure~\ref{fig:pt_composition} (right column) shows the volume mixing ratios of the detected species as a function of pressure, compared to Sonora Elf Owl models \citep{mukherjeeSonoraSubstellarAtmosphere2024,woganSonoraSubstellarAtmosphere2025}. Both objects exhibit moderate vertical mixing, with $\log K_{\mathrm{zz}} \approx 3-4$~dex, required to reconcile the observed CO absorption features and volume mixing ratios. The balance of chemical timescales at lower pressures drives the atmosphere out of chemical equilibrium, thereby inhibiting the conversion of CO into CH$_4$ and preserving a relatively high abundance of CO compared to chemical equilibrium predictions \citep{zahnleMethaneCarbonMonoxide2014}. CO$_2$ is also significantly sensitive to $K_{zz}$: vertical mixing can transport CO$_2$ from deeper, hotter layers into the photosphere, potentially enhancing its observable abundance \citep{woganSonoraSubstellarAtmosphere2025,beilerTaleTwoMolecules2024}. We detect \COtwo\ in both objects, with abundances consistent with the updated Sonora Elf Owl model atmospheres \citep{beilerTaleTwoMolecules2024,woganSonoraSubstellarAtmosphere2025}.

We report elemental ratios including the empirical correction for oxygen sequestration (see \Cref{sec:atmospheric_modelling}):
\begin{equation}
    \begin{aligned}
        \mathrm{[C/H]}_{\rm HD~19467~B} &= 0.14^{+0.04}_{-0.04}, &
        \mathrm{[C/H]}_{\rm 2M~0415} &= 0.15^{+0.09}_{-0.07} \\
        \mathrm{[O/H]}_{\rm HD~19467~B} &= 0.18^{+0.05}_{-0.04}, &
        \mathrm{[O/H]}_{\rm 2M~0415} &= 0.11^{+0.09}_{-0.08} \\
        \mathrm{C/O}_{\rm HD~19467~B} &= 0.389^{+0.012}_{-0.012}, &
        \mathrm{C/O}_{\rm 2M~0415} &= 0.451^{+0.005}_{-0.006}
    \end{aligned}
\end{equation}
The carbon and oxygen abundances are slightly super-solar for both objects, while the C/O ratios are sub-solar (C/O$_{\odot}=0.59 \pm 0.08$; \citealt{asplundChemicalMakeupSun2021}). Our C/O value for HD~19467~B is consistent with previous ground-based measurements from \citet{mesaCharacterizingBrownDwarf2020} (C/O=$0.36 \pm 0.03$). For 2M~0415, our C/O value is lower than the full-SED retrieval from \citet{hoodHighPrecisionAtmosphericConstraints2024} (C/O=$0.53 \pm 0.01$; corrected for oxygen condensation similar to our empirical correction) but we note that in the same study the authors also found lower C/O values when using Sonora Elf Owl models (C/O=$0.36 \pm 0.002$).

\section{Discussion}\label{sec:discussion}

\subsection{Speckle contamination in HD~19467~B}\label{sec:speckles}

The high-contrast nature of the HD~19467~B system necessitates careful treatment of residual stellar contamination in the extracted companion spectrum. At a separation of 1.6 arcseconds, the flux ratio between the companion and host star ranges from $10^{-5}$ to $10^{-6}$ across 3–5 $\mu$m. Despite the PSF-subtraction approach described in \citealt{ruffioJWSTTSTHighContrast2023}, speckle contamination remains a significant systematic that must be addressed to correctly fit the spectrum and interpret atmospheric parameters. We identified this contamination after numerous attempts to extend our atmospheric model to include additional effects that could account for the clear mismatch between data and model around 3.0-3.7 $\mu$m. This discrepancy was independently noted by \citealt{hochJWSTTSTHighContrast2024}, who fitted the same dataset with four different self-consistent models and found a similar mismatch in all cases (see \Cref{fig:hd19467b_spec_comparison_hoch24_free}). To mitigate this issue, \citet{hochJWSTTSTHighContrast2024} removed the continuum from both data and model, but this approach results in information loss and may not eliminate the speckle contamination on the spectral features.

Speckle contamination predominantly affects the low-frequency spectral continuum, with the most severe residuals concentrated in the 3.0--3.7~\micron\ interval (see \Cref{fig:hd19467b_best_fit_spec}). Speckle residuals are smaller at 4.3--5.2~\micron\ but remain non-negligible and can still bias the inferred abundances. Omitting the speckle component from our retrieval yields a dramatically worse fit (disfavoured at $\sim\!60\sigma$; \Cref{fig:hd19467b_spec_comparison_hoch24_free}) and introduces systematic biases in the retrieved parameters, including a smaller inferred radius, non-detection of NH$_3$, a spurious H$_2$S detection, and inaccurate isotopic abundance ratios (\Cref{fig:cornerplot_hd19467b}). Simply excluding the 3.0--3.7~\micron\ region reduces the apparent residuals but does not prevent these biases, as speckle contamination persists across the full wavelength range.

\subsection{Fundamental parameters}\label{sec:fundamental_parameters}
We compute \teff\ from the bolometric flux inferred by integrating low-resolution model spectra over 0.3--28~\micron\ (see \Cref{sec:atmospheric_modelling}). Retrieved radii and masses are summarised in Table~\ref{tab:HD19467B_J0415_parameters}. The quoted radii primarily capture the statistical uncertainties from the retrieval. An additional systematic uncertainty arises from the absolute flux calibration of NIRSpec, which is typically at the 5--10\% level for spectroscopy\footnote{\url{https://jwst-docs.stsci.edu/jwst-calibration-status/jwst-absolute-flux-calibration}}.

For HD~19467~B, our default retrieval uses the dynamical mass of \citet{hochJWSTTSTHighContrast2024} as a Gaussian prior, yielding a surface gravity $\log g=5.45\pm0.03$ consistent with an old, high-gravity brown dwarf and near-solar metallicity (Table~\ref{tab:fundamental_parameters_summary}).

Comparing to previous analyses of HD~19467~B (Table~\ref{tab:fundamental_parameters_summary}), \citet{greenbaumFirstObservationsBrown2023} found a significantly smaller radius (and correspondingly lower luminosity) than expected for an old brown dwarf, while \citet{hochJWSTTSTHighContrast2024} derived a dynamical mass consistent with earlier orbit-based estimates and argued for stronger vertical mixing. Medium-resolution long-slit spectroscopy of HD~19467~B with SPHERE/IRDIS by \citet{mesaCharacterizingBrownDwarf2020} yielded an independent spectral characterisation (T6\(\pm\)1) and atmospheric parameters broadly consistent with a cool (\Teff{} $\sim$ 1000~K), high-gravity (\(\log g\sim5\)) object. Our inferred \Teff\(\sim 1080\)~K is in good agreement with the effective-temperature scale found across these studies.

For completeness, we also performed a retrieval without the mass prior: this leads to a lower surface gravity and a mass significantly lower than the dynamical constraint (around 20~\Mjup) with a slightly (${\sim}4\sigma$) improved fit, but this is likely a way for the model to compensate for other mismatches between data and model; all parameters are consistent within 1$\sigma$ except for the surface gravity and the abundances, which are shifted accordingly due to the metallicity--surface gravity degeneracy, resulting in a very low metallicity (${\sim}-0.50$), an effect also observed in this type of analysis of other objects. Including the mass prior therefore adds relevant information to correctly interpret the abundances and fundamental parameters, and we adopt the mass-constrained retrieval as our default throughout.

For 2M~0415, the inferred fundamental parameters and literature comparisons are summarised in Table~\ref{tab:fundamental_parameters_summary}. Despite the limited wavelength coverage, our G395H-only inference yields \(L_{\rm bol}\) and \Teff{} consistent with full-SED analyses \citep{hoodHighPrecisionAtmosphericConstraints2024,merchanDiversityColdWorlds2025}. In particular, the near-complete SED analysis of \citet{merchanDiversityColdWorlds2025} should provide the most direct constraints on fundamental parameters owing to its extended wavelength coverage, which are consistent with our inferred parameters within 1$\sigma$. This agreement suggests that the 2.87--5.27~\micron\ window already captures much of the information needed to constrain the bolometric flux and effective temperature of this object. However, the smaller uncertainties of our analysis do not account for the assumptions made outside the wavelength range of the G395H window, which may introduce additional systematic uncertainties. We recommend the use of full-SED analyses for more precise constraints on fundamental parameters.

\begin{table*}[t]
    \caption{Summary of fundamental parameters.}
    \label{tab:fundamental_parameters_summary}
    \centering
    \setlength{\tabcolsep}{3.5pt}
    \renewcommand{\arraystretch}{1.15}
    \begin{tabular}{@{}lllllll@{}}
    \toprule
    Object & Source & \Teff{} [K] & $\log(L/L_{\odot})$ & $R$ [\Rjup] & $\log g$ & $M$ [\Mjup] \\
    \midrule
    \multicolumn{7}{@{}l}{HD~19467~B} \\
    \midrule
     & This work & $1081^{+28}_{-29}$ & $-5.09^{+0.03}_{-0.04}$ & $0.790^{+0.012}_{-0.013}$ & $5.45\pm0.03$ & $71.6^{+5.3}_{-4.6}$ \\
     & \citet{hochJWSTTSTHighContrast2024} & $\approx1103$ & \ldots & \ldots & $\approx4.5$ & $71.6^{+5.3}_{-4.6}$ \\
     & \citet{greenbaumFirstObservationsBrown2023} & $1080\pm22$ & $-5.32\pm0.02$ & $0.62\pm0.03$ & $4.60^{+0.2}_{-0.1}$ & $62\pm1$ \\
     & \citet{mesaCharacterizingBrownDwarf2020} & $\sim1000$ & \ldots & $0.83\pm0.06$ & $\sim5.0$ & $78.0^{+1.6}_{-3.1}$ \\
    \midrule
    \multicolumn{7}{@{}l}{2MASS~J0415$-$0935} \\
    \midrule
     & This work (Free) & $728\pm8$ & $-5.719^{+0.016}_{-0.018}$ & $0.846\pm0.005$ & $4.97^{+0.11}_{-0.08}$ & $27^{+7}_{-5}$ \\
     & \citet{merchanDiversityColdWorlds2025} & $729^{+47}_{-10}$ & $-5.71\pm0.01$ & $0.855^{+0.110}_{-0.020}$ & $5.13^{+0.12}_{-0.35}$ & $37^{+10}_{-12}$ \\
     & \citet{hoodHighPrecisionAtmosphericConstraints2024} & $758^{+18}_{-3}$ & $-5.70^{+0.04}_{-0.01}$ & $0.80\pm0.04$ & $5.14\pm0.03$ & $36\pm3$ \\
    \bottomrule
    \end{tabular}
    \vspace{0.6ex}
    \begin{minipage}{\linewidth}
    \footnotesize
    Notes. Fundamental parameters for HD~19467~B and 2M 0415 from this work and selected literature studies. Quantities are quoted as reported in the cited works.
    \end{minipage}
\end{table*}

\subsection{Isotopic ratios}\label{sec:isotopes}

We infer carbon and oxygen isotope ratios from CO and \water\ isotopologues (Figure~\ref{fig:isotope_ratios}). We quote the CO-based ratios (higher sensitivity) and find consistent constraints from \water. We find:
\begin{equation}
    \begin{aligned}
        \rm \Cratio_{\rm HD~19467~B} &= 154^{+19}_{-17}, & \rm \Oratio_{\rm HD~19467~B} &= 711^{+204}_{-135},\\
        \rm \Cratio_{\rm 2M~0415} &= 85 \pm 5, & \rm \Oratio_{\rm 2M~0415} &= 469^{+65}_{-53}.
    \end{aligned}
\end{equation}
Both objects show higher ratios than Solar (\Cratio$_{\odot}=93 \pm 3$, \Oratio$_{\odot}=525 \pm 21$; \citealt{ayresSUNLIGHTEREARTH2013,lyonsLightCarbonIsotope2018}), and HD~19467~B is higher than 2M~0415. This is qualitatively consistent with galactic chemical evolution: older systems are expected to show higher \Cratio\ and \Oratio\ as the ISM becomes enriched in \thirteenC\ and \eighteenO\ \citep{prantzosEvolutionCarbonOxygen1996,romanoEvolutionCNOElements2022}. Given the broad age estimates for 2M~0415 (1--8~Gyr; \citealt{hsuBrownDwarfKinematics2021,hoodHighPrecisionAtmosphericConstraints2024,merchanDiversityColdWorlds2025}), our measurements suggest it is younger than the 8--10~Gyr HD~19467 system and plausibly closer to Solar age \citep{bouvierAgeSolarSystem2010}. Similar high ratios have been reported for old M dwarfs \citep{crossfieldUnusualIsotopicAbundances2019,gonzalezpicosChemicalEvolutionImprints2025a}.

\begin{figure}[htbp]
    \centering
    \includegraphics[width=\columnwidth]{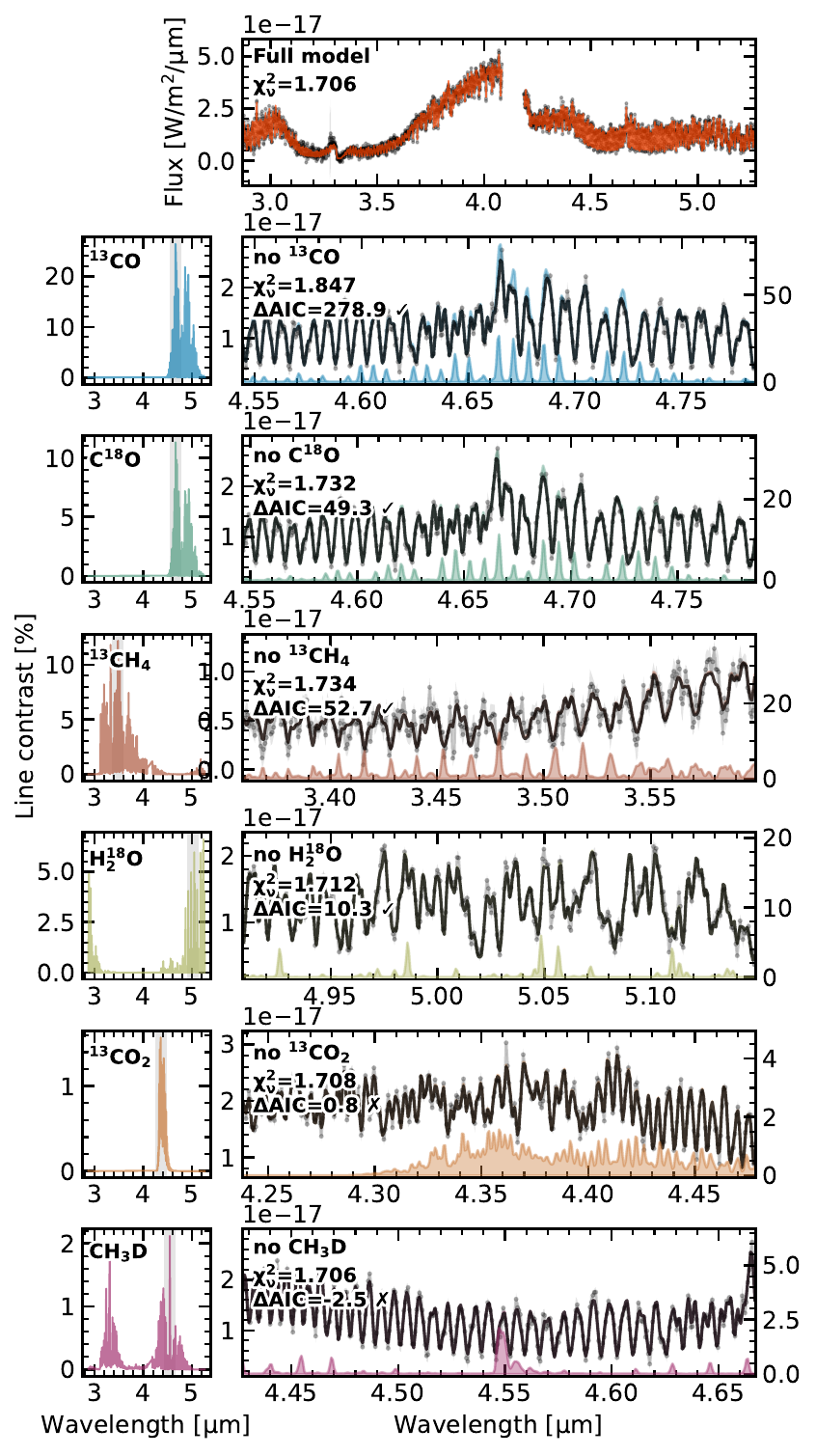}
    \caption{Detection assessment of minor isotopologues in HD~19467~B. Left: Model contribution of each species across the wavelength range. Right: Data (grey) and best-fit model without the given isotopologue (solid black); shaded regions show the change when adding that isotopologue and the line contrast of each species (right axis). The \(\Delta\)AIC values are shown in each panel; positive values favour inclusion.}
    \label{fig:app_minor_isotopologues}
\end{figure}

We also test \thirteenmethane, \thirteenCOtwo\ and \dmethane. \thirteenmethane\ is only tentatively detected; for HD~19467~B the strongest feature (3.32~\micron) overlaps strong speckle residuals (Figure~\ref{fig:app_minor_isotopologues}), likely biasing the inferred \Cratio. \thirteenCOtwo\ is not detected in HD~19467~B and is weakly constrained in 2M~0415 (log(\twelveCOtwo/\thirteenCOtwo)=2.11$^{+0.43}_{-0.23}$). For \CseventeenO, the posteriors rail against the upper prior wall in both objects ($\log\,^{12}$CO/\CseventeenO $> 3.2$; Table~\ref{tab:HD19467B_J0415_parameters}), indicating that this isotopologue is not detected with the current data.

Finally, we find no evidence for \dmethane\ in either object and place 3$\sigma$ upper limits on the methane D/H ratio of D/H $< 5.8 \times 10^{-4}$ for HD~19467~B and D/H $< 1.29 \times 10^{-4}$ for 2M~0415. These constraints are consistent with the protosolar value (D/H $= 2 \times 10^{-5}$; \citealt{geissAbundancesDeuteriumHelium31998}) and with the lower values inferred for Jupiter \citep{pierelRatiosSaturnJupiter2017}. The recent JWST/NIRSpec measurement of D/H $= (1.27^{+0.22}_{-0.19}) \times 10^{-4}$ in a Y dwarf \citep{rowlandProtosolarDtoHAbundance2024} suggests that D/H constraints may be more readily accessible for colder brown dwarfs, where methane features are stronger due to lower CO abundances.

\subsection{Additional trace molecules}\label{sec:trace_molecules}

Silicon oxide (SiO) can be abundant in the deep atmospheres of substellar objects in thermochemical equilibrium, but is expected to decrease above the silicate cloud base due to condensation \citep{allardLimitingEffectsDust2001,gaoAerosolsExoplanetAtmospheres2021}. At the photospheric temperatures of HD~19467~B and 2M~0415, Si-bearing species are expected to be largely sequestered by silicate condensation, and SiO is therefore not expected to be readily observable. While our retrievals return seemingly constrained posteriors for SiO in both objects (Table~\ref{tab:HD19467B_J0415_parameters}), we do not find significant SiO cross-correlation peaks (SNR $<3$; \Cref{fig:cross_correlation_molecules}), and we treat these constraints as tentative/non-detections.

Hydrogen sulfide (H$_2$S) is expected to contain most of the sulfur inventory in giant-planet and brown-dwarf atmospheres \citep{visscherAtmosphericChemistryGiant2006}, and has been detected in comparably cool atmospheres with high-dispersion techniques (e.g. \citealt{ruffioJupiterlikeUniformMetal2026}). In our retrievals, H$_2$S can appear constrained in 2M~0415, but we do not obtain a significant H$_2$S cross-correlation detection (SNR $<3$; \Cref{fig:cross_correlation_molecules}), and we therefore treat it as a tentative constraint at best. A key practical limitation is that the strongest H$_2$S opacity across the G395H window overlaps the NRS1/NRS2 detector gap (see \Cref{fig:hd19467b_spectrum}), reducing the effective leverage on the broad 3.6--4.0~\micron\ band; for HD~19467~B, sensitivity is further impacted because this band overlaps regions of strong speckle residuals (\Cref{sec:speckles,fig:speckle_field}).

Phosphine (PH$_3$) is expected in reducing, H$_2$-dominated atmospheres and its observability depends on metallicity, thermal structure, and disequilibrium chemistry \citep{visscherAtmosphericChemistryGiant2006,beilerTaleTwoMolecules2024,rowlandProtosolarDtoHAbundance2024,burgasserObservationUndepletedPhosphine2025}. We find no evidence for PH$_3$ in either target and report \(3\sigma\) upper limits of \(\log X_{\mathrm{PH}_3}<-9.3\) (HD~19467~B) and \(\log X_{\mathrm{PH}_3}<-8.4\) (2M~0415; Table~\ref{tab:HD19467B_J0415_parameters}).

\section{Conclusion}
We presented native-resolution (2.87--5.2~\micron; \(R\sim2700\)) JWST/NIRSpec retrievals for HD~19467~B and 2MASS~J0415$-$0935, with modelling of residual stellar contamination. Our main conclusions are:

\begin{enumerate}
    \item Speckle modelling is required for robust inference of directly-imaged companions in the presence of residual starlight: for HD~19467~B, speckles primarily affect continuum regions and the 3.0--3.7~\micron\ interval (\Cref{sec:speckles,fig:hd19467b_spec_comparison_hoch24_free}), and neglecting them biases retrieved parameters (e.g. NH$_3$ abundance; \Cref{fig:cornerplot_hd19467b}).

    \item We detect \water, \methane, CO, \COtwo, and NH$_3$ in both atmospheres (\Cref{sec:composition,fig:pt_composition}), and assess the detection of minor isotopologues of \water, \methane, CO and \COtwo\ (\Cref{fig:app_minor_isotopologues}).

    \item The observed CO absorption requires moderate vertical mixing with \(\log K_{\mathrm{zz}} \approx 3\)–\(4\)~dex (\Cref{sec:composition}), consistent with inhibited CO$\rightarrow$CH$_4$ conversion at low pressures.

    \item From CO isotopologues we infer \(^{12}\mathrm{C}/^{13}\mathrm{C}=154^{+19}_{-17}\) (HD~19467~B) and \(85\pm5\) (2M~0415), and \(^{16}\mathrm{O}/^{18}\mathrm{O}=711^{+204}_{-135}\) (HD~19467~B) and \(469^{+65}_{-53}\) (2M~0415) (\Cref{sec:isotopes,fig:isotope_ratios}). The higher ratios for the older HD~19467 system are qualitatively consistent with galactic chemical evolution.

    \item For SiO and H$_2$S, our retrievals can in some cases return seemingly constrained posteriors, but the lack of significant cross-correlation peaks implies these are tentative/non-detections (SNR $<3$; \Cref{sec:trace_molecules,fig:cross_correlation_molecules}). This is consistent with physical expectations for SiO (condensation at these photospheric temperatures) and with reduced H$_2$S coverage due to the NRS1/NRS2 detector gap (\Cref{fig:hd19467b_spectrum}) and speckle residuals around 3.6--4.0~\micron\ for HD~19467~B (\Cref{sec:speckles,fig:speckle_field}). We find no evidence for PH$_3$ and provide upper limits for both objects (see \Cref{tab:HD19467B_J0415_parameters}).

    \item We do not detect \dmethane\ and place 3$\sigma$ upper limits on methane D/H of D/H \(< 5.8 \times 10^{-4}\) (HD~19467~B) and D/H \(< 1.29 \times 10^{-4}\) (2M~0415), consistent with protosolar and Jovian values (\Cref{sec:isotopes}).

\end{enumerate}

Our results demonstrate that native-resolution retrievals of directly imaged companions with JWST/NIRSpec require careful treatment of residual stellar contamination to obtain unbiased atmospheric inferences. Over 2.87--5.2~\micron, we constrain a rich molecular inventory in both T dwarfs and access isotopic composition via CO isotopologues in the 4.6~\micron\ band. Our G395H-only fundamental-parameter estimates are broadly consistent with full-SED analyses for 2M~0415 \citep{merchanDiversityColdWorlds2025}, and for HD~19467~B the dynamical-mass prior is key to obtain physically meaningful \logg{} and abundances from this restricted wavelength window.

Looking ahead, extending this analysis to broader SEDs (e.g. combining NIRSpec with MIRI) will tighten constraints on radius, and \Teff{}, while improved treatments of instrumental systematics (including absolute flux calibration and correlated noise) will reduce biases in derived masses and trace-species abundances. Applying similar speckle-aware retrieval frameworks to larger samples of high-contrast companions will enable population-level tests of condensation, disequilibrium chemistry, and formation pathways through joint constraints on elemental and isotopic composition. As future ground- and space-based facilities move toward directly imaging temperate planets at small angular separations (e.g. ELT; \citealt{brandlMETISMidinfraredELT2021}, HWO; \citealt{mennessonCurrentLaboratoryPerformance2024}), their scientific return will be determined not only by raw contrast but also by advances in post-processing and forward-modelling frameworks that incorporate speckles into the atmospheric inference, as demonstrated here.

Data availability. The model spectra for HD 19467 B (including the speckle model), 2M 0415, and the temperature profiles presented in this work are available on Zenodo at \url{https://zenodo.org/records/19130087} \citep{gonzalezpicosJWSTHighcontrastSpectroscopy2026}.

\begin{acknowledgements}
We thank the referee for their careful reading of the manuscript and for their helpful comments and suggestions. We thank J.\,B.~Ruffio for making the extracted JWST/NIRSpec spectra of HD 19467 B and the speckles publicly accessible \footnote{\url{https://github.com/jruffio/HD_19467_B} commit hash b853a66}. We thank K.~Hoch for insightful discussions on atmospheric modelling. D.G.P and I.S. acknowledge NWO grant OCENW.M.21.010. Support for this work was provided by the NL-NWO Spinoza (SPI.2022.004). This work used the Dutch national e-infrastructure with the support of the SURF Cooperative using grant no. EINF-4556.
This work is based in part on observations made with the NASA/ESA/CSA James Webb Space Telescope. The data were obtained from the Mikulski Archive for Space Telescopes (MAST) at the Space Telescope Science Institute, which is operated by the Association of Universities for Research in Astronomy, Inc., under NASA contract NAS~5-03127 for JWST. These observations are associated with programme \#1414. This data can be found in MAST: \url{https://doi.org/10.17909/q524-zn59} (DOI creator: Ruffio, Jean-Baptiste).
\newline
Software: NumPy \citep{harrisArrayProgrammingNumPy2020}, SciPy \citep{virtanenSciPyFundamentalAlgorithms2020}, Matplotlib \citep{hunterMatplotlib2DGraphics2007}, jwst \citep{bushouseJWSTCalibrationPipeline2025}, petitRADTRANS \citep{mollierePetitRADTRANSPythonRadiative2019}, fastchem \citep{kitzmannFastchemCondEquilibrium2023}, ultranest \citep{buchnerUltraNestRobustGeneral2021}, PyMultiNest \citep{buchnerPyMultiNestPythonInterface2016}, Astropy \citep{collaborationAstropyProjectSustaining2022}, corner \citep{foreman-mackeyCornerPyScatterplot2016}, ExoMol \citep{tennyson2024ReleaseExoMol2024}, HITEMP \citep{rothmanHITEMPHightemperatureMolecular2010}, pyROX \citep{regtPyROXRapidOpacity2025}, and Claude Sonnet 4.x \citep{anthropic_claude_sonnet4_2026}.
\end{acknowledgements}

\bibliography{autolibrary}

@string{june = {June}}

@article{ackermanPrecipitatingCondensationClouds2001,
 adsurl = {https://ui.adsabs.harvard.edu/abs/2001ApJ...556..872A},
 archiveprefix = {arXiv},
 author = {{Ackerman}, Andrew S. and {Marley}, Mark S.},
 doi = {10.1086/321540},
 eprint = {astro-ph/0103423},
 journal = {\apj},
 month = {August},
 number = {2},
 pages = {872-884},
 primaryclass = {astro-ph},
 title = {{Precipitating Condensation Clouds in Substellar Atmospheres}},
 volume = {556},
 year = {2001}
}

@article{akaikeNewLookStatistical1974,
 adsurl = {https://ui.adsabs.harvard.edu/abs/1974ITAC...19..716A},
 author = {{Akaike}, H.},
 doi = {10.1109/TAC.1974.1100705},
 journal = {IEEE Transactions on Automatic Control},
 month = {January},
 pages = {716-723},
 title = {{A New Look at the Statistical Model Identification}},
 volume = {19},
 year = {1974}
}

@article{aldersonEarlyReleaseScience2023,
 adsurl = {https://ui.adsabs.harvard.edu/abs/2023Natur.614..664A},
 archiveprefix = {arXiv},
 author = {{Alderson}, Lili and {Wakeford}, Hannah R. and {Alam}, Munazza K. and {Batalha}, Natasha E. and {Lothringer}, Joshua D. and {Adams Redai}, Jea and {Barat}, Saugata and {Brande}, Jonathan and {Damiano}, Mario and {Daylan}, Tansu and {Espinoza}, N{\'e}stor and {Flagg}, Laura and {Goyal}, Jayesh M. and {Grant}, David and {Hu}, Renyu and {Inglis}, Julie and {Lee}, Elspeth K.~H. and {Mikal-Evans}, Thomas and {Ramos-Rosado}, Lakeisha and {Roy}, Pierre-Alexis and {Wallack}, Nicole L. and {Batalha}, Natalie M. and {Bean}, Jacob L. and {Benneke}, Bj{\"o}rn and {Berta-Thompson}, Zachory K. and {Carter}, Aarynn L. and {Changeat}, Quentin and {Col{\'o}n}, Knicole D. and {Crossfield}, Ian J.~M. and {D{\'e}sert}, Jean-Michel and {Foreman-Mackey}, Daniel and {Gibson}, Neale P. and {Kreidberg}, Laura and {Line}, Michael R. and {L{\'o}pez-Morales}, Mercedes and {Molaverdikhani}, Karan and {Moran}, Sarah E. and {Morello}, Giuseppe and {Moses}, Julianne I. and {Mukherjee}, Sagnick and {Schlawin}, Everett and {Sing}, David K. and {Stevenson}, Kevin B. and {Taylor}, Jake and {Aggarwal}, Keshav and {Ahrer}, Eva-Maria and {Allen}, Natalie H. and {Barstow}, Joanna K. and {Bell}, Taylor J. and {Blecic}, Jasmina and {Casewell}, Sarah L. and {Chubb}, Katy L. and {Crouzet}, Nicolas and {Cubillos}, Patricio E. and {Decin}, Leen and {Feinstein}, Adina D. and {Fortney}, Joanthan J. and {Harrington}, Joseph and {Heng}, Kevin and {Iro}, Nicolas and {Kempton}, Eliza M.-R. and {Kirk}, James and {Knutson}, Heather A. and {Krick}, Jessica and {Leconte}, J{\'e}r{\'e}my and {Lendl}, Monika and {MacDonald}, Ryan J. and {Mancini}, Luigi and {Mansfield}, Megan and {May}, Erin M. and {Mayne}, Nathan J. and {Miguel}, Yamila and {Nikolov}, Nikolay K. and {Ohno}, Kazumasa and {Palle}, Enric and {Parmentier}, Vivien and {Petit dit de la Roche}, Dominique J.~M. and {Piaulet}, Caroline and {Powell}, Diana and {Rackham}, Benjamin V. and {Redfield}, Seth and {Rogers}, Laura K. and {Rustamkulov}, Zafar and {Tan}, Xianyu and {Tremblin}, P. and {Tsai}, Shang-Min and {Turner}, Jake D. and {de Val-Borro}, Miguel and {Venot}, Olivia and {Welbanks}, Luis and {Wheatley}, Peter J. and {Zhang}, Xi},
 doi = {10.1038/s41586-022-05591-3},
 eprint = {2211.10488},
 journal = {\nat},
 month = {February},
 number = {7949},
 pages = {664-669},
 primaryclass = {astro-ph.EP},
 title = {{Early Release Science of the exoplanet WASP-39b with JWST NIRSpec G395H}},
 volume = {614},
 year = {2023}
}

@article{allardLimitingEffectsDust2001,
 adsurl = {https://ui.adsabs.harvard.edu/abs/2001ApJ...556..357A},
 archiveprefix = {arXiv},
 author = {{Allard}, France and {Hauschildt}, Peter H. and {Alexander}, David R. and {Tamanai}, Akemi and {Schweitzer}, Andreas},
 doi = {10.1086/321547},
 eprint = {astro-ph/0104256},
 journal = {\apj},
 month = {July},
 number = {1},
 pages = {357-372},
 primaryclass = {astro-ph},
 title = {{The Limiting Effects of Dust in Brown Dwarf Model Atmospheres}},
 volume = {556},
 year = {2001}
}

@article{allardModelsVerylowmassStars2012,
 adsurl = {https://ui.adsabs.harvard.edu/abs/2012RSPTA.370.2765A},
 archiveprefix = {arXiv},
 author = {{Allard}, F. and {Homeier}, D. and {Freytag}, B.},
 doi = {10.1098/rsta.2011.0269},
 eprint = {1112.3591},
 journal = {Philosophical Transactions of the Royal Society of London Series A},
 month = {June},
 number = {1968},
 pages = {2765-2777},
 primaryclass = {astro-ph.SR},
 title = {{Models of very-low-mass stars, brown dwarfs and exoplanets}},
 volume = {370},
 year = {2012}
}

@misc{anthropic_claude_sonnet4_2026,
 author = {{Anthropic}},
 howpublished = {\url{https://www.anthropic.com/claude}},
 note = {Large language model; accessed 2026-04-20},
 title = {{Claude Sonnet 4.x}},
 year = {2026}
}

@article{asplundChemicalMakeupSun2021,
 adsurl = {https://ui.adsabs.harvard.edu/abs/2021A&A...653A.141A},
 archiveprefix = {arXiv},
 author = {{Asplund}, M. and {Amarsi}, A.~M. and {Grevesse}, N.},
 doi = {10.1051/0004-6361/202140445},
 eid = {A141},
 eprint = {2105.01661},
 journal = {\aap},
 month = {September},
 pages = {A141},
 primaryclass = {astro-ph.SR},
 title = {{The chemical make-up of the Sun: A 2020 vision}},
 volume = {653},
 year = {2021}
}

@article{ayresSUNLIGHTEREARTH2013,
 adsurl = {https://ui.adsabs.harvard.edu/abs/2013ApJ...765...46A},
 archiveprefix = {arXiv},
 author = {{Ayres}, Thomas R. and {Lyons}, J.~R. and {Ludwig}, H.-G. and {Caffau}, E. and {Wedemeyer-B{\"o}hm}, S.},
 doi = {10.1088/0004-637X/765/1/46},
 eid = {46},
 eprint = {1301.5281},
 journal = {\apj},
 month = {March},
 number = {1},
 pages = {46},
 primaryclass = {astro-ph.SR},
 title = {{Is the Sun Lighter than the Earth? Isotopic CO in the Photosphere, Viewed through the Lens of Three-dimensional Spectrum Synthesis}},
 volume = {765},
 year = {2013}
}

@article{azzamExoMolMolecularLine2016,
 adsurl = {https://ui.adsabs.harvard.edu/abs/2016MNRAS.460.4063A},
 archiveprefix = {arXiv},
 author = {{Azzam}, Ala'a. A.~A. and {Tennyson}, Jonathan and {Yurchenko}, Sergei N. and {Naumenko}, Olga V.},
 doi = {10.1093/mnras/stw1133},
 eprint = {1607.00499},
 journal = {\mnras},
 month = {August},
 number = {4},
 pages = {4063-4074},
 primaryclass = {astro-ph.EP},
 title = {{ExoMol molecular line lists - XVI. The rotation-vibration spectrum of hot H$_{2}$S}},
 volume = {460},
 year = {2016}
}

@article{beilerFirstJWSTSpectral2023,
 adsurl = {https://ui.adsabs.harvard.edu/abs/2023ApJ...951L..48B},
 archiveprefix = {arXiv},
 author = {{Beiler}, Samuel A. and {Cushing}, Michael C. and {Kirkpatrick}, J. Davy and {Schneider}, Adam C. and {Mukherjee}, Sagnick and {Marley}, Mark S.},
 doi = {10.3847/2041-8213/ace32c},
 eid = {L48},
 eprint = {2306.11807},
 journal = {\apjl},
 month = {July},
 number = {2},
 pages = {L48},
 primaryclass = {astro-ph.SR},
 title = {{The First JWST Spectral Energy Distribution of a Y Dwarf}},
 volume = {951},
 year = {2023}
}

@article{beilerTaleTwoMolecules2024,
 adsurl = {https://ui.adsabs.harvard.edu/abs/2024ApJ...973...60B},
 archiveprefix = {arXiv},
 author = {{Beiler}, Samuel A. and {Mukherjee}, Sagnick and {Cushing}, Michael C. and {Kirkpatrick}, J. Davy and {Schneider}, Adam C. and {Kothari}, Harshil and {Marley}, Mark S. and {Visscher}, Channon},
 doi = {10.3847/1538-4357/ad6759},
 eid = {60},
 eprint = {2407.15950},
 journal = {\apj},
 month = {September},
 number = {1},
 pages = {60},
 primaryclass = {astro-ph.EP},
 title = {{A Tale of Two Molecules: The Underprediction of CO$_{2}$ and Overprediction of PH$_{3}$ in Late T and Y Dwarf Atmospheric Models}},
 volume = {973},
 year = {2024}
}

@article{bezardCarbonMonoxideJupiter2002,
 adsurl = {https://ui.adsabs.harvard.edu/abs/2002Icar..159...95B},
 author = {{B{\'e}zard}, Bruno and {Lellouch}, Emmanuel and {Strobel}, Darrell and {Maillard}, Jean-Pierre and {Drossart}, Pierre},
 doi = {10.1006/icar.2002.6917},
 journal = {\icarus},
 month = {September},
 number = {1},
 pages = {95-111},
 title = {{Carbon Monoxide on Jupiter: Evidence for Both Internal and External Sources}},
 volume = {159},
 year = {2002}
}

@article{blainSpectralModelHighresolutionFramework2024,
 adsurl = {https://ui.adsabs.harvard.edu/abs/2024JOSS....9.7028B},
 author = {{Blain}, Doriann and {Molli{\`e}re}, Paul and {Nasedkin}, Evert},
 doi = {10.21105/joss.07028},
 eid = {7028},
 journal = {The Journal of Open Source Software},
 month = {October},
 number = {102},
 pages = {7028},
 title = {{SpectralModel: a high-resolution framework for petitRADTRANS 3}},
 volume = {9},
 year = {2024}
}

@article{bouvierAgeSolarSystem2010,
 adsurl = {https://ui.adsabs.harvard.edu/abs/2010NatGe...3..637B},
 author = {{Bouvier}, Audrey and {Wadhwa}, Meenakshi},
 doi = {10.1038/ngeo941},
 journal = {Nature Geoscience},
 month = {September},
 number = {9},
 pages = {637-641},
 title = {{The age of the Solar System redefined by the oldest Pb-Pb age of a meteoritic inclusion}},
 volume = {3},
 year = {2010}
}

@article{brandlMETISMidinfraredELT2021,
 adsurl = {https://ui.adsabs.harvard.edu/abs/2021Msngr.182...22B},
 archiveprefix = {arXiv},
 author = {{Brandl}, B. and {Bettonvil}, F. and {van Boekel}, R. and {Glauser}, A. and {Quanz}, S. and {Absil}, O. and {Amorim}, A. and {Feldt}, M. and {Glasse}, A. and {G{\"u}del}, M. and {Ho}, P. and {Labadie}, L. and {Meyer}, M. and {Pantin}, E. and {van Winckel}, H. and {METIS Consortium}},
 doi = {10.18727/0722-6691/5218},
 eprint = {2103.11208},
 journal = {The Messenger},
 month = {March},
 pages = {22-26},
 primaryclass = {astro-ph.IM},
 title = {{METIS: The Mid-infrared ELT Imager and Spectrograph}},
 volume = {182},
 year = {2021}
}

@inproceedings{buchnerComparisonStepSamplers2023,
 adsurl = {https://ui.adsabs.harvard.edu/abs/2022PSFor...5...46B},
 archiveprefix = {arXiv},
 author = {{Buchner}, Johannes},
 booktitle = {Physical Sciences Forum},
 doi = {10.3390/psf2022005046},
 eid = {46},
 eprint = {2211.09426},
 month = {December},
 pages = {46},
 primaryclass = {stat.CO},
 series = {Physical Sciences Forum},
 title = {{Comparison of Step Samplers for Nested Sampling}},
 volume = {5},
 year = {2022}
}

@article{buchnerNestedSamplingMethods2023,
 adsurl = {https://ui.adsabs.harvard.edu/abs/2023StSur..17..169B},
 archiveprefix = {arXiv},
 author = {{Buchner}, Johannes},
 doi = {10.1214/23-SS144},
 eprint = {2101.09675},
 journal = {Statistics Surveys},
 month = {January},
 pages = {169-215},
 primaryclass = {stat.CO},
 title = {{Nested Sampling Methods}},
 volume = {17},
 year = {2023}
}

@software{buchnerPyMultiNestPythonInterface2016,
 adsurl = {https://ui.adsabs.harvard.edu/abs/2016ascl.soft06005B},
 archiveprefix = {ascl},
 author = {{Buchner}, Johannes},
 eid = {ascl:1606.005},
 eprint = {1606.005},
 howpublished = {Astrophysics Source Code Library, record ascl:1606.005},
 month = {June},
 title = {{PyMultiNest: Python interface for MultiNest}},
 year = {2016}
}

@article{buchnerUltraNestRobustGeneral2021,
 adsurl = {https://ui.adsabs.harvard.edu/abs/2021JOSS....6.3001B},
 archiveprefix = {arXiv},
 author = {{Buchner}, Johannes},
 doi = {10.21105/joss.03001},
 eid = {3001},
 eprint = {2101.09604},
 journal = {The Journal of Open Source Software},
 month = {April},
 number = {60},
 pages = {3001},
 primaryclass = {stat.CO},
 title = {{UltraNest - a robust, general purpose Bayesian inference engine}},
 volume = {6},
 year = {2021}
}

@article{burgasserEvidenceCloudDisruption2002,
 adsurl = {https://ui.adsabs.harvard.edu/abs/2002ApJ...571L.151B},
 archiveprefix = {arXiv},
 author = {{Burgasser}, Adam J. and {Marley}, Mark S. and {Ackerman}, Andrew S. and {Saumon}, Didier and {Lodders}, Katharina and {Dahn}, Conard C. and {Harris}, Hugh C. and {Kirkpatrick}, J. Davy},
 doi = {10.1086/341343},
 eprint = {astro-ph/0205051},
 journal = {\apjl},
 month = {June},
 number = {2},
 pages = {L151-L154},
 primaryclass = {astro-ph},
 title = {{Evidence of Cloud Disruption in the L/T Dwarf Transition}},
 volume = {571},
 year = {2002}
}

@article{burgasserObservationUndepletedPhosphine2025,
 adsurl = {https://ui.adsabs.harvard.edu/abs/2025Sci...390..697B},
 archiveprefix = {arXiv},
 author = {{Burgasser}, Adam J. and {Gonzales}, Eileen C. and {Beiler}, Samuel A. and {Visscher}, Channon and {Burningham}, Ben and {Mace}, Gregory N. and {Faherty}, Jacqueline K. and {Zhang}, Zenghua and {Sousa-Silva}, Clara and {Lodieu}, Nicolas and {Metchev}, Stanimir A. and {Meisner}, Aaron and {Cushing}, Michael and {Schneider}, Adam C. and {Suarez}, Genaro and {Hsu}, Chih-Chun and {Gerasimov}, Roman and {Aganze}, Christian and {Theissen}, Christopher A.},
 doi = {10.1126/science.adu0401},
 eprint = {2510.03916},
 journal = {Science},
 month = {November},
 number = {6774},
 pages = {697-701},
 primaryclass = {astro-ph.SR},
 title = {{Observation of undepleted phosphine in the atmosphere of a low-temperature brown dwarf}},
 volume = {390},
 year = {2025}
}

@article{burgasserSpectraDwarfsNearInfrared2002,
 adsurl = {https://ui.adsabs.harvard.edu/abs/2002ApJ...564..421B},
 archiveprefix = {arXiv},
 author = {{Burgasser}, Adam J. and {Kirkpatrick}, J. Davy and {Brown}, Michael E. and {Reid}, I. Neill and {Burrows}, Adam and {Liebert}, James and {Matthews}, Keith and {Gizis}, John E. and {Dahn}, Conard C. and {Monet}, David G. and {Cutri}, Roc M. and {Skrutskie}, Michael F.},
 doi = {10.1086/324033},
 eprint = {astro-ph/0108452},
 journal = {\apj},
 month = {January},
 number = {1},
 pages = {421-451},
 primaryclass = {astro-ph},
 title = {{The Spectra of T Dwarfs. I. Near-Infrared Data and Spectral Classification}},
 volume = {564},
 year = {2002}
}

@article{burninghamRetrievalAtmosphericProperties2017,
 adsurl = {https://ui.adsabs.harvard.edu/abs/2017MNRAS.470.1177B},
 archiveprefix = {arXiv},
 author = {{Burningham}, Ben and {Marley}, M.~S. and {Line}, M.~R. and {Lupu}, R. and {Visscher}, C. and {Morley}, C.~V. and {Saumon}, D. and {Freedman}, R.},
 doi = {10.1093/mnras/stx1246},
 eprint = {1701.01257},
 journal = {\mnras},
 month = {September},
 number = {1},
 pages = {1177-1197},
 primaryclass = {astro-ph.SR},
 title = {{Retrieval of atmospheric properties of cloudy L dwarfs}},
 volume = {470},
 year = {2017}
}

@article{burrowsChemicalEquilibriumAbundances1999,
 adsurl = {https://ui.adsabs.harvard.edu/abs/1999ApJ...512..843B},
 archiveprefix = {arXiv},
 author = {{Burrows}, Adam and {Sharp}, C.~M.},
 doi = {10.1086/306811},
 eprint = {astro-ph/9807055},
 journal = {\apj},
 month = {February},
 number = {2},
 pages = {843-863},
 primaryclass = {astro-ph},
 title = {{Chemical Equilibrium Abundances in Brown Dwarf and Extrasolar Giant Planet Atmospheres}},
 volume = {512},
 year = {1999}
}

@article{burrowsDwarfModelsTransition2006,
 adsurl = {https://ui.adsabs.harvard.edu/abs/2006ApJ...640.1063B},
 archiveprefix = {arXiv},
 author = {{Burrows}, Adam and {Sudarsky}, David and {Hubeny}, Ivan},
 doi = {10.1086/500293},
 eprint = {astro-ph/0509066},
 journal = {\apj},
 month = {April},
 number = {2},
 pages = {1063-1077},
 primaryclass = {astro-ph},
 title = {{L and T Dwarf Models and the L to T Transition}},
 volume = {640},
 year = {2006}
}

@article{burrowsDwarfsTheoreticalSpectra2003,
 adsurl = {https://ui.adsabs.harvard.edu/abs/2003ApJ...596..587B},
 archiveprefix = {arXiv},
 author = {{Burrows}, Adam and {Sudarsky}, David and {Lunine}, Jonathan I.},
 doi = {10.1086/377709},
 eprint = {astro-ph/0304226},
 journal = {\apj},
 month = {October},
 number = {1},
 pages = {587-596},
 primaryclass = {astro-ph},
 title = {{Beyond the T Dwarfs: Theoretical Spectra, Colors, and Detectability of the Coolest Brown Dwarfs}},
 volume = {596},
 year = {2003}
}

@article{burrowsTheoryBrownDwarfs2001,
 adsurl = {https://ui.adsabs.harvard.edu/abs/2001RvMP...73..719B},
 archiveprefix = {arXiv},
 author = {{Burrows}, Adam and {Hubbard}, W.~B. and {Lunine}, J.~I. and {Liebert}, James},
 doi = {10.1103/RevModPhys.73.719},
 eprint = {astro-ph/0103383},
 journal = {Reviews of Modern Physics},
 month = {July},
 number = {3},
 pages = {719-765},
 primaryclass = {astro-ph},
 title = {{The theory of brown dwarfs and extrasolar giant planets}},
 volume = {73},
 year = {2001}
}

@software{bushouseJWSTCalibrationPipeline2025,
 adsurl = {https://ui.adsabs.harvard.edu/abs/2025zndo..15178003B},
 author = {{Bushouse}, Howard and {Eisenhamer}, Jonathan and {Dencheva}, Nadia and {Davies}, James and {Greenfield}, Perry and {Morrison}, Jane and {Hodge}, Phil and {Simon}, Bernie and {Grumm}, David and {Droettboom}, Michael and {Slavich}, Edward and {Sosey}, Megan and {Pauly}, Tyler and {Miller}, Todd and {Jedrzejewski}, Robert and {Hack}, Warren and {Davis}, David and {Crawford}, Steven and {Law}, David and {Gordon}, Karl and {Regan}, Michael and {Cara}, Mihai and {MacDonald}, Ken and {Bradley}, Larry and {Shanahan}, Clare and {Jamieson}, William and {Teodoro}, Mairan and {Williams}, Thomas and {Pena-Guerrero}, Maria and {Graham}, Brett and {Molter}, Edward and {Brandt}, Timothy and {Hayes}, Christian and {Cooper}, Rachel and {Clarke}, Melanie and {Filippazzo}, Joseph},
 doi = {10.5281/zenodo.15178003},
 eid = {10.5281/zenodo.15178003},
 month = {April},
 publisher = {Zenodo},
 title = {{JWST Calibration Pipeline}},
 version = {1.18.0},
 year = {2025}
}

@article{calamariAtmosphericRetrievalBrown2022,
 adsurl = {https://ui.adsabs.harvard.edu/abs/2022ApJ...940..164C},
 archiveprefix = {arXiv},
 author = {{Calamari}, Emily and {Faherty}, Jacqueline K. and {Burningham}, Ben and {Gonzales}, Eileen and {Bardalez-Gagliuffi}, Daniella and {Vos}, Johanna M. and {Gemma}, Marina and {Whiteford}, Niall and {Gaarn}, Josefine},
 doi = {10.3847/1538-4357/ac9cc9},
 eid = {164},
 eprint = {2210.13614},
 journal = {\apj},
 month = {December},
 number = {2},
 pages = {164},
 primaryclass = {astro-ph.SR},
 title = {{An Atmospheric Retrieval of the Brown Dwarf Gliese 229B}},
 volume = {940},
 year = {2022}
}

@article{calamariPredictingCloudConditions2024,
 adsurl = {https://ui.adsabs.harvard.edu/abs/2024ApJ...963...67C},
 archiveprefix = {arXiv},
 author = {{Calamari}, Emily and {Faherty}, Jacqueline K. and {Visscher}, Channon and {Gemma}, Marina E. and {Burningham}, Ben and {Rothermich}, Austin},
 doi = {10.3847/1538-4357/ad1f6d},
 eid = {67},
 eprint = {2401.11038},
 journal = {\apj},
 month = {March},
 number = {1},
 pages = {67},
 primaryclass = {astro-ph.SR},
 title = {{Predicting Cloud Conditions in Substellar Mass Objects Using Ultracool Dwarf Companions}},
 volume = {963},
 year = {2024}
}

@article{chubbMarvelAnalysisMeasured2018,
 adsurl = {https://ui.adsabs.harvard.edu/abs/2018JQSRT.218..178C},
 archiveprefix = {arXiv},
 author = {{Chubb}, Katy L. and {Naumenko}, Olga and {Keely}, Stefan and {Bartolotto}, Sebestiano and {Macdonald}, Skye and {Mukhtar}, Mahmoud and {Grachov}, Andrey and {White}, Joe and {Coleman}, Eden and {Liu}, Anwen and {Fazliev}, Alexander Z. and {Polovtseva}, Elena R. and {Horneman}, Veli-Matti and {Campargue}, Alain and {Furtenbacher}, Tibor and {Cs{\'a}sz{\'a}r}, Attila G. and {Yurchenko}, Sergei N. and {Tennyson}, Jonathan},
 doi = {10.1016/j.jqsrt.2018.07.012},
 eprint = {1812.10503},
 journal = {\jqsrt},
 month = {October},
 pages = {178-186},
 primaryclass = {astro-ph.EP},
 title = {{MARVEL analysis of the measured high-resolution rovibrational spectra of H$_{2}$$^{32}$S}},
 volume = {218},
 year = {2018}
}

@article{colesExoMolMolecularLine2019,
 adsurl = {https://ui.adsabs.harvard.edu/abs/2019MNRAS.490.4638C},
 archiveprefix = {arXiv},
 author = {{Coles}, Phillip A. and {Yurchenko}, Sergei N. and {Tennyson}, Jonathan},
 doi = {10.1093/mnras/stz2778},
 eprint = {1911.10369},
 journal = {\mnras},
 month = {December},
 number = {4},
 pages = {4638-4647},
 primaryclass = {astro-ph.SR},
 title = {{ExoMol molecular line lists - XXXV. A rotation-vibration line list for hot ammonia}},
 volume = {490},
 year = {2019}
}

@article{collaborationAstropyProjectSustaining2022,
 adsurl = {https://ui.adsabs.harvard.edu/abs/2022ApJ...935..167A},
 archiveprefix = {arXiv},
 author = {{Astropy Collaboration} and {Price-Whelan}, Adrian M. and {Lim}, Pey Lian and {Earl}, Nicholas and {Starkman}, Nathaniel and {Bradley}, Larry and {Shupe}, David L. and {Patil}, Aarya A. and {Corrales}, Lia and {Brasseur}, C.~E. and {N{\"o}the}, Maximilian and {Donath}, Axel and {Tollerud}, Erik and {Morris}, Brett M. and {Ginsburg}, Adam and {Vaher}, Eero and {Weaver}, Benjamin A. and {Tocknell}, James and {Jamieson}, William and {van Kerkwijk}, Marten H. and {Robitaille}, Thomas P. and {Merry}, Bruce and {Bachetti}, Matteo and {G{\"u}nther}, H. Moritz and {Aldcroft}, Thomas L. and {Alvarado-Montes}, Jaime A. and {Archibald}, Anne M. and {B{\'o}di}, Attila and {Bapat}, Shreyas and {Barentsen}, Geert and {Baz{\'a}n}, Juanjo and {Biswas}, Manish and {Boquien}, M{\'e}d{\'e}ric and {Burke}, D.~J. and {Cara}, Daria and {Cara}, Mihai and {Conroy}, Kyle E. and {Conseil}, Simon and {Craig}, Matthew W. and {Cross}, Robert M. and {Cruz}, Kelle L. and {D'Eugenio}, Francesco and {Dencheva}, Nadia and {Devillepoix}, Hadrien A.~R. and {Dietrich}, J{\"o}rg P. and {Eigenbrot}, Arthur Davis and {Erben}, Thomas and {Ferreira}, Leonardo and {Foreman-Mackey}, Daniel and {Fox}, Ryan and {Freij}, Nabil and {Garg}, Suyog and {Geda}, Robel and {Glattly}, Lauren and {Gondhalekar}, Yash and {Gordon}, Karl D. and {Grant}, David and {Greenfield}, Perry and {Groener}, Austen M. and {Guest}, Steve and {Gurovich}, Sebastian and {Handberg}, Rasmus and {Hart}, Akeem and {Hatfield-Dodds}, Zac and {Homeier}, Derek and {Hosseinzadeh}, Griffin and {Jenness}, Tim and {Jones}, Craig K. and {Joseph}, Prajwel and {Kalmbach}, J. Bryce and {Karamehmetoglu}, Emir and {Ka{\l}uszy{\'n}ski}, Miko{\l}aj and {Kelley}, Michael S.~P. and {Kern}, Nicholas and {Kerzendorf}, Wolfgang E. and {Koch}, Eric W. and {Kulumani}, Shankar and {Lee}, Antony and {Ly}, Chun and {Ma}, Zhiyuan and {MacBride}, Conor and {Maljaars}, Jakob M. and {Muna}, Demitri and {Murphy}, N.~A. and {Norman}, Henrik and {O'Steen}, Richard and {Oman}, Kyle A. and {Pacifici}, Camilla and {Pascual}, Sergio and {Pascual-Granado}, J. and {Patil}, Rohit R. and {Perren}, Gabriel I. and {Pickering}, Timothy E. and {Rastogi}, Tanuj and {Roulston}, Benjamin R. and {Ryan}, Daniel F. and {Rykoff}, Eli S. and {Sabater}, Jose and {Sakurikar}, Parikshit and {Salgado}, Jes{\'u}s and {Sanghi}, Aniket and {Saunders}, Nicholas and {Savchenko}, Volodymyr and {Schwardt}, Ludwig and {Seifert-Eckert}, Michael and {Shih}, Albert Y. and {Jain}, Anany Shrey and {Shukla}, Gyanendra and {Sick}, Jonathan and {Simpson}, Chris and {Singanamalla}, Sudheesh and {Singer}, Leo P. and {Singhal}, Jaladh and {Sinha}, Manodeep and {Sip{\H{o}}cz}, Brigitta M. and {Spitler}, Lee R. and {Stansby}, David and {Streicher}, Ole and {{\v{S}}umak}, Jani and {Swinbank}, John D. and {Taranu}, Dan S. and {Tewary}, Nikita and {Tremblay}, Grant R. and {de Val-Borro}, Miguel and {Van Kooten}, Samuel J. and {Vasovi{\'c}}, Zlatan and {Verma}, Shresth and {de Miranda Cardoso}, Jos{\'e} Vin{\'\i}cius and {Williams}, Peter K.~G. and {Wilson}, Tom J. and {Winkel}, Benjamin and {Wood-Vasey}, W.~M. and {Xue}, Rui and {Yoachim}, Peter and {Zhang}, Chen and {Zonca}, Andrea and {Astropy Project Contributors}},
 doi = {10.3847/1538-4357/ac7c74},
 eid = {167},
 eprint = {2206.14220},
 journal = {\apj},
 month = {August},
 number = {2},
 pages = {167},
 primaryclass = {astro-ph.IM},
 title = {{The Astropy Project: Sustaining and Growing a Community-oriented Open-source Project and the Latest Major Release (v5.0) of the Core Package}},
 volume = {935},
 year = {2022}
}

@article{creppDIRECTSPECTRUMBENCHMARK2015,
 adsurl = {https://ui.adsabs.harvard.edu/abs/2015ApJ...798L..43C},
 archiveprefix = {arXiv},
 author = {{Crepp}, Justin R. and {Rice}, Emily L. and {Veicht}, Aaron and {Aguilar}, Jonathan and {Pueyo}, Laurent and {Giorla}, Paige and {Nilsson}, Ricky and {Luszcz-Cook}, Statia H. and {Oppenheimer}, Rebecca and {Hinkley}, Sasha and {Brenner}, Douglas and {Vasisht}, Gautam and {Cady}, Eric and {Beichman}, Charles A. and {Hillenbrand}, Lynne A. and {Lockhart}, Thomas and {Matthews}, Christopher T. and {Roberts}, Jr., Lewis C. and {Sivaramakrishnan}, Anand and {Soummer}, Remi and {Zhai}, Chengxing},
 doi = {10.1088/2041-8205/798/2/L43},
 eid = {L43},
 eprint = {1412.6101},
 journal = {\apjl},
 month = {January},
 number = {2},
 pages = {L43},
 primaryclass = {astro-ph.SR},
 title = {{Direct Spectrum of the Benchmark T Dwarf HD 19467 B}},
 volume = {798},
 year = {2015}
}

@article{creppTRENDSHIGHCONTRASTIMAGING2014,
 adsurl = {https://ui.adsabs.harvard.edu/abs/2014ApJ...781...29C},
 archiveprefix = {arXiv},
 author = {{Crepp}, Justin R. and {Johnson}, John Asher and {Howard}, Andrew W. and {Marcy}, Geoffrey W. and {Brewer}, John and {Fischer}, Debra A. and {Wright}, Jason T. and {Isaacson}, Howard},
 doi = {10.1088/0004-637X/781/1/29},
 eid = {29},
 eprint = {1311.0280},
 journal = {\apj},
 month = {January},
 number = {1},
 pages = {29},
 primaryclass = {astro-ph.EP},
 title = {{The TRENDS High-contrast Imaging Survey. V. Discovery of an Old and Cold Benchmark T-dwarf Orbiting the Nearby G-star HD 19467}},
 volume = {781},
 year = {2014}
}

@article{crossfieldUnusualIsotopicAbundances2019,
 adsurl = {https://ui.adsabs.harvard.edu/abs/2019ApJ...871L...3C},
 archiveprefix = {arXiv},
 author = {{Crossfield}, I.~J.~M. and {Lothringer}, J.~D. and {Flores}, B. and {Mills}, E.~A.~C. and {Freedman}, R. and {Valverde}, J. and {Miles}, B. and {Guo}, X. and {Skemer}, A.},
 doi = {10.3847/2041-8213/aaf9b6},
 eid = {L3},
 eprint = {1901.02607},
 journal = {\apjl},
 month = {January},
 number = {1},
 pages = {L3},
 primaryclass = {astro-ph.SR},
 title = {{Unusual Isotopic Abundances in a Fully Convective Stellar Binary}},
 volume = {871},
 year = {2019}
}

@inproceedings{currieDirectImagingSpectroscopy2023,
 adsurl = {https://ui.adsabs.harvard.edu/abs/2023ASPC..534..799C},
 archiveprefix = {arXiv},
 author = {{Currie}, T. and {Biller}, B. and {Lagrange}, A. and {Marois}, C. and {Guyon}, O. and {Nielsen}, E.~L. and {Bonnefoy}, M. and {De Rosa}, R.~J.},
 booktitle = {Protostars and Planets VII},
 doi = {10.48550/arXiv.2205.05696},
 editor = {{Inutsuka}, S. and {Aikawa}, Y. and {Muto}, T. and {Tomida}, K. and {Tamura}, M.},
 eprint = {2205.05696},
 month = {July},
 pages = {799},
 primaryclass = {astro-ph.EP},
 series = {Astronomical Society of the Pacific Conference Series},
 title = {{Direct Imaging and Spectroscopy of Extrasolar Planets}},
 volume = {534},
 year = {2023}
}

@article{cushingInfraredSpectroscopicSequence2005,
 adsurl = {https://ui.adsabs.harvard.edu/abs/2005ApJ...623.1115C},
 archiveprefix = {arXiv},
 author = {{Cushing}, Michael C. and {Rayner}, John T. and {Vacca}, William D.},
 doi = {10.1086/428040},
 eprint = {astro-ph/0412313},
 journal = {\apj},
 month = {April},
 number = {2},
 pages = {1115-1140},
 primaryclass = {astro-ph},
 title = {{An Infrared Spectroscopic Sequence of M, L, and T Dwarfs}},
 volume = {623},
 year = {2005}
}

@article{dupuyHAWAIIINFRAREDPARALLAX2012,
 adsurl = {https://ui.adsabs.harvard.edu/abs/2012ApJS..201...19D},
 archiveprefix = {arXiv},
 author = {{Dupuy}, Trent J. and {Liu}, Michael C.},
 doi = {10.1088/0067-0049/201/2/19},
 eid = {19},
 eprint = {1201.2465},
 journal = {\apjs},
 month = {August},
 number = {2},
 pages = {19},
 primaryclass = {astro-ph.SR},
 title = {{The Hawaii Infrared Parallax Program. I. Ultracool Binaries and the L/T Transition}},
 volume = {201},
 year = {2012}
}

@article{fahertyMethaneEmissionCool2024,
 adsurl = {https://ui.adsabs.harvard.edu/abs/2024Natur.628..511F},
 archiveprefix = {arXiv},
 author = {{Faherty}, Jacqueline K. and {Burningham}, Ben and {Gagn{\'e}}, Jonathan and {Su{\'a}rez}, Genaro and {Vos}, Johanna M. and {Alejandro Merchan}, Sherelyn and {Morley}, Caroline V. and {Rowland}, Melanie and {Lacy}, Brianna and {Kiman}, Rocio and {Caselden}, Dan and {Kirkpatrick}, J. Davy and {Meisner}, Aaron and {Schneider}, Adam C. and {Kuchner}, Marc Jason and {Bardalez Gagliuffi}, Daniella Carolina and {Beichman}, Charles and {Eisenhardt}, Peter and {Gelino}, Christopher R. and {Gharib-Nezhad}, Ehsan and {Gonzales}, Eileen and {Marocco}, Federico and {Rothermich}, Austin James and {Whiteford}, Niall},
 doi = {10.1038/s41586-024-07190-w},
 eprint = {2404.10977},
 journal = {\nat},
 month = {April},
 number = {8008},
 pages = {511-514},
 primaryclass = {astro-ph.SR},
 title = {{Methane emission from a cool brown dwarf}},
 volume = {628},
 year = {2024}
}

@article{fahertyPOPULATIONPROPERTIESBROWN2016,
 adsurl = {https://ui.adsabs.harvard.edu/abs/2016ApJS..225...10F},
 archiveprefix = {arXiv},
 author = {{Faherty}, Jacqueline K. and {Riedel}, Adric R. and {Cruz}, Kelle L. and {Gagne}, Jonathan and {Filippazzo}, Joseph C. and {Lambrides}, Erini and {Fica}, Haley and {Weinberger}, Alycia and {Thorstensen}, John R. and {Tinney}, C.~G. and {Baldassare}, Vivienne and {Lemonier}, Emily and {Rice}, Emily L.},
 doi = {10.3847/0067-0049/225/1/10},
 eid = {10},
 eprint = {1605.07927},
 journal = {\apjs},
 month = {July},
 number = {1},
 pages = {10},
 primaryclass = {astro-ph.SR},
 title = {{Population Properties of Brown Dwarf Analogs to Exoplanets}},
 volume = {225},
 year = {2016}
}

@article{fegleyAtmosphericChemistryBrown1996,
 adsurl = {https://ui.adsabs.harvard.edu/abs/1996ApJ...472L..37F},
 author = {{Fegley}, Jr., Bruce and {Lodders}, Katharina},
 doi = {10.1086/310356},
 journal = {\apjl},
 month = {November},
 pages = {L37},
 title = {{Atmospheric Chemistry of the Brown Dwarf Gliese 229B: Thermochemical Equilibrium Predictions}},
 volume = {472},
 year = {1996}
}

@article{filippazzoFUNDAMENTALPARAMETERSSPECTRAL2015,
 adsurl = {https://ui.adsabs.harvard.edu/abs/2015ApJ...810..158F},
 archiveprefix = {arXiv},
 author = {{Filippazzo}, Joseph C. and {Rice}, Emily L. and {Faherty}, Jacqueline and {Cruz}, Kelle L. and {Van Gordon}, Mollie M. and {Looper}, Dagny L.},
 doi = {10.1088/0004-637X/810/2/158},
 eid = {158},
 eprint = {1508.01767},
 journal = {\apj},
 month = {September},
 number = {2},
 pages = {158},
 primaryclass = {astro-ph.SR},
 title = {{Fundamental Parameters and Spectral Energy Distributions of Young and Field Age Objects with Masses Spanning the Stellar to Planetary Regime}},
 volume = {810},
 year = {2015}
}

@article{foreman-mackeyCornerPyScatterplot2016,
 adsurl = {https://ui.adsabs.harvard.edu/abs/2016JOSS....1...24F},
 author = {{Foreman-Mackey}, Daniel},
 doi = {10.21105/joss.00024},
 journal = {The Journal of Open Source Software},
 month = {June},
 pages = {24},
 title = {{corner.py: Scatterplot matrices in Python}},
 volume = {1},
 year = {2016}
}

@article{fortneyCARBONTOOXYGENRATIOMEASUREMENT2012,
 adsurl = {https://ui.adsabs.harvard.edu/abs/2012ApJ...747L..27F},
 archiveprefix = {arXiv},
 author = {{Fortney}, Jonathan J.},
 doi = {10.1088/2041-8205/747/2/L27},
 eid = {L27},
 eprint = {1201.1504},
 journal = {\apjl},
 month = {March},
 number = {2},
 pages = {L27},
 primaryclass = {astro-ph.SR},
 title = {{On the Carbon-to-oxygen Ratio Measurement in nearby Sun-like Stars: Implications for Planet Formation and the Determination of Stellar Abundances}},
 volume = {747},
 year = {2012}
}

@article{gandhiJWSTMeasurements132023,
 adsurl = {https://ui.adsabs.harvard.edu/abs/2023ApJ...957L..36G},
 archiveprefix = {arXiv},
 author = {{Gandhi}, Siddharth and {de Regt}, Sam and {Snellen}, Ignas and {Zhang}, Yapeng and {Rugers}, Benson and {van Leur}, Niels and {Bosschaart}, Quincy},
 doi = {10.3847/2041-8213/ad07e2},
 eid = {L36},
 eprint = {2311.05349},
 journal = {\apjl},
 month = {November},
 number = {2},
 pages = {L36},
 primaryclass = {astro-ph.EP},
 title = {{JWST Measurements of $^{13}$C, $^{18}$O, and $^{17}$O in the Atmosphere of Super-Jupiter VHS 1256 b}},
 volume = {957},
 year = {2023}
}

@article{gaoAerosolsExoplanetAtmospheres2021,
 adsurl = {https://ui.adsabs.harvard.edu/abs/2021JGRE..12606655G},
 archiveprefix = {arXiv},
 author = {{Gao}, Peter and {Wakeford}, Hannah R. and {Moran}, Sarah E. and {Parmentier}, Vivien},
 doi = {10.1029/2020JE006655},
 eid = {e06655},
 eprint = {2102.03480},
 journal = {Journal of Geophysical Research (Planets)},
 month = {April},
 number = {4},
 pages = {e06655},
 primaryclass = {astro-ph.EP},
 title = {{Aerosols in Exoplanet Atmospheres}},
 volume = {126},
 year = {2021}
}

@article{gaoSedimentationEfficiencyCondensation2018,
 adsurl = {https://ui.adsabs.harvard.edu/abs/2018ApJ...855...86G},
 archiveprefix = {arXiv},
 author = {{Gao}, Peter and {Marley}, Mark S. and {Ackerman}, Andrew S.},
 doi = {10.3847/1538-4357/aab0a1},
 eid = {86},
 eprint = {1802.06241},
 journal = {\apj},
 month = {March},
 number = {2},
 pages = {86},
 primaryclass = {astro-ph.EP},
 title = {{Sedimentation Efficiency of Condensation Clouds in Substellar Atmospheres}},
 volume = {855},
 year = {2018}
}

@article{geissAbundancesDeuteriumHelium31998,
 adsurl = {https://ui.adsabs.harvard.edu/abs/1998SSRv...84..239G},
 author = {{Geiss}, J. and {Gloeckler}, G.},
 journal = {\ssr},
 month = {April},
 pages = {239-250},
 title = {{Abundances of Deuterium and Helium-3 in the Protosolar Cloud}},
 volume = {84},
 year = {1998}
}

@article{gonzalezpicosChemicalEvolutionImprints2025a,
 adsurl = {https://ui.adsabs.harvard.edu/abs/2025NatAs...9.1692G},
 archiveprefix = {arXiv},
 author = {{Gonz{\'a}lez Picos}, Dar{\'\i}o and {Snellen}, Ignas and {de Regt}, Sam},
 doi = {10.1038/s41550-025-02641-4},
 eprint = {2508.18424},
 journal = {Nature Astronomy},
 month = {November},
 number = {11},
 pages = {1692-1700},
 primaryclass = {astro-ph.SR},
 title = {{Chemical evolution imprints in the rare isotopes of nearby M dwarfs}},
 volume = {9},
 year = {2025}
}

@article{gonzalezpicosDisentanglingDiscAtmospheric2025,
 adsurl = {https://ui.adsabs.harvard.edu/abs/2025A&A...703A..65G},
 archiveprefix = {arXiv},
 author = {{Gonz{\'a}lez Picos}, D. and {de Regt}, S. and {Gandhi}, S. and {Grasser}, N. and {Snellen}, I.~A.~G.},
 doi = {10.1051/0004-6361/202555669},
 eid = {A65},
 eprint = {2509.11694},
 journal = {\aap},
 month = {November},
 pages = {A65},
 primaryclass = {astro-ph.EP},
 title = {{Disentangling disc and atmospheric signatures of young brown dwarfs with JWST/NIRSpec}},
 volume = {703},
 year = {2025}
}

@article{gonzalezpicosJWSTHighcontrastSpectroscopy2026,
 author = {Gonz{\'a}lez Picos, Dar{\'i}o and {van der Post}, Tessa and {de Regt}, Sam and Ruffio, Jean-Baptiste and Grasser, Natalie and Snellen, Ignas},
 doi = {10.5281/zenodo.19130087},
 month = {March},
 publisher = {Zenodo},
 shorttitle = {{{JWST}} High-Contrast Spectroscopy with Speckle Modelling},
 title = {{{JWST}} High-Contrast Spectroscopy with Speckle Modelling: {{Atmospheric}} Retrievals of the {{T}} Dwarf Companion {{HD}} 19467 {{B}}},
 url = {https://zenodo.org/records/19130087},
 urldate = {2026-04-20},
 year = {2026}
}

@article{gordonHITRAN2020MolecularSpectroscopic2022,
 adsurl = {https://ui.adsabs.harvard.edu/abs/2022JQSRT.27707949G},
 author = {{Gordon}, I.~E. and {Rothman}, L.~S. and {Hargreaves}, R.~J. and {Hashemi}, R. and {Karlovets}, E.~V. and {Skinner}, F.~M. and {Conway}, E.~K. and {Hill}, C. and {Kochanov}, R.~V. and {Tan}, Y. and {Wcis{\l}o}, P. and {Finenko}, A.~A. and {Nelson}, K. and {Bernath}, P.~F. and {Birk}, M. and {Boudon}, V. and {Campargue}, A. and {Chance}, K.~V. and {Coustenis}, A. and {Drouin}, B.~J. and {Flaud}, J.-M. and {Gamache}, R.~R. and {Hodges}, J.~T. and {Jacquemart}, D. and {Mlawer}, E.~J. and {Nikitin}, A.~V. and {Perevalov}, V.~I. and {Rotger}, M. and {Tennyson}, J. and {Toon}, G.~C. and {Tran}, H. and {Tyuterev}, V.~G. and {Adkins}, E.~M. and {Baker}, A. and {Barbe}, A. and {Can{\`e}}, E. and {Cs{\'a}sz{\'a}r}, A.~G. and {Dudaryonok}, A. and {Egorov}, O. and {Fleisher}, A.~J. and {Fleurbaey}, H. and {Foltynowicz}, A. and {Furtenbacher}, T. and {Harrison}, J.~J. and {Hartmann}, J.-M. and {Horneman}, V.-M. and {Huang}, X. and {Karman}, T. and {Karns}, J. and {Kassi}, S. and {Kleiner}, I. and {Kofman}, V. and {Kwabia-Tchana}, F. and {Lavrentieva}, N.~N. and {Lee}, T.~J. and {Long}, D.~A. and {Lukashevskaya}, A.~A. and {Lyulin}, O.~M. and {Makhnev}, V. Yu. and {Matt}, W. and {Massie}, S.~T. and {Melosso}, M. and {Mikhailenko}, S.~N. and {Mondelain}, D. and {M{\"u}ller}, H.~S.~P. and {Naumenko}, O.~V. and {Perrin}, A. and {Polyansky}, O.~L. and {Raddaoui}, E. and {Raston}, P.~L. and {Reed}, Z.~D. and {Rey}, M. and {Richard}, C. and {T{\'o}bi{\'a}s}, R. and {Sadiek}, I. and {Schwenke}, D.~W. and {Starikova}, E. and {Sung}, K. and {Tamassia}, F. and {Tashkun}, S.~A. and {Vander Auwera}, J. and {Vasilenko}, I.~A. and {Vigasin}, A.~A. and {Villanueva}, G.~L. and {Vispoel}, B. and {Wagner}, G. and {Yachmenev}, A. and {Yurchenko}, S.~N.},
 doi = {10.1016/j.jqsrt.2021.107949},
 eid = {107949},
 journal = {\jqsrt},
 month = {January},
 pages = {107949},
 title = {{The HITRAN2020 molecular spectroscopic database}},
 volume = {277},
 year = {2022}
}

@article{grasserESOSupJupSurvey2025,
 adsurl = {https://ui.adsabs.harvard.edu/abs/2025A&A...698A.252G},
 archiveprefix = {arXiv},
 author = {{Grasser}, N. and {Snellen}, I.~A.~G. and {de Regt}, S. and {Gonz{\'a}lez Picos}, D. and {Zhang}, Y. and {Stolker}, T. and {Gandhi}, S. and {Nasedkin}, E. and {Landman}, R. and {Kesseli}, A.~Y. and {Mulder}, W.},
 doi = {10.1051/0004-6361/202554195},
 eid = {A252},
 eprint = {2507.02706},
 journal = {\aap},
 month = {June},
 pages = {A252},
 primaryclass = {astro-ph.EP},
 title = {{The ESO SupJup Survey: VIII. Chemical fingerprints of young L dwarf twins}},
 volume = {698},
 year = {2025}
}

@article{grecoMEASUREMENTTREATMENTIMPACT2016,
 adsurl = {https://ui.adsabs.harvard.edu/abs/2016ApJ...833..134G},
 archiveprefix = {arXiv},
 author = {{Greco}, Johnny P. and {Brandt}, Timothy D.},
 doi = {10.3847/1538-4357/833/2/134},
 eid = {134},
 eprint = {1602.00691},
 journal = {\apj},
 month = {December},
 number = {2},
 pages = {134},
 primaryclass = {astro-ph.EP},
 title = {{The Measurement, Treatment, and Impact of Spectral Covariance and Bayesian Priors in Integral-field Spectroscopy of Exoplanets}},
 volume = {833},
 year = {2016}
}

@article{greenbaumFirstObservationsBrown2023,
 adsurl = {https://ui.adsabs.harvard.edu/abs/2023ApJ...945..126G},
 archiveprefix = {arXiv},
 author = {{Greenbaum}, Alexandra Z. and {Llop-Sayson}, Jorge and {Lew}, Ben W.~P. and {Bryden}, Geoffrey and {Roellig}, Thomas L. and {Ygouf}, Marie and {Fulton}, B.~J. and {Hey}, Daniel R. and {Huber}, Daniel and {Mukherjee}, Sagnick and {Meyer}, Michael and {Leisenring}, Jarron and {Rieke}, Marcia and {Boyer}, Martha and {Green}, Joseph J. and {Kelly}, Doug and {Misselt}, Karl and {Serabyn}, Eugene and {Stansberry}, John and {Chu}, Laurie E.~U. and {De Furio}, Matthew and {Johnstone}, Doug and {Schlieder}, Joshua E. and {Beichman}, Charles},
 doi = {10.3847/1538-4357/acb68b},
 eid = {126},
 eprint = {2301.11455},
 journal = {\apj},
 month = {March},
 number = {2},
 pages = {126},
 primaryclass = {astro-ph.SR},
 title = {{First Observations of the Brown Dwarf HD 19467 B with JWST}},
 volume = {945},
 year = {2023}
}

@article{hargreavesUpdatingCarbonDioxide2025,
 adsurl = {https://ui.adsabs.harvard.edu/abs/2025JQSRT.33309324H},
 author = {{Hargreaves}, Robert J. and {Gordon}, Iouli E. and {Huang}, Xinchuan and {Toon}, Geoffrey C. and {Rothman}, Laurence S.},
 doi = {10.1016/j.jqsrt.2024.109324},
 eid = {109324},
 journal = {\jqsrt},
 month = {March},
 pages = {109324},
 title = {{Updating the carbon dioxide line list in HITEMP}},
 volume = {333},
 year = {2025}
}

@article{harrisArrayProgrammingNumPy2020,
 adsurl = {https://ui.adsabs.harvard.edu/abs/2020Natur.585..357H},
 archiveprefix = {arXiv},
 author = {{Harris}, Charles R. and {Millman}, K. Jarrod and {van der Walt}, St{\'e}fan J. and {Gommers}, Ralf and {Virtanen}, Pauli and {Cournapeau}, David and {Wieser}, Eric and {Taylor}, Julian and {Berg}, Sebastian and {Smith}, Nathaniel J. and {Kern}, Robert and {Picus}, Matti and {Hoyer}, Stephan and {van Kerkwijk}, Marten H. and {Brett}, Matthew and {Haldane}, Allan and {del R{\'\i}o}, Jaime Fern{\'a}ndez and {Wiebe}, Mark and {Peterson}, Pearu and {G{\'e}rard-Marchant}, Pierre and {Sheppard}, Kevin and {Reddy}, Tyler and {Weckesser}, Warren and {Abbasi}, Hameer and {Gohlke}, Christoph and {Oliphant}, Travis E.},
 doi = {10.1038/s41586-020-2649-2},
 eprint = {2006.10256},
 journal = {\nat},
 month = {September},
 number = {7825},
 pages = {357-362},
 primaryclass = {cs.MS},
 title = {{Array programming with NumPy}},
 volume = {585},
 year = {2020}
}

@article{hauschildtNewEraModelGrid2025,
 adsurl = {https://ui.adsabs.harvard.edu/abs/2025A&A...698A..47H},
 archiveprefix = {arXiv},
 author = {{Hauschildt}, P.~H. and {Barman}, T. and {Baron}, E. and {Aufdenberg}, J.~P. and {Schweitzer}, A.},
 doi = {10.1051/0004-6361/202554171},
 eid = {A47},
 eprint = {2504.17597},
 journal = {\aap},
 month = {June},
 pages = {A47},
 primaryclass = {astro-ph.SR},
 title = {{The NewEra model grid}},
 volume = {698},
 year = {2025}
}

@article{hochJWSTTSTHighContrast2024,
 adsurl = {https://ui.adsabs.harvard.edu/abs/2024AJ....168..187H},
 archiveprefix = {arXiv},
 author = {{Hoch}, Kielan K.~W. and {Theissen}, Christopher A. and {Barman}, Travis S. and {Perrin}, Marshall D. and {Ruffio}, Jean-Baptiste and {Rickman}, Emily and {Konopacky}, Quinn M. and {Manjavacas}, Elena and {Balmer}, William O. and {Pueyo}, Laurent and {Kammerer}, Jens and {van der Marel}, Roeland P. and {Lewis}, Nikole K. and {Girard}, Julien H. and {Seager}, Sara and {Clampin}, Mark and {Mountain}, C. Matt},
 doi = {10.3847/1538-3881/ad6cd3},
 eid = {187},
 eprint = {2408.03830},
 journal = {\aj},
 month = {October},
 number = {4},
 pages = {187},
 primaryclass = {astro-ph.SR},
 title = {{JWST-TST High Contrast: Spectroscopic Characterization of the Benchmark Brown Dwarf HD 19467 B with the NIRSpec Integral Field Spectrograph}},
 volume = {168},
 year = {2024}
}

@article{hochModerateresolutionKbandSpectroscopy2020,
 adsurl = {https://ui.adsabs.harvard.edu/abs/2020AJ....160..207W},
 archiveprefix = {arXiv},
 author = {{Wilcomb}, Kielan K. and {Konopacky}, Quinn M. and {Barman}, Travis S. and {Theissen}, Christopher A. and {Ruffio}, Jean-Baptiste and {Brock}, Laci and {Macintosh}, Bruce and {Marois}, Christian},
 doi = {10.3847/1538-3881/abb9b1},
 eid = {207},
 eprint = {2009.08959},
 journal = {\aj},
 month = {November},
 number = {5},
 pages = {207},
 primaryclass = {astro-ph.EP},
 title = {{Moderate-resolution K-band Spectroscopy of Substellar Companion {\ensuremath{\kappa}} Andromedae b}},
 volume = {160},
 year = {2020}
}

@article{hoeijmakersMediumresolutionIntegralfieldSpectroscopy2018,
 adsurl = {https://ui.adsabs.harvard.edu/abs/2018A&A...617A.144H},
 archiveprefix = {arXiv},
 author = {{Hoeijmakers}, H.~J. and {Schwarz}, H. and {Snellen}, I.~A.~G. and {de Kok}, R.~J. and {Bonnefoy}, M. and {Chauvin}, G. and {Lagrange}, A.~M. and {Girard}, J.~H.},
 doi = {10.1051/0004-6361/201832902},
 eid = {A144},
 eprint = {1802.09721},
 journal = {\aap},
 month = {October},
 pages = {A144},
 primaryclass = {astro-ph.EP},
 title = {{Medium-resolution integral-field spectroscopy for high-contrast exoplanet imaging. Molecule maps of the {\ensuremath{\beta}} Pictoris system with SINFONI}},
 volume = {617},
 year = {2018}
}

@article{hoodHighPrecisionAtmosphericConstraints2024,
 adsurl = {https://ui.adsabs.harvard.edu/abs/2024arXiv240205345H},
 archiveprefix = {arXiv},
 author = {{Hood}, Callie E. and {Mukherjee}, Sagnick and {Fortney}, Jonathan J. and {Line}, Michael R. and {Faherty}, Jacqueline K. and {Merchan}, Sherelyn Alejandro and {Burningham}, Ben and {Su{\'a}rez}, Genaro and {Kiman}, Rocio and {Gagn{\'e}}, Jonathan and {Beichman}, Charles A. and {Vos}, Johanna M. and {Bardalez Gagliuffi}, Daniella and {Meisner}, Aaron M. and {Gonzales}, Eileen C.},
 doi = {10.48550/arXiv.2402.05345},
 eid = {arXiv:2402.05345},
 eprint = {2402.05345},
 journal = {arXiv e-prints},
 month = {February},
 pages = {arXiv:2402.05345},
 primaryclass = {astro-ph.SR},
 title = {{High-Precision Atmospheric Constraints for a Cool T Dwarf from JWST Spectroscopy}},
 year = {2024}
}

@article{hsuBrownDwarfKinematics2021,
 adsurl = {https://ui.adsabs.harvard.edu/abs/2021ApJS..257...45H},
 archiveprefix = {arXiv},
 author = {{Hsu}, Chih-Chun and {Burgasser}, Adam J. and {Theissen}, Christopher A. and {Gelino}, Christopher R. and {Birky}, Jessica L. and {Diamant}, Sharon J.~M. and {Bardalez Gagliuffi}, Daniella C. and {Aganze}, Christian and {Blake}, Cullen H. and {Faherty}, Jacqueline K.},
 doi = {10.3847/1538-4365/ac1c7d},
 eid = {45},
 eprint = {2107.01222},
 journal = {\apjs},
 month = {December},
 number = {2},
 pages = {45},
 primaryclass = {astro-ph.SR},
 title = {{The Brown Dwarf Kinematics Project (BDKP). V. Radial and Rotational Velocities of T Dwarfs from Keck/NIRSPEC High-resolution Spectroscopy}},
 volume = {257},
 year = {2021}
}

@article{hunterMatplotlib2DGraphics2007,
 adsurl = {https://ui.adsabs.harvard.edu/abs/2007CSE.....9...90H},
 author = {{Hunter}, John D.},
 doi = {10.1109/MCSE.2007.55},
 journal = {Computing in Science and Engineering},
 month = {January},
 number = {3},
 pages = {90-95},
 title = {{Matplotlib: A 2D Graphics Environment}},
 volume = {9},
 year = {2007}
}

@article{irwinNEMESISPlanetaryAtmosphere2008a,
 adsurl = {https://ui.adsabs.harvard.edu/abs/2008JQSRT.109.1136I},
 author = {{Irwin}, P.~G.~J. and {Teanby}, N.~A. and {de Kok}, R. and {Fletcher}, L.~N. and {Howett}, C.~J.~A. and {Tsang}, C.~C.~C. and {Wilson}, C.~F. and {Calcutt}, S.~B. and {Nixon}, C.~A. and {Parrish}, P.~D.},
 doi = {10.1016/j.jqsrt.2007.11.006},
 journal = {\jqsrt},
 month = {April},
 pages = {1136-1150},
 title = {{The NEMESIS planetary atmosphere radiative transfer and retrieval tool}},
 volume = {109},
 year = {2008}
}

@article{kirkpatrickDwarfsCoolerDefinition1999,
 adsurl = {https://ui.adsabs.harvard.edu/abs/1999ApJ...519..802K},
 author = {{Kirkpatrick}, J. Davy and {Reid}, I. Neill and {Liebert}, James and {Cutri}, Roc M. and {Nelson}, Brant and {Beichman}, Charles A. and {Dahn}, Conard C. and {Monet}, David G. and {Gizis}, John E. and {Skrutskie}, Michael F.},
 doi = {10.1086/307414},
 journal = {\apj},
 month = {July},
 number = {2},
 pages = {802-833},
 title = {{Dwarfs Cooler than ``M``: The Definition of Spectral Type ``L'' Using Discoveries from the 2 Micron All-Sky Survey (2MASS)}},
 volume = {519},
 year = {1999}
}

@article{kitzmannFastchemCondEquilibrium2023,
 adsurl = {https://ui.adsabs.harvard.edu/abs/2024MNRAS.527.7263K},
 archiveprefix = {arXiv},
 author = {{Kitzmann}, Daniel and {Stock}, Joachim W. and {Patzer}, A. Beate C.},
 doi = {10.1093/mnras/stad3515},
 eprint = {2309.02337},
 journal = {\mnras},
 month = {January},
 number = {3},
 pages = {7263-7283},
 primaryclass = {astro-ph.EP},
 title = {{FASTCHEM COND: equilibrium chemistry with condensation and rainout for cool planetary and stellar environments}},
 volume = {527},
 year = {2024}
}

@article{kothariComprehensiveAtmosphericRetrieval2026,
 adsurl = {https://ui.adsabs.harvard.edu/abs/2026arXiv260405104K},
 archiveprefix = {arXiv},
 author = {{Kothari}, Harshil and {Cushing}, Michael C. and {Beiler}, Samuel A. and {Visscher}, Channon and {Marley}, Mark S. and {Burningham}, Ben and {Schneider}, Adam C. and {Kirkpatrick}, J. Davy},
 doi = {10.48550/arXiv.2604.05104},
 eid = {arXiv:2604.05104},
 eprint = {2604.05104},
 journal = {arXiv e-prints},
 month = {April},
 pages = {arXiv:2604.05104},
 primaryclass = {astro-ph.SR},
 title = {{A Comprehensive Atmospheric Retrieval Analysis of 22 James Webb Space Telescope Spectral Energy Distributions of Cool Brown Dwarfs}},
 year = {2026}
}

@article{landmanPictorisEyesUpgraded2024,
 adsurl = {https://ui.adsabs.harvard.edu/abs/2024A&A...682A..48L},
 archiveprefix = {arXiv},
 author = {{Landman}, R. and {Stolker}, T. and {Snellen}, I.~A.~G. and {Costes}, J. and {de Regt}, S. and {Zhang}, Y. and {Gandhi}, S. and {Molliere}, P. and {Kesseli}, A. and {Vigan}, A. and {Sanchez-L{\'o}pez}, A.},
 doi = {10.1051/0004-6361/202347846},
 eid = {A48},
 eprint = {2311.13527},
 journal = {\aap},
 month = {February},
 pages = {A48},
 primaryclass = {astro-ph.EP},
 title = {{{\ensuremath{\beta}} Pictoris b through the eyes of the upgraded CRIRES+. Atmospheric composition, spin rotation, and radial velocity}},
 volume = {682},
 year = {2024}
}

@article{leggettPhysicalSpectralCharacteristics2007,
 adsurl = {https://ui.adsabs.harvard.edu/abs/2007ApJ...667..537L},
 archiveprefix = {arXiv},
 author = {{Leggett}, S.~K. and {Marley}, M.~S. and {Freedman}, R. and {Saumon}, D. and {Liu}, Michael C. and {Geballe}, T.~R. and {Golimowski}, D.~A. and {Stephens}, D.~C.},
 doi = {10.1086/519948},
 eprint = {0705.2602},
 journal = {\apj},
 month = {September},
 number = {1},
 pages = {537-548},
 primaryclass = {astro-ph},
 title = {{Physical and Spectral Characteristics of the T8 and Later Type Dwarfs}},
 volume = {667},
 year = {2007}
}

@article{lineUniformAtmosphericRetrieval2015,
 adsurl = {https://ui.adsabs.harvard.edu/abs/2015ApJ...807..183L},
 archiveprefix = {arXiv},
 author = {{Line}, Michael R. and {Teske}, Johanna and {Burningham}, Ben and {Fortney}, Jonathan J. and {Marley}, Mark S.},
 doi = {10.1088/0004-637X/807/2/183},
 eid = {183},
 eprint = {1504.06670},
 journal = {\apj},
 month = {July},
 number = {2},
 pages = {183},
 primaryclass = {astro-ph.SR},
 title = {{Uniform Atmospheric Retrieval Analysis of Ultracool Dwarfs. I. Characterizing Benchmarks, Gl 570D and HD 3651B}},
 volume = {807},
 year = {2015}
}

@article{lineUniformAtmosphericRetrieval2017,
 adsurl = {https://ui.adsabs.harvard.edu/abs/2017ApJ...848...83L},
 archiveprefix = {arXiv},
 author = {{Line}, Michael R. and {Marley}, Mark S. and {Liu}, Michael C. and {Burningham}, Ben and {Morley}, Caroline V. and {Hinkel}, Natalie R. and {Teske}, Johanna and {Fortney}, Jonathan J. and {Freedman}, Richard and {Lupu}, Roxana},
 doi = {10.3847/1538-4357/aa7ff0},
 eid = {83},
 eprint = {1612.02809},
 journal = {\apj},
 month = {October},
 number = {2},
 pages = {83},
 primaryclass = {astro-ph.SR},
 title = {{Uniform Atmospheric Retrieval Analysis of Ultracool Dwarfs. II. Properties of 11 T dwarfs}},
 volume = {848},
 year = {2017}
}

@article{liROVIBRATIONALLINELISTS2015,
 adsurl = {https://ui.adsabs.harvard.edu/abs/2015ApJS..216...15L},
 author = {{Li}, Gang and {Gordon}, Iouli E. and {Rothman}, Laurence S. and {Tan}, Yan and {Hu}, Shui-Ming and {Kassi}, Samir and {Campargue}, Alain and {Medvedev}, Emile S.},
 doi = {10.1088/0067-0049/216/1/15},
 eid = {15},
 journal = {\apjs},
 month = {January},
 number = {1},
 pages = {15},
 title = {{Rovibrational Line Lists for Nine Isotopologues of the CO Molecule in the X $^{1}${\ensuremath{\Sigma}}$^{+}$ Ground Electronic State}},
 volume = {216},
 year = {2015}
}

@article{lyonsLightCarbonIsotope2018,
 adsurl = {https://ui.adsabs.harvard.edu/abs/2018NatCo...9..908L},
 author = {{Lyons}, James R. and {Gharib-Nezhad}, Ehsan and {Ayres}, Thomas R.},
 doi = {10.1038/s41467-018-03093-3},
 eid = {908},
 journal = {Nature Communications},
 month = {March},
 pages = {908},
 title = {{A light carbon isotope composition for the Sun}},
 volume = {9},
 year = {2018}
}

@article{madhusudhanExoplanetaryAtmospheresKey2019,
 adsurl = {https://ui.adsabs.harvard.edu/abs/2019ARA&A..57..617M},
 archiveprefix = {arXiv},
 author = {{Madhusudhan}, Nikku},
 doi = {10.1146/annurev-astro-081817-051846},
 eprint = {1904.03190},
 journal = {\araa},
 month = {August},
 pages = {617-663},
 primaryclass = {astro-ph.EP},
 title = {{Exoplanetary Atmospheres: Key Insights, Challenges, and Prospects}},
 volume = {57},
 year = {2019}
}

@article{maireOrbitalSpectralCharacterization2020,
 adsurl = {https://ui.adsabs.harvard.edu/abs/2020A&A...639A..47M},
 archiveprefix = {arXiv},
 author = {{Maire}, A.-L. and {Molaverdikhani}, K. and {Desidera}, S. and {Trifonov}, T. and {Molli{\`e}re}, P. and {D'Orazi}, V. and {Frankel}, N. and {Baudino}, J.-L. and {Messina}, S. and {M{\"u}ller}, A. and {Charnay}, B. and {Cheetham}, A.~C. and {Delorme}, P. and {Ligi}, R. and {Bonnefoy}, M. and {Brandner}, W. and {Mesa}, D. and {Cantalloube}, F. and {Galicher}, R. and {Henning}, T. and {Biller}, B.~A. and {Hagelberg}, J. and {Lagrange}, A.-M. and {Lavie}, B. and {Rickman}, E. and {S{\'e}gransan}, D. and {Udry}, S. and {Chauvin}, G. and {Gratton}, R. and {Langlois}, M. and {Vigan}, A. and {Meyer}, M.~R. and {Beuzit}, J.-L. and {Bhowmik}, T. and {Boccaletti}, A. and {Lazzoni}, C. and {Perrot}, C. and {Schmidt}, T. and {Zurlo}, A. and {Gluck}, L. and {Pragt}, J. and {Ramos}, J. and {Roelfsema}, R. and {Roux}, A. and {Sauvage}, J.-F.},
 doi = {10.1051/0004-6361/202037984},
 eid = {A47},
 eprint = {2005.10312},
 journal = {\aap},
 month = {July},
 pages = {A47},
 primaryclass = {astro-ph.EP},
 title = {{Orbital and spectral characterization of the benchmark T-type brown dwarf HD 19467B}},
 volume = {639},
 year = {2020}
}

@article{marleySonoraBrownDwarf2021,
 adsurl = {https://ui.adsabs.harvard.edu/abs/2021ApJ...920...85M},
 archiveprefix = {arXiv},
 author = {{Marley}, Mark S. and {Saumon}, Didier and {Visscher}, Channon and {Lupu}, Roxana and {Freedman}, Richard and {Morley}, Caroline and {Fortney}, Jonathan J. and {Seay}, Christopher and {Smith}, Adam J.~R.~W. and {Teal}, D.~J. and {Wang}, Ruoyan},
 doi = {10.3847/1538-4357/ac141d},
 eid = {85},
 eprint = {2107.07434},
 journal = {\apj},
 month = {October},
 number = {2},
 pages = {85},
 primaryclass = {astro-ph.SR},
 title = {{The Sonora Brown Dwarf Atmosphere and Evolution Models. I. Model Description and Application to Cloudless Atmospheres in Rainout Chemical Equilibrium}},
 volume = {920},
 year = {2021}
}

@article{mennessonCurrentLaboratoryPerformance2024,
 adsurl = {https://ui.adsabs.harvard.edu/abs/2024JATIS..10c5004M},
 archiveprefix = {arXiv},
 author = {{Mennesson}, Bertrand and {Belikov}, Ruslan and {Por}, Emiel and {Serabyn}, Eugene and {Ruane}, Garreth and {Riggs}, A.~J. Eldorado and {Sirbu}, Dan and {Pueyo}, Laurent and {Soummer}, Remi and {Kasdin}, Jeremy and {Shaklan}, Stuart and {Seo}, Byoung-Joon and {Stark}, Christopher and {Cady}, Eric and {Chen}, Pin and {Crill}, Brendan and {Fogarty}, Kevin and {Greenbaum}, Alexandra and {Guyon}, Olivier and {Juanola-Parramon}, Roser and {Kern}, Brian and {Krist}, John and {Macintosh}, Bruce and {Marx}, David and {Mawet}, Dimitri and {Prada}, Camilo Mejia and {Morgan}, Rhonda and {Nemati}, Bijan and {Pogorelyuk}, Leonid and {Redmond}, Susan and {Seager}, Sara and {Siegler}, Nicholas and {Stapelfeldt}, Karl and {Steiger}, Sarah and {Trauger}, John and {Wallace}, James K. and {Ygouf}, Marie and {Zimmerman}, Neil},
 doi = {10.1117/1.JATIS.10.3.035004},
 eid = {035004},
 eprint = {2404.18036},
 journal = {Journal of Astronomical Telescopes, Instruments, and Systems},
 month = {July},
 pages = {035004},
 primaryclass = {astro-ph.IM},
 title = {{Current laboratory performance of starlight suppression systems and potential pathways to desired Habitable Worlds Observatory exoplanet science capabilities}},
 volume = {10},
 year = {2024}
}

@article{merchanDiversityColdWorlds2025,
 adsurl = {https://ui.adsabs.harvard.edu/abs/2025ApJ...989...80A},
 archiveprefix = {arXiv},
 author = {{Alejandro Merchan}, Sherelyn and {Faherty}, Jacqueline K. and {Su{\'a}rez}, Genaro and {Cruz}, Kelle L. and {Burgasser}, Adam J. and {Gagn{\'e}}, Jonathan and {Hood}, Callie E. and {Gonzales}, Eileen C. and {Bardalez Gagliuffi}, Daniella C. and {L'Heureux}, Jolie and {Vos}, Johanna M. and {Schneider}, Adam C. and {Meisner}, Aaron M. and {Morley}, Caroline and {Kirkpatrick}, J. Davy and {Marocco}, Federico and {Kiman}, Rocio and {Beichman}, Charles A. and {Burningham}, Ben and {Caselden}, Dan and {Eisenhardt}, Peter R. and {Gelino}, Christopher R. and {Gharib-Nezhad}, Ehsan and {Kuchner}, Marc J. and {Lacy}, Brianna and {Rothermich}, Austin and {Rowland}, Melanie J. and {Whiteford}, Niall},
 doi = {10.3847/1538-4357/ade3d7},
 eid = {80},
 eprint = {2505.24591},
 journal = {\apj},
 month = {August},
 number = {1},
 pages = {80},
 primaryclass = {astro-ph.SR},
 title = {{Diversity of Cold Worlds: A Near-complete Spectral Energy Distribution for 2MASS J04151954{\ensuremath{-}}0935066 Using JWST}},
 volume = {989},
 year = {2025}
}

@article{mesaCharacterizingBrownDwarf2020,
 adsurl = {https://ui.adsabs.harvard.edu/abs/2020MNRAS.495.4279M},
 archiveprefix = {arXiv},
 author = {{Mesa}, D. and {D'Orazi}, V. and {Vigan}, A. and {Kitzmann}, D. and {Heng}, K. and {Gratton}, R. and {Desidera}, S. and {Bonnefoy}, M. and {Lavie}, B. and {Maire}, A.-L. and {Peretti}, S. and {Boccaletti}, A.},
 doi = {10.1093/mnras/staa1444},
 eprint = {2005.10077},
 journal = {\mnras},
 month = {July},
 number = {4},
 pages = {4279-4290},
 primaryclass = {astro-ph.EP},
 title = {{Characterizing brown dwarf companions with IRDIS long-slit spectroscopy: HD 1160 B and HD 19467 B}},
 volume = {495},
 year = {2020}
}

@article{milam1213Isotope2005,
 adsurl = {https://ui.adsabs.harvard.edu/abs/2005ApJ...634.1126M},
 author = {{Milam}, S.~N. and {Savage}, C. and {Brewster}, M.~A. and {Ziurys}, L.~M. and {Wyckoff}, S.},
 doi = {10.1086/497123},
 journal = {\apj},
 month = {December},
 number = {2},
 pages = {1126-1132},
 title = {{The $^{12}$C/$^{13}$C Isotope Gradient Derived from Millimeter Transitions of CN: The Case for Galactic Chemical Evolution}},
 volume = {634},
 year = {2005}
}

@article{milesJWSTEarlyreleaseScience2023,
 adsurl = {https://ui.adsabs.harvard.edu/abs/2023ApJ...946L...6M},
 archiveprefix = {arXiv},
 author = {{Miles}, Brittany E. and {Biller}, Beth A. and {Patapis}, Polychronis and {Worthen}, Kadin and {Rickman}, Emily and {Hoch}, Kielan K.~W. and {Skemer}, Andrew and {Perrin}, Marshall D. and {Whiteford}, Niall and {Chen}, Christine H. and {Sargent}, B. and {Mukherjee}, Sagnick and {Morley}, Caroline V. and {Moran}, Sarah E. and {Bonnefoy}, Mickael and {Petrus}, Simon and {Carter}, Aarynn L. and {Choquet}, Elodie and {Hinkley}, Sasha and {Ward-Duong}, Kimberly and {Leisenring}, Jarron M. and {Millar-Blanchaer}, Maxwell A. and {Pueyo}, Laurent and {Ray}, Shrishmoy and {Sallum}, Steph and {Stapelfeldt}, Karl R. and {Stone}, Jordan M. and {Wang}, Jason J. and {Absil}, Olivier and {Balmer}, William O. and {Boccaletti}, Anthony and {Bonavita}, Mariangela and {Booth}, Mark and {Bowler}, Brendan P. and {Chauvin}, Gael and {Christiaens}, Valentin and {Currie}, Thayne and {Danielski}, Camilla and {Fortney}, Jonathan J. and {Girard}, Julien H. and {Grady}, Carol A. and {Greenbaum}, Alexandra Z. and {Henning}, Thomas and {Hines}, Dean C. and {Janson}, Markus and {Kalas}, Paul and {Kammerer}, Jens and {Kennedy}, Grant M. and {Kenworthy}, Matthew A. and {Kervella}, Pierre and {Lagage}, Pierre-Olivier and {Lew}, Ben W.~P. and {Liu}, Michael C. and {Macintosh}, Bruce and {Marino}, Sebastian and {Marley}, Mark S. and {Marois}, Christian and {Matthews}, Elisabeth C. and {Matthews}, Brenda C. and {Mawet}, Dimitri and {McElwain}, Michael W. and {Metchev}, Stanimir and {Meyer}, Michael R. and {Molliere}, Paul and {Pantin}, Eric and {Quirrenbach}, Andreas and {Rebollido}, Isabel and {Ren}, Bin B. and {Schneider}, Glenn and {Vasist}, Malavika and {Wyatt}, Mark C. and {Zhou}, Yifan and {Briesemeister}, Zackery W. and {Bryan}, Marta L. and {Calissendorff}, Per and {Cantalloube}, Faustine and {Cugno}, Gabriele and {De Furio}, Matthew and {Dupuy}, Trent J. and {Factor}, Samuel M. and {Faherty}, Jacqueline K. and {Fitzgerald}, Michael P. and {Franson}, Kyle and {Gonzales}, Eileen C. and {Hood}, Callie E. and {Howe}, Alex R. and {Kraus}, Adam L. and {Kuzuhara}, Masayuki and {Lagrange}, Anne-Marie and {Lawson}, Kellen and {Lazzoni}, Cecilia and {Liu}, Pengyu and {Llop-Sayson}, Jorge and {Lloyd}, James P. and {Martinez}, Raquel A. and {Mazoyer}, Johan and {Quanz}, Sascha P. and {Redai}, Jea Adams and {Samland}, Matthias and {Schlieder}, Joshua E. and {Tamura}, Motohide and {Tan}, Xianyu and {Uyama}, Taichi and {Vigan}, Arthur and {Vos}, Johanna M. and {Wagner}, Kevin and {Wolff}, Schuyler G. and {Ygouf}, Marie and {Zhang}, Xi and {Zhang}, Keming and {Zhang}, Zhoujian},
 doi = {10.3847/2041-8213/acb04a},
 eid = {L6},
 eprint = {2209.00620},
 journal = {\apjl},
 month = {March},
 number = {1},
 pages = {L6},
 primaryclass = {astro-ph.EP},
 title = {{The JWST Early-release Science Program for Direct Observations of Exoplanetary Systems II: A 1 to 20 {\ensuremath{\mu}}m Spectrum of the Planetary-mass Companion VHS 1256-1257 b}},
 volume = {946},
 year = {2023}
}

@article{molliereDetectingIsotopologuesExoplanet2019,
 adsurl = {https://ui.adsabs.harvard.edu/abs/2019A&A...622A.139M},
 archiveprefix = {arXiv},
 author = {{Molli{\`e}re}, P. and {Snellen}, I.~A.~G.},
 doi = {10.1051/0004-6361/201834169},
 eid = {A139},
 eprint = {1809.01156},
 journal = {\aap},
 month = {February},
 pages = {A139},
 primaryclass = {astro-ph.EP},
 title = {{Detecting isotopologues in exoplanet atmospheres using ground-based high-dispersion spectroscopy}},
 volume = {622},
 year = {2019}
}

@article{molliereInterpretingAtmosphericComposition2022,
 adsurl = {https://ui.adsabs.harvard.edu/abs/2022ApJ...934...74M},
 archiveprefix = {arXiv},
 author = {{Molli{\`e}re}, Paul and {Molyarova}, Tamara and {Bitsch}, Bertram and {Henning}, Thomas and {Schneider}, Aaron and {Kreidberg}, Laura and {Eistrup}, Christian and {Burn}, Remo and {Nasedkin}, Evert and {Semenov}, Dmitry and {Mordasini}, Christoph and {Schlecker}, Martin and {Schwarz}, Kamber R. and {Lacour}, Sylvestre and {Nowak}, Mathias and {Schulik}, Matth{\"a}us},
 doi = {10.3847/1538-4357/ac6a56},
 eid = {74},
 eprint = {2204.13714},
 journal = {\apj},
 month = {July},
 number = {1},
 pages = {74},
 primaryclass = {astro-ph.EP},
 title = {{Interpreting the Atmospheric Composition of Exoplanets: Sensitivity to Planet Formation Assumptions}},
 volume = {934},
 year = {2022}
}

@article{mollierePetitRADTRANSPythonRadiative2019,
 adsurl = {https://ui.adsabs.harvard.edu/abs/2019A&A...627A..67M},
 archiveprefix = {arXiv},
 author = {{Molli{\`e}re}, P. and {Wardenier}, J.~P. and {van Boekel}, R. and {Henning}, Th. and {Molaverdikhani}, K. and {Snellen}, I.~A.~G.},
 doi = {10.1051/0004-6361/201935470},
 eid = {A67},
 eprint = {1904.11504},
 journal = {\aap},
 month = {July},
 pages = {A67},
 primaryclass = {astro-ph.EP},
 title = {{petitRADTRANS. A Python radiative transfer package for exoplanet characterization and retrieval}},
 volume = {627},
 year = {2019}
}

@article{molliereRetrievingScatteringClouds2020a,
 adsurl = {https://ui.adsabs.harvard.edu/abs/2020A&A...640A.131M},
 archiveprefix = {arXiv},
 author = {{Molli{\`e}re}, P. and {Stolker}, T. and {Lacour}, S. and {Otten}, G.~P.~P.~L. and {Shangguan}, J. and {Charnay}, B. and {Molyarova}, T. and {Nowak}, M. and {Henning}, Th. and {Marleau}, G.-D. and {Semenov}, D.~A. and {van Dishoeck}, E. and {Eisenhauer}, F. and {Garcia}, P. and {Garcia Lopez}, R. and {Girard}, J.~H. and {Greenbaum}, A.~Z. and {Hinkley}, S. and {Kervella}, P. and {Kreidberg}, L. and {Maire}, A.-L. and {Nasedkin}, E. and {Pueyo}, L. and {Snellen}, I.~A.~G. and {Vigan}, A. and {Wang}, J. and {de Zeeuw}, P.~T. and {Zurlo}, A.},
 doi = {10.1051/0004-6361/202038325},
 eid = {A131},
 eprint = {2006.09394},
 journal = {\aap},
 month = {August},
 pages = {A131},
 primaryclass = {astro-ph.EP},
 title = {{Retrieving scattering clouds and disequilibrium chemistry in the atmosphere of HR 8799e}},
 volume = {640},
 year = {2020}
}

@article{mosesDisequilibriumCarbonOxygen2011,
 adsurl = {https://ui.adsabs.harvard.edu/abs/2011ApJ...737...15M},
 archiveprefix = {arXiv},
 author = {{Moses}, Julianne I. and {Visscher}, C. and {Fortney}, J.~J. and {Showman}, A.~P. and {Lewis}, N.~K. and {Griffith}, C.~A. and {Klippenstein}, S.~J. and {Shabram}, M. and {Friedson}, A.~J. and {Marley}, M.~S. and {Freedman}, R.~S.},
 doi = {10.1088/0004-637X/737/1/15},
 eid = {15},
 eprint = {1102.0063},
 journal = {\apj},
 month = {August},
 number = {1},
 pages = {15},
 primaryclass = {astro-ph.EP},
 title = {{Disequilibrium Carbon, Oxygen, and Nitrogen Chemistry in the Atmospheres of HD 189733b and HD 209458b}},
 volume = {737},
 year = {2011}
}

@article{mukherjeeSonoraSubstellarAtmosphere2024,
 adsurl = {https://ui.adsabs.harvard.edu/abs/2024ApJ...963...73M},
 archiveprefix = {arXiv},
 author = {{Mukherjee}, Sagnick and {Fortney}, Jonathan J. and {Morley}, Caroline V. and {Batalha}, Natasha E. and {Marley}, Mark S. and {Karalidi}, Theodora and {Visscher}, Channon and {Lupu}, Roxana and {Freedman}, Richard and {Gharib-Nezhad}, Ehsan},
 doi = {10.3847/1538-4357/ad18c2},
 eid = {73},
 eprint = {2402.00756},
 journal = {\apj},
 month = {March},
 number = {1},
 pages = {73},
 primaryclass = {astro-ph.EP},
 title = {{The Sonora Substellar Atmosphere Models. IV. Elf Owl: Atmospheric Mixing and Chemical Disequilibrium with Varying Metallicity and C/O Ratios}},
 volume = {963},
 year = {2024}
}

@article{obergEFFECTSSNOWLINESPLANETARY2011,
 adsurl = {https://ui.adsabs.harvard.edu/abs/2011ApJ...743L..16O},
 archiveprefix = {arXiv},
 author = {{{\"O}berg}, Karin I. and {Murray-Clay}, Ruth and {Bergin}, Edwin A.},
 doi = {10.1088/2041-8205/743/1/L16},
 eid = {L16},
 eprint = {1110.5567},
 journal = {\apjl},
 month = {December},
 number = {1},
 pages = {L16},
 primaryclass = {astro-ph.GA},
 title = {{The Effects of Snowlines on C/O in Planetary Atmospheres}},
 volume = {743},
 year = {2011}
}

@article{pierelRatiosSaturnJupiter2017,
 adsurl = {https://ui.adsabs.harvard.edu/abs/2017AJ....154..178P},
 author = {{Pierel}, J.~D.~R. and {Nixon}, C.~A. and {Lellouch}, E. and {Fletcher}, L.~N. and {Bjoraker}, G.~L. and {Achterberg}, R.~K. and {B{\'e}zard}, B. and {Hesman}, B.~E. and {Irwin}, P.~G.~J. and {Flasar}, F.~M.},
 doi = {10.3847/1538-3881/aa899d},
 eid = {178},
 journal = {\aj},
 month = {November},
 number = {5},
 pages = {178},
 title = {{D/H Ratios on Saturn and Jupiter from Cassini CIRS}},
 volume = {154},
 year = {2017}
}

@article{polyanskyExoMolMolecularLine2017,
 adsurl = {https://ui.adsabs.harvard.edu/abs/2017MNRAS.466.1363P},
 archiveprefix = {arXiv},
 author = {{Polyansky}, Oleg L. and {Kyuberis}, Aleksandra A. and {Lodi}, Lorenzo and {Tennyson}, Jonathan and {Yurchenko}, Sergei N. and {Ovsyannikov}, Roman I. and {Zobov}, Nikolai F.},
 doi = {10.1093/mnras/stw3125},
 eprint = {1702.03570},
 journal = {\mnras},
 month = {April},
 number = {2},
 pages = {1363-1371},
 primaryclass = {astro-ph.EP},
 title = {{ExoMol molecular line lists XIX: high-accuracy computed hot line lists for H$_{2}$$^{18}$O and H$_{2}$$^{17}$O}},
 volume = {466},
 year = {2017}
}

@article{polyanskyExoMolMolecularLine2018,
 adsurl = {https://ui.adsabs.harvard.edu/abs/2018MNRAS.480.2597P},
 archiveprefix = {arXiv},
 author = {{Polyansky}, Oleg L. and {Kyuberis}, Aleksandra A. and {Zobov}, Nikolai F. and {Tennyson}, Jonathan and {Yurchenko}, Sergei N. and {Lodi}, Lorenzo},
 doi = {10.1093/mnras/sty1877},
 eprint = {1807.04529},
 journal = {\mnras},
 month = {October},
 number = {2},
 pages = {2597-2608},
 primaryclass = {astro-ph.EP},
 title = {{ExoMol molecular line lists XXX: a complete high-accuracy line list for water}},
 volume = {480},
 year = {2018}
}

@article{prantzosEvolutionCarbonOxygen1996,
 adsurl = {https://ui.adsabs.harvard.edu/abs/1996A&A...309..760P},
 author = {{Prantzos}, N. and {Aubert}, O. and {Audouze}, J.},
 journal = {\aap},
 month = {May},
 pages = {760-774},
 title = {{Evolution of the carbon and oxygen isotopes in the Galaxy.}},
 volume = {309},
 year = {1996}
}

@article{prinnCarbonMonoxideJupiter1977,
 adsurl = {https://ui.adsabs.harvard.edu/abs/1977Sci...198.1031P},
 author = {{Prinn}, R.~G. and {Barshay}, S.~S.},
 doi = {10.1126/science.198.4321.1031},
 journal = {Science},
 month = {December},
 number = {4321},
 pages = {1031-1034},
 title = {{Carbon Monoxide on Jupiter and Implications for Atmospheric Convection}},
 volume = {198},
 year = {1977}
}

@article{regtESOSupJupSurvey2024,
 adsurl = {https://ui.adsabs.harvard.edu/abs/2024A&A...688A.116D},
 archiveprefix = {arXiv},
 author = {{de Regt}, S. and {Gandhi}, S. and {Snellen}, I.~A.~G. and {Zhang}, Y. and {Ginski}, C. and {Gonz{\'a}lez Picos}, D. and {Kesseli}, A.~Y. and {Landman}, R. and {Molli{\`e}re}, P. and {Nasedkin}, E. and {S{\'a}nchez-L{\'o}pez}, A. and {Stolker}, T.},
 doi = {10.1051/0004-6361/202348508},
 eid = {A116},
 eprint = {2405.10841},
 journal = {\aap},
 month = {August},
 pages = {A116},
 primaryclass = {astro-ph.EP},
 title = {{The ESO SupJup Survey. I. Chemical and isotopic characterisation of the late L-dwarf DENIS J0255-4700 with CRIRES$^{+}$}},
 volume = {688},
 year = {2024}
}

@article{regtPyROXRapidOpacity2025,
 adsurl = {https://ui.adsabs.harvard.edu/abs/2025arXiv251020870D},
 archiveprefix = {arXiv},
 author = {{de Regt}, Sam and {Gandhi}, Siddharth and {Siebenaler}, Louis and {Gonz{\'a}lez Picos}, Dar{\'\i}o},
 doi = {10.48550/arXiv.2510.20870},
 eid = {arXiv:2510.20870},
 eprint = {2510.20870},
 journal = {arXiv e-prints},
 month = {October},
 pages = {arXiv:2510.20870},
 primaryclass = {astro-ph.IM},
 title = {{pyROX: Rapid Opacity X-sections}},
 year = {2025}
}

@article{rigbySciencePerformanceJWST2023,
 adsurl = {https://ui.adsabs.harvard.edu/abs/2023PASP..135d8001R},
 archiveprefix = {arXiv},
 author = {{Rigby}, Jane and {Perrin}, Marshall and {McElwain}, Michael and {Kimble}, Randy and {Friedman}, Scott and {Lallo}, Matt and {Doyon}, Ren{\'e} and {Feinberg}, Lee and {Ferruit}, Pierre and {Glasse}, Alistair and {Rieke}, Marcia and {Rieke}, George and {Wright}, Gillian and {Willott}, Chris and {Colon}, Knicole and {Milam}, Stefanie and {Neff}, Susan and {Stark}, Christopher and {Valenti}, Jeff and {Abell}, Jim and {Abney}, Faith and {Abul-Huda}, Yasin and {Acton}, D. Scott and {Adams}, Evan and {Adler}, David and {Aguilar}, Jonathan and {Ahmed}, Nasif and {Albert}, Lo{\"\i}c and {Alberts}, Stacey and {Aldridge}, David and {Allen}, Marsha and {Altenburg}, Martin and {{\'A}lvarez-M{\'a}rquez}, Javier and {Alves de Oliveira}, Catarina and {Andersen}, Greg and {Anderson}, Harry and {Anderson}, Sara and {Argyriou}, Ioannis and {Armstrong}, Amber and {Arribas}, Santiago and {Artigau}, Etienne and {Arvai}, Amanda and {Atkinson}, Charles and {Bacon}, Gregory and {Bair}, Thomas and {Banks}, Kimberly and {Barrientes}, Jaclyn and {Barringer}, Bruce and {Bartosik}, Peter and {Bast}, William and {Baudoz}, Pierre and {Beatty}, Thomas and {Bechtold}, Katie and {Beck}, Tracy and {Bergeron}, Eddie and {Bergkoetter}, Matthew and {Bhatawdekar}, Rachana and {Birkmann}, Stephan and {Blazek}, Ronald and {Blome}, Claire and {Boccaletti}, Anthony and {B{\"o}ker}, Torsten and {Boia}, John and {Bonaventura}, Nina and {Bond}, Nicholas and {Bosley}, Kari and {Boucarut}, Ray and {Bourque}, Matthew and {Bouwman}, Jeroen and {Bower}, Gary and {Bowers}, Charles and {Boyer}, Martha and {Bradley}, Larry and {Brady}, Greg and {Braun}, Hannah and {Breda}, David and {Bresnahan}, Pamela and {Bright}, Stacey and {Britt}, Christopher and {Bromenschenkel}, Asa and {Brooks}, Brian and {Brooks}, Keira and {Brown}, Bob and {Brown}, Matthew and {Brown}, Patricia and {Bunker}, Andy and {Burger}, Matthew and {Bushouse}, Howard and {Cale}, Steven and {Cameron}, Alex and {Cameron}, Peter and {Canipe}, Alicia and {Caplinger}, James and {Caputo}, Francis and {Cara}, Mihai and {Carey}, Larkin and {Carniani}, Stefano and {Carrasquilla}, Maria and {Carruthers}, Margaret and {Case}, Michael and {Catherine}, Riggs and {Chance}, Don and {Chapman}, George and {Charlot}, St{\'e}phane and {Charlow}, Brian and {Chayer}, Pierre and {Chen}, Bin and {Cherinka}, Brian and {Chichester}, Sarah and {Chilton}, Zack and {Chonis}, Taylor and {Clampin}, Mark and {Clark}, Charles and {Clark}, Kerry and {Coe}, Dan and {Coleman}, Benee and {Comber}, Brian and {Comeau}, Tom and {Connolly}, Dennis and {Cooper}, James and {Cooper}, Rachel and {Coppock}, Eric and {Correnti}, Matteo and {Cossou}, Christophe and {Coulais}, Alain and {Coyle}, Laura and {Cracraft}, Misty and {Curti}, Mirko and {Cuturic}, Steven and {Davis}, Katherine and {Davis}, Michael and {Dean}, Bruce and {DeLisa}, Amy and {deMeester}, Wim and {Dencheva}, Nadia and {Dencheva}, Nadezhda and {DePasquale}, Joseph and {Deschenes}, Jeremy and {Hunor Detre}, {\"O}rs and {Diaz}, Rosa and {Dicken}, Dan and {DiFelice}, Audrey and {Dillman}, Matthew and {Dixon}, William and {Doggett}, Jesse and {Donaldson}, Tom and {Douglas}, Rob and {DuPrie}, Kimberly and {Dupuis}, Jean and {Durning}, John and {Easmin}, Nilufar and {Eck}, Weston and {Edeani}, Chinwe and {Egami}, Eiichi and {Ehrenwinkler}, Ralf and {Eisenhamer}, Jonathan and {Eisenhower}, Michael and {Elie}, Michelle and {Elliott}, James and {Elliott}, Kyle and {Ellis}, Tracy and {Engesser}, Michael and {Espinoza}, Nestor and {Etienne}, Odessa and {Etxaluze}, Mireya and {Falini}, Patrick and {Feeney}, Matthew and {Ferry}, Malcolm and {Filippazzo}, Joseph and {Fincham}, Brian and {Fix}, Mees and {Flagey}, Nicolas and {Florian}, Michael and {Flynn}, Jim and {Fontanella}, Erin and {Ford}, Terrance and {Forshay}, Peter and {Fox}, Ori and {Franz}, David and {Fu}, Henry and {Fullerton}, Alexander and {Galkin}, Sergey and {Galyer}, Anthony and {Garc{\'\i}a Mar{\'\i}n}, Macarena and {Gardner}, Jonathan P. and {Gardner}, Lisa and {Garland}, Dennis and {Garrett}, Bruce and {Gasman}, Danny and {Gaspar}, Andras and {Gaudreau}, Daniel and {Gauthier}, Peter and {Geers}, Vincent and {Geithner}, Paul and {Gennaro}, Mario and {Giardino}, Giovanna and {Girard}, Julien and {Giuliano}, Mark and {Glassmire}, Kirk and {Glauser}, Adrian},
 doi = {10.1088/1538-3873/acb293},
 eid = {048001},
 eprint = {2207.05632},
 journal = {\pasp},
 month = {April},
 number = {1046},
 pages = {048001},
 primaryclass = {astro-ph.IM},
 title = {{The Science Performance of JWST as Characterized in Commissioning}},
 volume = {135},
 year = {2023}
}

@book{rodgersInverseMethodsAtmospheric2000,
 author = {Rodgers, Clive D.},
 googlebooks = {Xv7sCgAAQBAJ},
 isbn = {978-981-4498-68-5},
 langid = {english},
 month = {July},
 publisher = {World Scientific},
 shorttitle = {Inverse {{Methods For Atmospheric Sounding}}},
 title = {Inverse {{Methods For Atmospheric Sounding}}: {{Theory And Practice}}},
 year = {2000}
}

@article{romanoEvolutionCNOElements2022,
 adsurl = {https://ui.adsabs.harvard.edu/abs/2022A&ARv..30....7R},
 archiveprefix = {arXiv},
 author = {{Romano}, Donatella},
 doi = {10.1007/s00159-022-00144-z},
 eid = {7},
 eprint = {2210.04350},
 journal = {\aapr},
 month = {December},
 number = {1},
 pages = {7},
 primaryclass = {astro-ph.GA},
 title = {{The evolution of CNO elements in galaxies}},
 volume = {30},
 year = {2022}
}

@article{rothmanHITEMPHightemperatureMolecular2010,
 adsurl = {https://ui.adsabs.harvard.edu/abs/2010JQSRT.111.2139R},
 author = {{Rothman}, L.~S. and {Gordon}, I.~E. and {Barber}, R.~J. and {Dothe}, H. and {Gamache}, R.~R. and {Goldman}, A. and {Perevalov}, V.~I. and {Tashkun}, S.~A. and {Tennyson}, J.},
 doi = {10.1016/j.jqsrt.2010.05.001},
 journal = {\jqsrt},
 month = {October},
 pages = {2139-2150},
 title = {{HITEMP, the high-temperature molecular spectroscopic database}},
 volume = {111},
 year = {2010}
}

@article{ruffioDataPostprocessingGain2026,
 adsurl = {https://ui.adsabs.harvard.edu/abs/2026arXiv260105598R},
 archiveprefix = {arXiv},
 author = {{Ruffio}, Jean-Baptiste and {Pueyo}, Laurent},
 doi = {10.48550/arXiv.2601.05598},
 eid = {arXiv:2601.05598},
 eprint = {2601.05598},
 journal = {arXiv e-prints},
 month = {January},
 pages = {arXiv:2601.05598},
 primaryclass = {astro-ph.IM},
 title = {{Data post-processing gain resulting from the patchy nature of speckles}},
 year = {2026}
}

\onecolumn
\appendix

\clearpage
\section{Best-fit model spectrum for 2M 0415}
\begin{figure}[htbp]
    \centering
    \includegraphics[width=\textwidth]{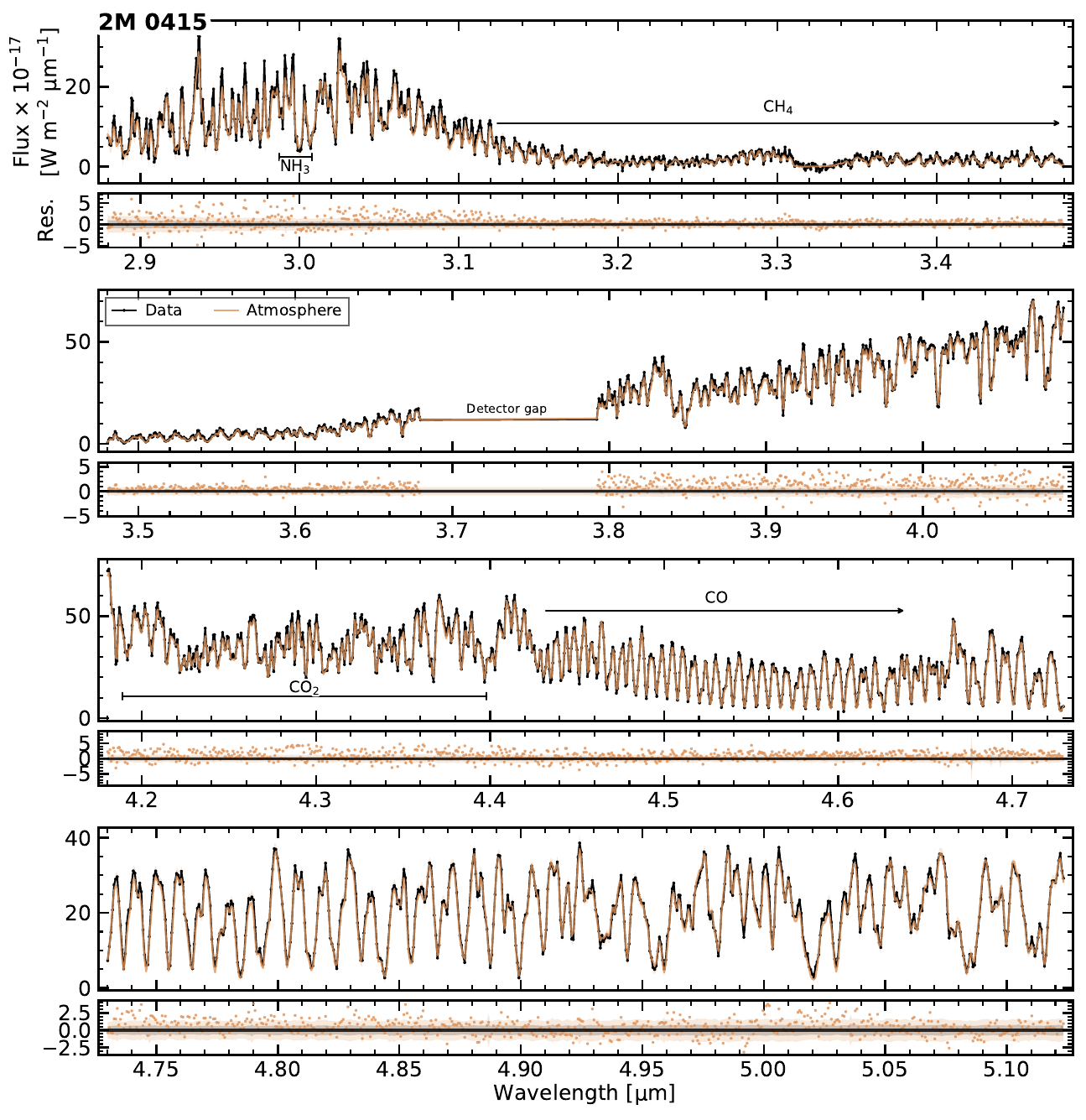}
    \caption{Best-fit model spectrum for 2M 0415.}
    \label{fig:best_fit_model_spectrum_2M0415}
\end{figure}

\clearpage
\section{Cross-correlation functions}
\begin{figure}[ht]
    \centering
    \includegraphics[width=\textwidth]{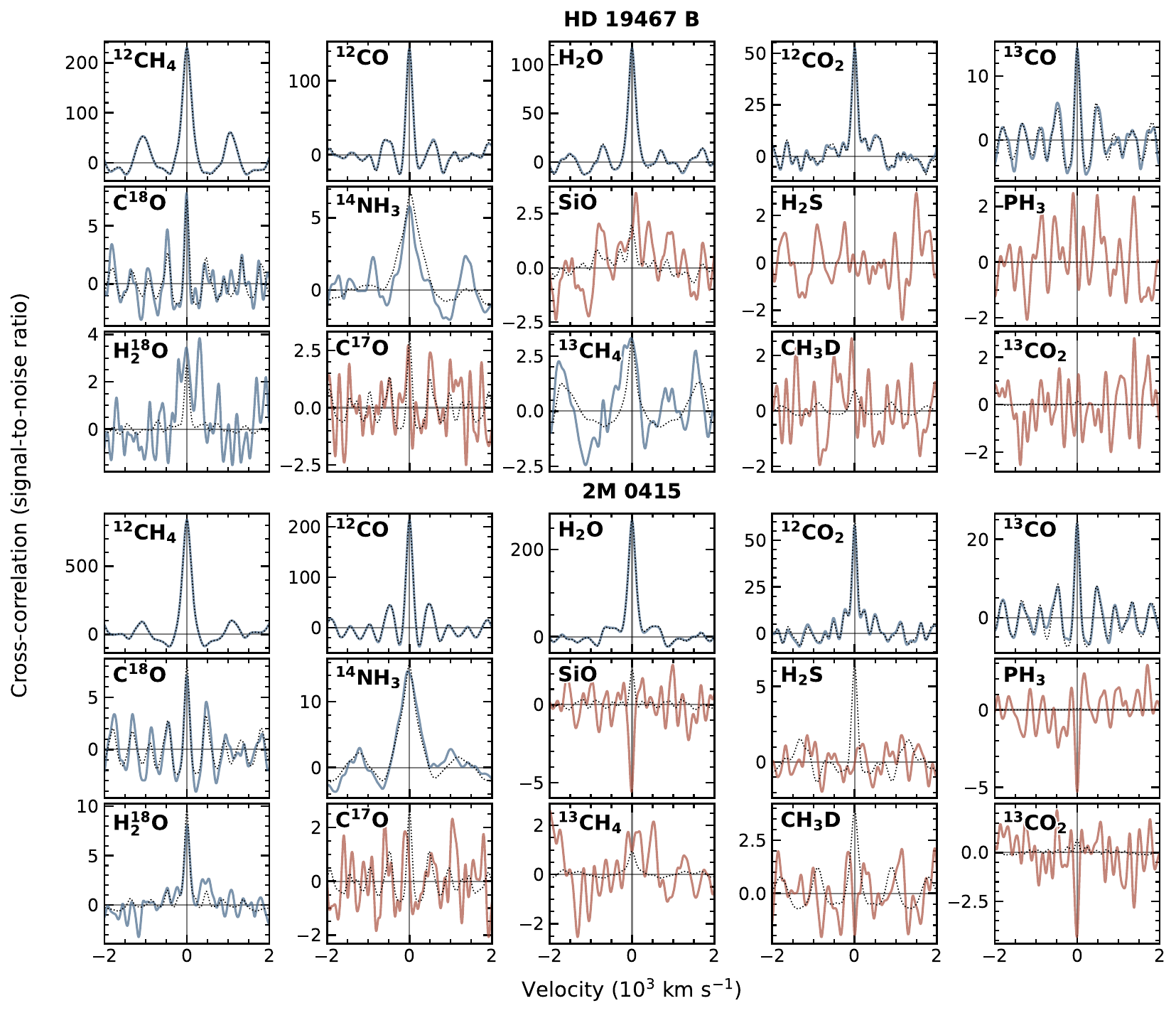}
    \caption{Cross-correlation functions for all species for HD 19467 B (top grid) and 2MASS 0415 (bottom grid). Detections (blue) are defined as peaks with S/N $> 3$, while non-detections (brown) fall below this threshold. Dotted lines show the model auto-correlation function.}
    \label{fig:cross_correlation_molecules}
\end{figure}

\clearpage
\section{Retrieved parameters}
\begin{table}[H]
    \centering
    \caption{Retrieved parameters for HD 19467 B and 2M 0415.}
    \label{tab:HD19467B_J0415_parameters}
    {\renewcommand{\arraystretch}{1.3}%
    \begin{tabular}{@{}lccc@{}}
    \toprule
    Parameter & Prior & HD 19467 B & 2M 0415 \\
    \midrule
    $R$ [R$_{\rm Jup}$] & $[0.4, 1.2]$ & $0.790_{-0.013}^{+0.012}$ & $0.846_{-0.005}^{+0.005}$ \\
    $b$ (noise scaling) & $[0, 2]$ & $0.194_{-0.007}^{+0.007}$ & $0.450_{-0.005}^{+0.005}$ \\
    $\log g$ & $[4, 6]$ & $\cdots$ & $4.97_{-0.08}^{+0.11}$ \\
    $M$ [M$_{\rm Jup}$] & $\mathcal{N}(71.6, 5.0)$ & $[66, 75]$ & $\cdots$ \\
    $v_{\rm rad,NRS1}$ [km/s] & $[-100, 100]$ & $1.2_{-0.9}^{+0.9}$ & $38.5_{-1.1}^{+1.2}$ \\
    $v_{\rm rad,NRS2}$ [km/s] & $[-100, 100]$ & $6.1_{-0.4}^{+0.5}$ & $42.41_{-0.23}^{+0.22}$ \\
    $T_0$ [K] & $[1000, 5000]$ & $2546_{-194}^{+180}$ & $2308_{-89}^{+85}$ \\
    $\log P_{\rm RCE}$ & $[-2, 1]$ & $-0.1_{-0.7}^{+0.2}$ & $-0.55_{-0.16}^{+0.28}$ \\
    $\Delta\log P$ & $[0.2, 1.8]$ & $>0.9$ & $>1.33$ \\
    $\nabla_{T,0}$ & $[0, 0.42]$ & $>0.17$ & $>0.24$ \\
    $\nabla_{T,1}$ & $[0, 0.42]$ & $0.29_{-0.05}^{+0.06}$ & $0.372_{-0.018}^{+0.017}$ \\
    $\nabla_{T,2}$ & $[0, 0.42]$ & $0.267_{-0.024}^{+0.028}$ & $0.248_{-0.009}^{+0.006}$ \\
    $\nabla_{T,3}$ & $[0, 0.42]$ & $<0.21$ & $0.251_{-0.008}^{+0.007}$ \\
    $\nabla_{T,4}$ & $[0, 0.42]$ & $<0.15$ & $[0.11, 0.35]$ \\
    $\nabla_{T,5}$ & $[0, 0.42]$ & $[0.12, 0.37]$ & $>0.18$ \\
    $\nabla_{T,6}$ & $[0, 0.42]$ & $<0.28$ & $[0.10, 0.35]$ \\
    $\log$ H$_2^{16}$O & $[-12, -1]$ & $-2.88_{-0.04}^{+0.05}$ & $-2.90_{-0.08}^{+0.09}$ \\
    $\log$ $^{12}$CO & $[-12, -1]$ & $-3.88_{-0.05}^{+0.05}$ & $-4.58_{-0.08}^{+0.09}$ \\
    $\log$ $^{12}$CH$_4$ & $[-12, -1]$ & $-3.28_{-0.04}^{+0.04}$ & $-3.17_{-0.08}^{+0.10}$ \\
    $\log$ $^{12}$CO$_2$ & $[-12, -1]$ & $-6.49_{-0.04}^{+0.04}$ & $-7.22_{-0.08}^{+0.09}$ \\
    $\log$ $^{14}$NH$_3$ & $[-12, -1]$ & $-5.05_{-0.08}^{+0.07}$ & $-5.11_{-0.09}^{+0.11}$ \\
    $\log$ H$_2$S & $[-12, -1]$ & $<-6.9$ & $-4.51_{-0.08}^{+0.09}$ \\
    $\log$ PH$_3$ & $[-12, -1]$ & $<-8.8$ & $<-8.4$ \\
    $\log$ SiO & $[-12, -1]$ & $-4.97_{-0.16}^{+0.14}$ & $-5.73_{-0.29}^{+0.20}$ \\
    $\log$ $^{12}$CO/$^{13}$CO & $[1, 3.4]$ & $2.18_{-0.05}^{+0.05}$ & $1.930_{-0.025}^{+0.027}$ \\
    $\log$ $^{12}$CO/C$^{18}$O & $[1, 3.4]$ & $2.79_{-0.08}^{+0.09}$ & $2.67_{-0.05}^{+0.06}$ \\
    $\log$ $^{12}$CO/C$^{17}$O & $[1, 3.4]$ & $>3.20$ & $>3.18$ \\
    $\log\,\mathrm{H}_2^{16}\mathrm{O}/\mathrm{H}_2^{18}\mathrm{O}$ & $[1, 3.4]$ & $2.68_{-0.13}^{+0.14}$ & $2.74_{-0.06}^{+0.07}$ \\
    $\log$ $^{12}$CH$_4$/$^{13}$CH$_4$ & $[1, 3.4]$ & $1.76_{-0.14}^{+0.19}$ & $2.6_{-0.2}^{+0.3}$ \\
    $\log$ $^{12}$CH$_4$/CH$_3$D & $[1, 3.4]$ & $>2.95$ & $>3.362$ \\
    $\log$ $^{12}$CO$_2$/$^{13}$CO$_2$ & $[1, 3.4]$ & $>2.2$ & $2.1_{-0.2}^{+0.4}$ \\
    \midrule
    \multicolumn{4}{c}{Derived parameters} \\
    \midrule
    $[\rm{C/H}]$ & $\cdots$ & $0.14_{-0.04}^{+0.04}$ & $0.15_{-0.07}^{+0.09}$ \\
    $[\rm{O/H}]$ & $\cdots$ & $0.18_{-0.04}^{+0.05}$ & $0.11_{-0.08}^{+0.09}$ \\
    $\rm{C/O}$ & $\cdots$ & $0.389_{-0.012}^{+0.012}$ & $0.451_{-0.006}^{+0.005}$ \\
    $v_{\rm rad}$ [km/s] & $\cdots$ & $3.7_{-0.5}^{+0.5}$ & $40.4_{-0.5}^{+0.6}$ \\
    $T_{\rm eff}$ [K] & $\cdots$ & $1081_{-29}^{+28}$ & $728_{-8}^{+8}$ \\
    $\log(L/L_{\odot})$ & $\cdots$ & $-5.09_{-0.04}^{+0.03}$ & $-5.719_{-0.018}^{+0.016}$ \\
    $\log g$ & $\cdots$ & $5.45_{-0.03}^{+0.03}$ & $\cdots$ \\
    \bottomrule
    \end{tabular}
    \vspace{0.6ex}
    \begin{minipage}{\linewidth}
    \footnotesize
    Notes. For HD~19467~B we adopt the dynamical mass as a Gaussian prior, $\mathcal{N}(71.6, 5.0)\,M_{\rm Jup}$ \citep{hochJWSTTSTHighContrast2024}, and derive \logg{} from the retrieved radius and mass. For 2M~0415 \logg{} is a free parameter and mass is derived. $b$ is the noise scaling factor (lower values indicate a better fit). Derived parameters are computed from posterior samples of the free parameters, and uncertainties denote the corresponding posterior intervals. One-sided limits ($<$ or $>$) indicate posteriors that rail against a prior boundary (prior-limited). Bracketed ranges indicate quantities that rail against both prior walls and are weakly constrained within the prior. The C/O and [O/H] represent bulk abundances, empirically corrected for oxygen condensation following \citet{calamariPredictingCloudConditions2024}. Elemental ratios [X/H] are logarithmic and Solar-normalised as described in \cite{asplundChemicalMakeupSun2021}.
    \end{minipage}
    }
\end{table}

\clearpage
\section{Posterior distributions of HD 19467 B}
\begin{figure}[htbp]
    \centering
    \includegraphics[width=\textwidth]{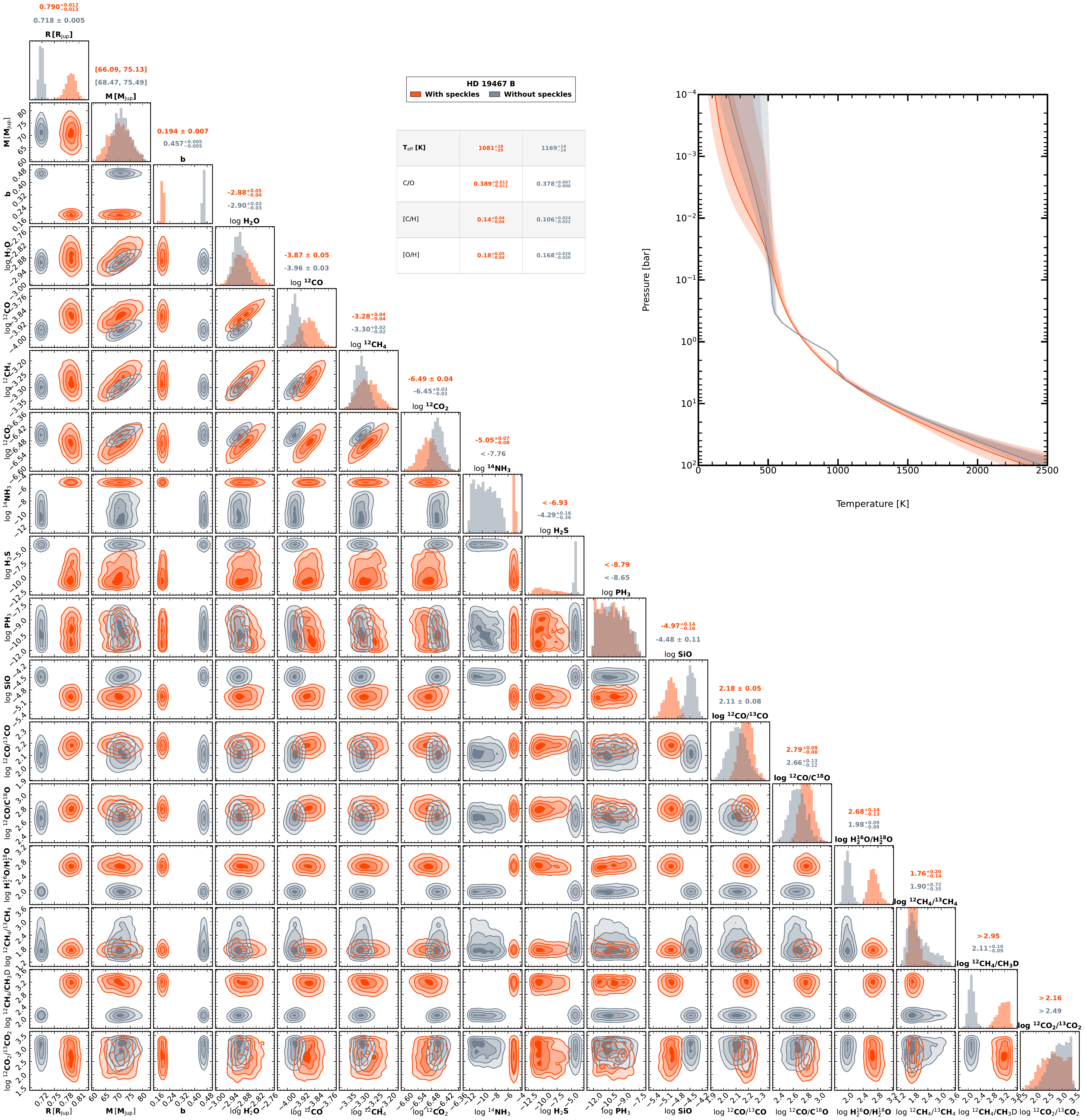}
    \caption{Posterior distributions of selected parameters for HD~19467~B, comparing the default retrieval (with dynamical-mass prior and speckle component) with a variant that omits the speckle component. Titles indicate the 16th, 50th and 84th percentiles of the 1D marginals; 2D contours show the 0.5, 1.0, 1.5 and 2.0-$\sigma$ credible intervals for a 2D Gaussian distribution. The upper-right panel shows the corresponding pressure--temperature profiles with their 1, 2, 3-$\sigma$ envelopes, and the inset table summarises the derived parameters.}
    \label{fig:cornerplot_hd19467b}
\end{figure}

\clearpage
\section{Posterior distributions of 2M 0415}
\begin{figure}[htbp]
    \centering
    \includegraphics[width=\textwidth]{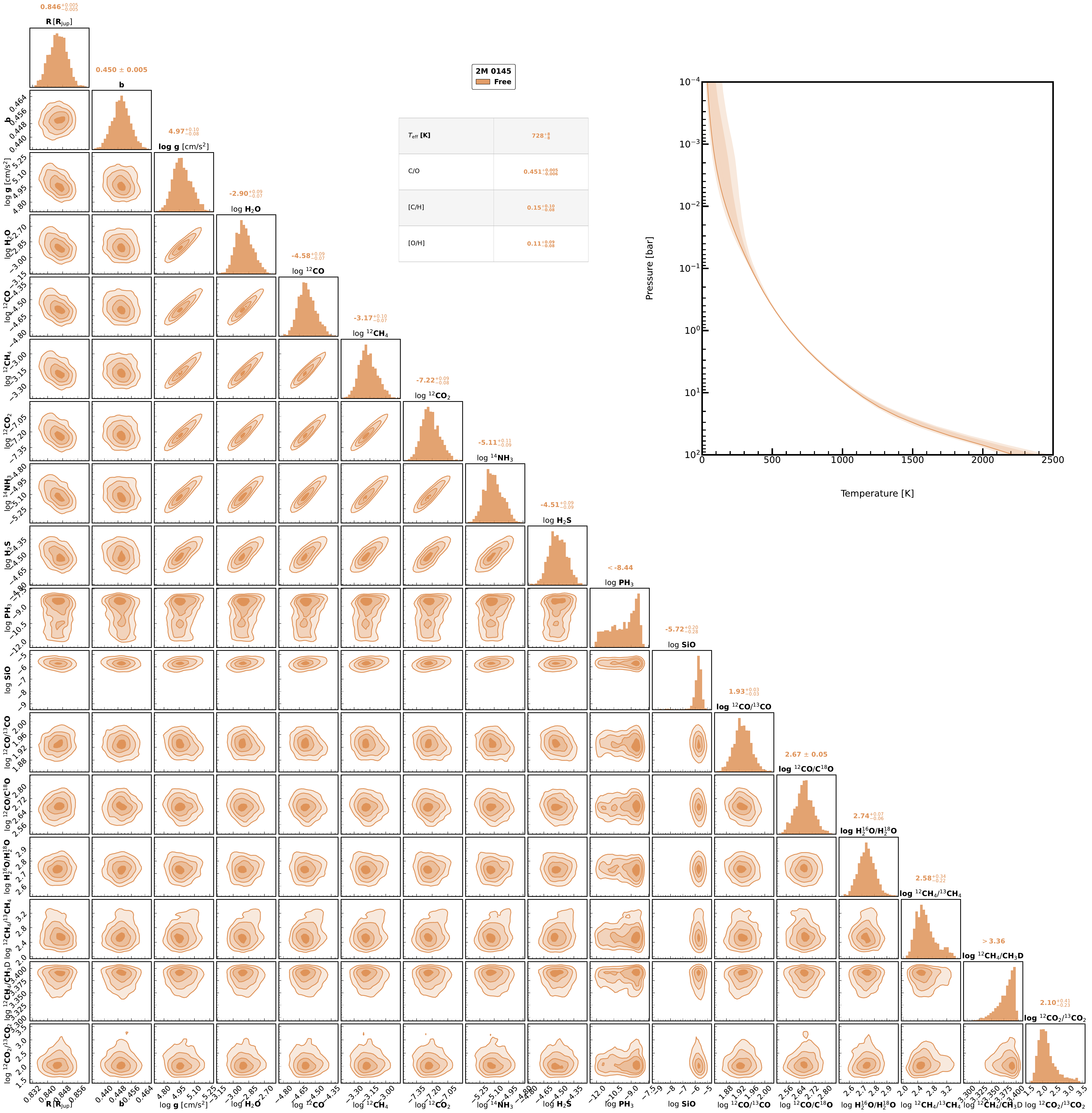}
    \caption{As \Cref{fig:cornerplot_hd19467b}, but for 2M~0415 (single retrieval, with \logg\ as a free parameter).}
    \label{fig:cornerplot_j0415}
\end{figure}

\clearpage
\section{Radial velocity}\label{sec:radial_velocity}
The high-quality observations and broad wavelength coverage allow us to measure radial velocities even at the modest resolving power of NIRSpec (\(R \sim 2700\)). In practice, the achievable precision is limited by uncertainties in the wavelength calibration for each detector, as noted by \citet{ruffioJWSTTSTHighContrast2023}. We therefore retrieve an independent radial velocity for each detector (NRS1 and NRS2), achieving typical precisions of \(\sim 1~\mathrm{km~s^{-1}}\) (NRS1) and \(0.5~\mathrm{km~s^{-1}}\) (NRS2). The NRS2 values are more precise, likely owing to its higher signal-to-noise ratio; however, we find a systematic offset of \(\sim 4\)–\(5~\mathrm{km~s^{-1}}\) between the two detectors. The mean radial velocities (with standard deviations across detectors) are:
\begin{equation}
    \begin{aligned}
        \rm RV_{\rm HD 19467 B} &= 3.6 \pm 2.5 \text{ km s}^{-1},\\
        \rm RV_{\rm 2M 0415} &= 40.2 \pm 2.2 \text{ km s}^{-1}.
    \end{aligned}
\end{equation}
For HD~19467~B, the detector-level radial velocities (NRS1: \(1.2 \pm 0.9~\mathrm{km~s^{-1}}\); NRS2: \(6.1^{+0.5}_{-0.4}~\mathrm{km~s^{-1}}\)) are consistent with the values reported by \citet{ruffioJWSTTSTHighContrast2023} for the same dataset (their Figure~21). They also broadly overlap the range of radial velocities inferred by \citet{hochJWSTTSTHighContrast2024}, with some model-dependent differences among their self-consistent grid fits (their Figures~12--16).

For 2M~0415, \citet{merchanDiversityColdWorlds2025} report \(v_{\rm rad}=47.1\pm1.8~\mathrm{km~s^{-1}}\) from the same NIRSpec/G395H dataset, but do not discuss the inter-detector offset. Our mean value (\(\rm RV_{\rm 2M~0415}=40.2\pm2.2~km~s^{-1}\); Eq.~above) is lower than both detector-level measurements at the \(\sim2\sigma\) level and also lower than the high-resolution measurement \(51.1\pm1.8~\mathrm{km~s^{-1}}\) from \citet{hsuBrownDwarfKinematics2021} (\(\sim3\sigma\)). We therefore caution that JWST/NIRSpec radial velocities can be limited by systematic wavelength-calibration uncertainties and detector-dependent offsets \citep{ruffioJWSTTSTHighContrast2023}.

\section{\texttt{ultranest} and \texttt{PyMultiNest} comparison for HD 19467 B}
\begin{figure}[htbp]
    \centering
    \includegraphics[width=\textwidth]{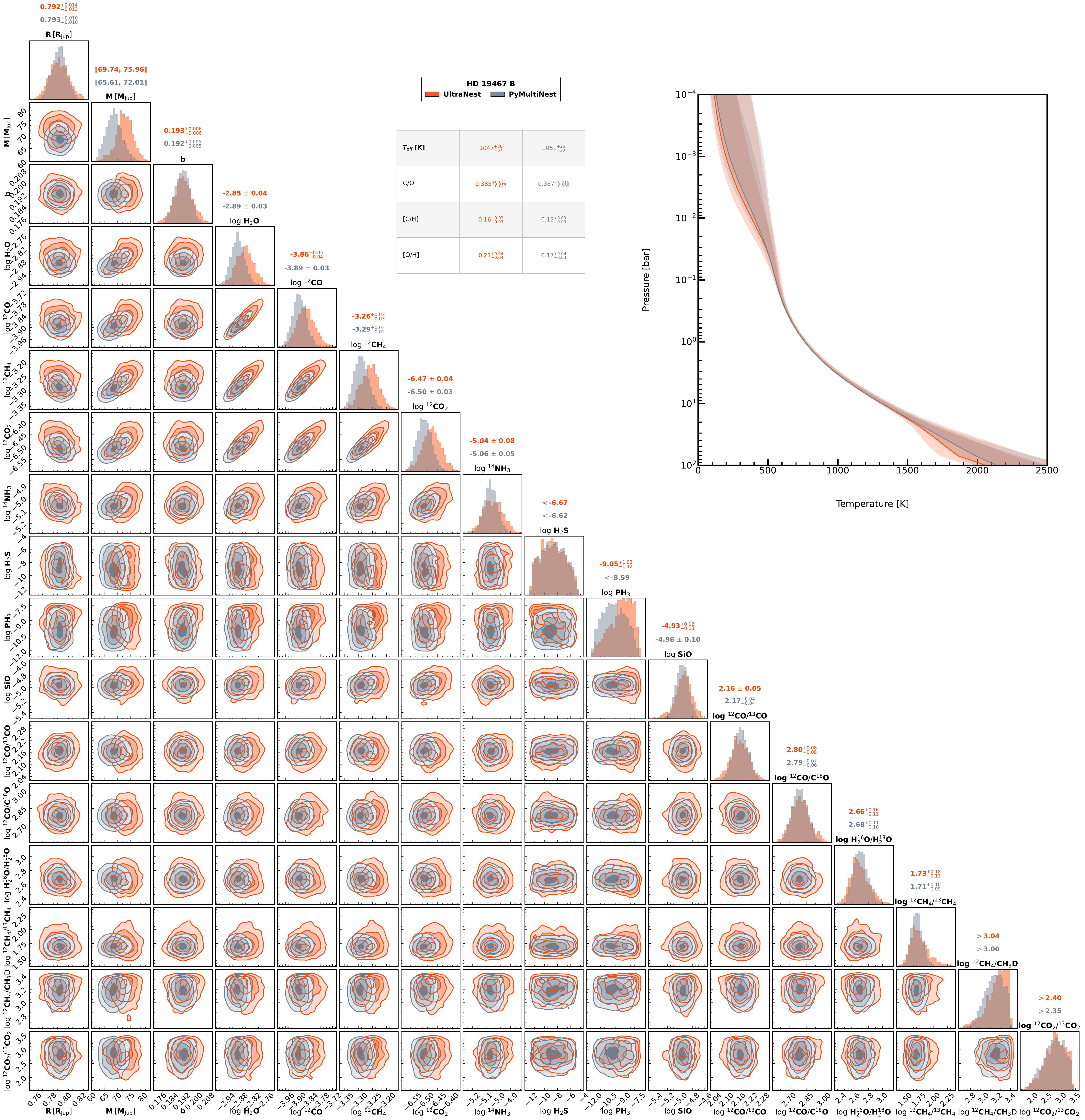}
    \caption{As \Cref{fig:cornerplot_hd19467b}, but comparing the \texttt{ultranest} and \texttt{PyMultiNest} samplers for the HD~19467~B default retrieval. Both samplers used 1000 live points and $\Delta\ln Z=0.5$; \texttt{ultranest} with a slice sampler ($n_{\rm steps}=n_{\rm dim}=30$) and \texttt{PyMultiNest} with a constant sampling efficiency of 5\%. This benchmark was run with an earlier setup than the final one, so the central values differ slightly from the adopted results quoted in the main text and \Cref{tab:HD19467B_J0415_parameters}.}
    \label{fig:corner_comparison_hd19467b_pmn}
\end{figure}

Nested sampling (NS) is widely used in atmospheric retrievals to explore high-dimensional parameter spaces and compute Bayesian evidences. \texttt{PyMultiNest} \citep{buchnerPyMultiNestPythonInterface2016} has been used extensively in the exoplanet literature (e.g. \citealt{aldersonEarlyReleaseScience2023,landmanPictorisEyesUpgraded2024}); in practice, many applications enable constant-efficiency mode to reduce runtime. This setting may bias the sampling, as it preferentially replaces live points in a way that can discard relevant posterior mass in complex problems. We therefore benchmark \texttt{PyMultiNest} against \texttt{ultranest} \citep{buchnerUltraNestRobustGeneral2021}, which provides modern step samplers designed to improve robustness (at the cost of speed). For the HD~19467~B free retrieval with mass-prior, we find consistent posteriors between the two samplers (\Cref{fig:corner_comparison_hd19467b_pmn}), with \texttt{PyMultiNest} yielding slightly tighter constraints; this may reflect an underestimation of uncertainties in constant-efficiency mode. This validation supports our use of NS for atmospheric retrievals, where the computational bottleneck is forward modelling (at \(R=100{,}000\) across 2.87--5.28~\micron). Further gains will likely come from faster radiative transfer or surrogate models (e.g. neural-network emulators; \citealt{tingPayneSelfconsistentInitio2019}) and amortised inference (e.g. neural posterior estimation; \citealt{vasistNeuralPosteriorEstimation2023}).

\end{document}